\documentclass[twocolumn,aps,superscriptaddress]{revtex4}
\usepackage{amsmath}
\usepackage{amssymb}
\usepackage{graphicx}
\usepackage{color}

\newcommand{\be}{\begin{equation}}
\newcommand{\ee}{\end{equation}}
\newcommand{\bea}{\begin{eqnarray}}
\newcommand{\eea}{\end{eqnarray}}
\newcommand{\bse}{\begin{subequations}}
\newcommand{\ese}{\end{subequations}}

\newcommand{\tcs}{${\rm ThCr_2Si_2}$}

\newcommand{\sca} {${\rm SrCo_2As_2}$}

\newcommand{\bca} {${\rm BaCo_2As_2}$}

\newcommand{\ecaa} {${\rm EuCo}_{2-y}{\rm As_2}$}
\newcommand{\ecp} {${\rm EuCo_2P_2}$}
\newcommand{\csca}{Ca$_{1-x}$Sr$_x$Co$_{2-y}$As$_2$}

\newcommand{\ecnaa}{${\rm Eu(Co}_{1-x}{\rm Ni}_x)_2{\rm As}_2$}
\newcommand{\ecna}{Eu(Co$_{1-x}$Ni$_x$)$_{2-y}$As$_2$}
\newcommand{\scna}{Sr(Co$_{1-x}$Ni$_x$)$_2$As$_2$}
\newcommand{\ena}{EuNi$_{1.95}$As$_2$}
\newcommand{\enaa}{EuNi$_{2-y}$As$_2$}

\begin{document}

\title{Magnetic phase transitions in Eu(Co$_{1-x}$Ni$_x$)$_{2-y}$As$_2$ single crystals}

\author{N. S. Sangeetha}
\affiliation{Ames Laboratory, Ames, Iowa 50011, USA}
\author{Santanu Pakhira}
\affiliation{Ames Laboratory, Ames, Iowa 50011, USA}
\author{D. H. Ryan}
\affiliation{Physics Department and Centre for the Physics of Materials, McGill University, 3600 University Street, Montreal, Quebec, H3A 2T8, Canada}
\author{V. Smetana}
\affiliation{Department of Materials and Environmental Chemistry, Stockholm University, Svante Arrhenius v\"ag 16 C, 106 91 Stockholm, Sweden}
\author{A.-V. Mudring}
\affiliation{Department of Materials and Environmental Chemistry, Stockholm University, Svante Arrhenius v\"ag 16 C, 106 91 Stockholm, Sweden}
\author{D. C. Johnston}
\affiliation{Ames Laboratory, Ames, Iowa 50011, USA}
\affiliation{Department of Physics and Astronomy, Iowa State University, Ames, Iowa 50011, USA}

\date{\today}

\begin{abstract}
The effects of Ni doping in \ecna\ single crystals with $x=0$ to~1 grown out of self flux are investigated via crystallographic, electronic transport, magnetic, and thermal measurements.  All compositions adopt the body-centered-tetragonal \tcs\ structure with space group $I4/mmm$. We also find 3--4\% of randomly-distributed vacancies on the Co/Ni site.  Anisotropic magnetic susceptibility $\chi_\alpha\,(\alpha=ab,\,c)$ data versus temperature~$T$ show clear signatures of an antiferromagnetic (AFM) $c$-axis helix structure associated with the Eu$^{+2}$ spins-7/2 for $x=0$ and~1 as previously reported.  The $\chi_\alpha(T)$ data for $x=0.03$ and~0.10 suggest an anomalous $2q$ magnetic structure containing two helix axes along the $c$~axis and in the $ab$~plane, respectively, whereas for $x = 0.75$ and~0.82, a $c$-axis helix is inferred as previously found for $x=0$ and~1.  At intermediate compositions $x=0.2$, 0.32, 0.42, 0.54, and 0.65 a magnetic structure with a large ferromagnetic (FM) $c$-axis component is found from magnetization versus field isotherms, suggested to be an incommensurate FM $c$-axis cone structure associated with the Eu spins, which consists of both AFM and FM components.  In addition, the $\chi(T)$ and heat capacity~$C_{\rm p}(T)$ data for $x=0.2$--0.65 indicate the occurrence of itinerant FM order associated with the Co/Ni atoms with Curie temperatures from 60 to 25~K, respectively.  Electrical resistivity $\rho(T)$ measurements indicate metallic character for all compositions with  abrupt increases in slope on cooling below the Eu AFM transition temperatures.  In addition to this panoply of magnetic transitions, $^{151}$Eu M\"ossbauer measurements indicate that ordering of the Eu moments proceeds via an incommensurate sine amplitude-modulated structure with additional transition temperatures associated with this effect.

\end{abstract}

\maketitle

\section{Introduction}

Many $RM_2X_2$ (abbreviated as 122-type) ternary compounds ($R=$ rare-earth or alkaline-earth metal, $M =$  transition metal, $X =$ pnictogen), crystallize in the body-centered tetragonal \tcs\ structure with space group $I4/mmm$.  They have been widely investigated owing to a variety of remarkable physical properties such as unconventional superconductivity, heavy fermion behavior, valence fluctuations, quantum criticality, and different types of magnetic transitions~\cite{Johnston2010, Canfield2010, Stewart2011, Scalapino2012, Fernandes2014, Dai2015, Inosov2016, Si2016, Pfisterer1980, Pfisterer1983}. The 122-type compounds are quite sensitive to the peculiarities of electronic band structure and the Fermi surface. In particular, tuning the system either by external stimuli (pressure, temperature) or chemical substitution leads to a variety of ground states~\cite{Singh2009, Pandey2012, Anand2014, Singh2009a, Sangeetha2016, Das2017, Jayasekara2017, Pandey2013, Sangeetha2017, Sangeetha2017a, Das2017a}. For example, ${\rm EuFe_2As_2}$ exhibits localized Eu$^{+2}$ spin-7/2 A-type antiferromagnetism below 19~K and an itinerant antiferromagnetic (AFM) spin-density-wave (SDW) ordering accompanied by a tetragonal-to-orthorhombic structural phase transition below 190~K\@. Superconductivity can be induced in ${\rm EuFe_2As_2}$ by suppressing the SDW ordering either by chemical substitution or by applying external pressure~\cite{Marchand1978, Xiao2009, Jeevan2008, Ren2009, Jiang2009, Miclea2009, Jeevan2011}.

The \tcs-type Co pnictides \ecp\ and \ecaa\ have also attracted considerable attention.  \ecp\ has an  uncollapsed-tetragonal  (ucT) structure~\cite{Anand2012}. It shows AFM ordering below $T\rm_N = 66$~K on the Eu$^{+2}$ sublattice and there is no evident contribution to the ordering from the Co atoms at ambient pressure \cite{Morsen1988}. Neutron-diffraction studies demonstrated that the AFM structure is a planar helix with the Eu ordered moments aligned in the $ab$ plane of the tetragonal structure, and with the helix axis being the $c$~axis with propagation vector ${\bf k} =  (0, 0, 0.85)(2\pi/c)$~\cite{Reehuis1992}. This compound undergoes a pressure-induced first-order ucT to collapsed-tetragonal (cT) transition at $\approx3$~GPa~ \cite{Huhnt1997}. $^{151}{\rm Eu}$ high-pressure M{\"o}ssbauer experiments ($0\leq p\leq 5$~GPa)~\cite{Chefki1998} reveal a valence transition of Eu from Eu$^{+2}$ with spin $S=7/2$ and angular momentum $L=0$ to nonmagnetic  Eu$^{+3}$ ($^7F_0$) at $p = 3$~GPa together with emergence of  itinerant 3$d$ magnetism on the Co sublattice which orders antiferromagnetically at $T\rm^ {Co}_N = 260$~K\@. The unique feature of \ecp\ is that it changes its magnetic character from Eu(4$f$) ordering to Co(3$d$) ordering depending on its lattice parameter. For this reason, \ecp\ is a model system for band-structure calculations studying 3$d$ moment formation in the \tcs-type structure~\cite{Chefki1998}. We showed that \ecp\ is a textbook example of a noncollinear helical antiferromagnet at ambient pressure for which the thermodynamic properties in the antiferromagnetic (AFM) state are well described~\cite{Sangeetha2016a} by our recent formulation of molecular field theory (MFT)~\cite{Johnston2012, Johnston2015}.

\ecaa\ is isostructural and isoelectronic to \ecp\ and exhibits AFM ordering of the Eu$^{+2}$ spins-7/2 below $T\rm_N = 47$~K~\cite{Marchand1978, Raffius1993, Bishop2010, Ballinger2012, Tan2016, Sangeetha2018}. Neutron-diffraction measurements revealed that the AFM structure is the same coplanar $c$-axis helical structure as in \ecp\ with propagation vector ${\bf k} = (0,0,0.79)(2\pi/c)$ and no evident participation of Co in the ordering~\cite{Tan2016}.

It is interesting to note the differences between \ecp\ and \ecaa\ as a consequence of their different Co-P and Co-As networks. Their $c/a$ ratios are similar, 3.01 for \ecp\ and 2.93 for \ecaa. However, \ecaa\ exhibits a continuous ucT to cT transition under applied pressure $p\approx5$~GPa that leads to an intermediate-valence state Eu$^{+2.25}$  at 12.6~GPa~\cite{Bishop2010}.  As a result, the electronic structure change eliminates the AFM ordering of the Eu spin sublattice and ferromagnetic (FM) ordering then arises in which both Eu (4$f$) and Co (3$d$) moments participate with a Curie temperature $T\rm_C = 125$~K, as confirmed by x-ray magnetic circular dichroism measurements and electronic structure calculations~\cite{Tan2016}.

Conventionally, the paramagnetic (PM) effective moment of Eu$^{+2}$ with spin $S= 7/2$ and spectroscopic splitting factor $g=2$ is calculated to be $\mu\rm_{eff}\approx 7.94~\mu_{\rm B}$/Eu. Instead, crystals of the helical antiferromagnet  \ecaa\ exhibit an enhanced effective moment $\mu\rm_{eff}\approx8.5~\mu_{\rm B}$/Eu, where $\mu_{\rm B}$ is the Bohr magneton~\cite{Sangeetha2018}.   This enhancement is consistent with the occurrence of an enhancement of the low-$T$ ordered moment to $7.26(8)\mu_{\rm B}$ found in the neutron diffraction study~\cite{Tan2016}.   {\it Ab~initio} calculations showed that the enhancements arise from FM polarization of the conduction electron spins around each Eu spin~\cite{Sangeetha2018}.  This was the second example of a helical antiferromagnet with thermal and magnetic properties that are fitted rather well by MFT~\cite{Johnston2012, Johnston2015}. In Ref~\cite{Sangeetha2018}, we also found a Co vacancy concentration up to $y = 0.1$ in \ecaa.

Studies of $^{153}$Eu, $^{75}$As, and $^{59}$Co NMR were also carried out on a crystal of \ecaa~\cite{Ding2017}, where NMR signals from all three isotopes were observed.  From the applied magnetic field dependence of the $^{153}$Eu and $^{75}$As NMR spectra at low temperature \mbox{$T\ll T_{\rm N} = 45$~K}, the signals were consistent with an incommensurate helical AFM structure with propagation vector \mbox{[0,\ 0, 0.73(7)]$2\pi/c$}, which is the same within the errors as the propagation vector found from the above neutron-           diffraction experiment~\cite{Tan2016}.  In addition, from  the $^{59}$Co NMR it was determined that the Co atoms do not participate in the AFM ordering, again consistent with the conclusion from the neutron-diffraction measurements~\cite{Tan2016}.  The temperature dependence of the $^{75}$As hyperfine field below $T_{\rm N}$ was fitted quite well by a Brillouin function for spin~$S=7/2$.

The electron-doped analog \enaa\ of \ecaa\ is metallic and from magnetic susceptibility and heat-capacity measurements exhibits an AFM transition at $T_{\rm N} \approx 14$--15~K~\cite{Ghadraoui1988, Bauer2008, Jin2019, Sangeetha2019a}.  The electron doping is inferred from the fact that Ni has one more $3d$~electron than Co does.  This compound has recently been found from neutron-diffraction measurements  to have a $c$-axis helical AFM structure with $T_{\rm N}=15$~K and AFM propagation vector $k = (0,0,0.9200)2\pi/c$, corresponding to a turn angle of 165.6(1)$^\circ$ along the $c$~axis between adjacent layers of Eu spins ferromagnetically aligned in the  $ab$~plane~\cite{Jin2019}.

We subsequently reported a comprehensive study of the properties of single crystals of \enaa\ \cite{Sangeetha2019a}, where the analysis of the anisotropic magnetic susceptibility $\chi$ data below $T_{\rm N}$ by MFT yielded a $c$-axis helical turn angle in good agreement with the neutron-diffraction result.  In addition, a 2.5\% vacancy concentration on the Ni sites was detected.  A good fit of the electrical resistivity $\rho(T)$ by the Bloch-Gr\"uneisen electron-phonon scattering theory was obtained.  The magnetic contribution to the heat capacity $C_{\rm p}(T)$ below $T_{\rm N}$ was compared with the MFT prediction.  The $ab$-plane isothermal magnetization $M_{ab}(H)$ exhibited two metamagnetic transitions in fields up to $H=14$~T at $T=2$~K, whereas $M_c(H)$ was linear up to the critical field, both in agreement with MFT\@.  A review of the crystallographic and magnetic properties of \tcs-type pnictides ($Pn$) Eu$M_2Pn_2$  was also presented in Ref.~\cite{Sangeetha2019a} in which only ferromagnetic and \mbox{$c$-axis} helical AFM structures are found to be present.  Here, an \mbox{A-type} antiferromagnet with FM $ab$-plane ordered-moment alignment was considered to be a $c$-axis helix with a 180$^\circ$ turn angle between the ordered moments in adjacent $ab$-plane layers.

In this work we report a comprehensive single-crystal study of the mixed system \ecna\ including the crystallographic, magnetic, thermal, and electronic-transport properties obtained using single-crystal x-ray diffraction (XRD), $\chi(T)$, $M(H)$, $^{151}$Eu M\"ossbauer, $C_{\rm p}(T)$, and $\rho(T)$ measurements.  Our major results are the discoveries of itinerant FM ordering associated with the Co/Ni spins and of a structure with both FM and AFM components at intermediate compositions from $M(H)$, $\chi(T)$, and $C_{\rm p}(T)$ measurements that we suggest is an Eu-spin cone phase, and the discovery of unexpected magnetic behaviors on approaching $T_{\rm N}$ in both undoped and Ni-doped \ecna\ from $^{151}$Eu M\"ossbauer measurements.

The experimental details are given in Sec.~\ref{Sec:ExpDetails}. The crystallographic data and composition analyses are presented in Sec.~\ref{Sec:crystalstudy}. The low-field $M(H)$ and the $\chi(T)$ data are presented and analyzed in Sec.~\ref{Sec:mag}, high-field $M(H)$ isotherms in Sec.~\ref{Sec:HighFieldM(H)}, $^{151}$Eu M\"ossbauer measurements  in Sec.~\ref{Sec:Mossbauer}, the $C_{\rm p}(T)$ data in Sec.~\ref{Sec:HC}, and the $\rho(T)$ measurements in Sec.~\ref{Sec:Res}.  Concluding remarks are given in Sec.~\ref{Sec:Summary}.

\section{\label{Sec:ExpDetails} Experimental Details}

Single crystals of  \ecna\ with measured compositions  $x = 0$, 0.03, 0.10, 0.20, 0.32, 0.42, 0.54, 0.65, 0.75, 0.82, and 0.93 were grown using self-flux. The starting materials were high-purity elemental Eu (Ames Laboratory) and Co (99.998\%), Ni (99.999\%)  and As (99.99999\%) (Alfa Aesar). The sample and flux  were taken in a Eu : 4($1-x$) CoAs : $4x$ NiAs molar ratio and placed in an alumina crucible that was sealed under $\approx 1/4$~atm high-purity argon in a silica tube. The sealed samples were preheated at 600~$^{\circ}$C for 5~h, and then heated to 1300~$^{\circ}$C at a rate of 50~$^{\circ}$C/h and held there for 15~h for homogenization. Then the furnace was slowly cooled at the rate of 6~$^{\circ}$C/h to 1180~$^{\circ}$C. The single crystals were separated by decanting the flux with a centrifuge at that temperature. Several 2--4~mm size shiny platelike single crystals were obtained from each growth. 

The phase purity and chemical composition of the \ecna\ single crystals were checked using an energy-dispersive x-ray spectroscopy (EDS) chemical analysis attachment to a JEOL scanning-electron microscope (SEM). SEM scans were taken on cleaved surfaces of the crystals which verified the single-phase nature of the crystals. The compositions of each side of a platelike crystal were measured at six or seven positions on each face, and the results were averaged and revealed good homogeneity in each crystal. The results of the composition analyses are given in Table~\ref{CrystalData} below.  The same crystals were utilized to perform the physical-property measurements.
 
Single-crystal x-ray diffraction (SCXRD) measurements were performed at room temperature on a Bruker D8 Venture diffractometer operating at a tube voltage of 50~kV and a current of 1~mA equipped with a Photon 100 CMOS detector, a flat graphite monochromator and a Mo~K$_\alpha$ I$\mu$S microfocus source ($\lambda = 0.71073$~\AA). The preliminary quality testing was performed on a set of 32 frames. The raw frame data were collected using the Bruker APEX3 software package~\cite{APEX2015}. The frames were integrated with the Bruker SAINT program~\cite{SAINT2015}  using a narrow-frame algorithm integration and the data were corrected for absorption effects using the multi-scan method (SADABS)~\cite{Krause2015} within the APEX3 package. The atomic positions were refined assuming partial occupancy of the Ni/Co sites and complete occupancy of the Eu and As sites. Since simultaneous refinement of the fraction of Ni/Co and total occupation of the position is not possible by means of SCXRD, only the total occupancies were refined based on the Ni/Co ratio taken from the EDS data.  The atomic displacement parameters were refined anisotropically. Initial models of the crystal structures were first obtained with the program SHELXT-2014~\cite{Sheldrick2015A} and refined using the program SHELXL-2014~\cite{Sheldrick2015C} within the APEX3 software package.

A polycrystalline sample of \ecaa\ was obtained for $^{151}$Eu M\"ossbauer measurements by crushing several \ecaa\ crystals grown in Sn flux;  the growth and properties of these crystals are well documented in Ref.~\cite{Sangeetha2018}. Polycrystalline samples of \ecnaa\ with $x$ = 0.2 and 0.65 for the M\"ossbauer measurements  were synthesized by solid-state reaction using high-purity Eu (Ames Laboratory), and  Co (99.998\%), Ni (99.999\%), and As (99.99999\%) from Alfa Aesar.  Stoichiometric mixtures of these elements were pelletized and placed in alumina crucibles that were sealed inside evacuated silica tubes. The samples were heated to 900~$^{\circ}$C at a rate of 50~$^{\circ}$C/h, held there for 24 h, and then cooled to room temperature at a rate of 50~$^{\circ}$C/h. The samples were then thoroughly ground, pelletized, and again sealed in evacuated quartz tubes. The samples were finally sintered at 900~$^{\circ}$C for 72~h followed by cooling to room temperature with a 100~$^{\circ}$C/h temperature ramping rate. Room-temperature powder XRD measurements were carried out using a Rigaku Geigerflex x-ray diffractometer with Cu-K$_\alpha$ radiation. Structural analyses were performed by Rietveld refinement using the {\tt Fullprof} software package~\cite{Carvajal1993}.

Magnetization data were obtained using a Quantum Design, Inc., SQUID-based magnetic-properties measurement system (MPMS) in magnetic fields up to 5.5~T, where 1~T~$\equiv10^4$~Oe, and using the vibrating-sample magnetometer (VSM) option in a Quantum Design, Inc., physical-properties measurement system (PPMS) in magnetic fields up to 14~T\@. The magnetic moment output of these instruments is expressed in Gaussian cgs electromagnetic units (emu), where 1~emu = 1~G\,cm$^3$ and 1~G = 1~Oe.  The heat capacity $C_{\rm p}(H,T)$ was measured with a relaxation technique using a PPMS\@. The $\rho(H,T)$ measurements were performed using a standard four-probe ac technique using the ac-transport option of the PPMS with the current in the $ab$~plane. Annealed platinum wire (25 $\mu$m diameter) electrical leads were attached to the crystals using silver epoxy.

The $^{151}$Eu M\"ossbauer spectroscopy measurements were carried out using
a 4~GBq $^{151}$SmF$_3$ source, driven in sine mode and
calibrated using a standard $^{57}$Co\underline{Rh}/$\alpha$-Fe foil.
Isomer shifts are quoted relative to EuF$_3$ at ambient temperature.
The sample was cooled in a vibration-isolated closed-cycle helium refrigerator
with the sample in helium exchange gas. Where line broadening was not observed
(typically at the lowest temperatures and again well above the magnetic
transitions) the spectra were fitted to a sum of Lorentzian lines with the
positions and intensities derived from a full solution to the nuclear
Hamiltonian \cite{voyer}. The tetragonal 4/$mmm$ point symmetry of the Eu site in these
materials puts significant constraints on the form of the electric field
gradient (efg) tensor, requiring it to be axially symmetric, with the principal
component (usually denoted $V_{zz}$) directed along the crystallographic
$c$~axis. These features of the efg mean that we are not sensitive to rotations
of the moments around the $c$~axis, only canting of the moments away from the
$c$~axis. For example, the incommensurate coplanar $c$-axis
helical magnetic structure reported for $\rm EuCo_2As_2$~\cite{Tan2016} has
all of the moments perpendicular to the $c$~axis with rotations within the
$ab$-plane as we move along the $c$~axis. This yields a single
effective environment for the europium atoms in the $^{151}$Eu M\"ossbauer
spectrum, so a sharp single-component spectrum is observed.

\begin{table*}
\caption{\label{CrystalData}Crystallographic data for \ecna\ single crystals  ($x=0.03$--0.93) at room temperature, including the fractional $c$-axis position $z_{\rm AS}$ of the As site, the tetragonal lattice parameters $a$ and~$c$, the $c/a$ ratio, the unit cell volume $V_{\rm cell}$ containing $Z=2$ formula units of \ecna, and the molar volume $V_{\rm M}$, where a mole is a mole of formula units. The Co and Ni compositions and total Co/Ni site occupation in the first column were obtained from EDS analyses, whereas the  unit cell parameters and fractional $z_{\rm As}$ coordinates were found from single-crystal structural analyses. The present work is denoted by PW.  Corresponding data from the literature for \ecaa\ and EuNi$_{2-y}$As$_2$ are also shown.}
\begin{ruledtabular}
\begin{tabular}{ cccccccc }
Compound  									& $z_{\rm As}$ 	& 	$a$   	& 	$c$  		&  $c/a$ 		& $V_{\rm cell}$ 	& 	$V_{\rm M}$		& 	Ref. \\
											&			&	(\AA)		&	(\AA)		&			&  (\AA$^3$)		& 	$({\rm cm^3/mol})$	&		\\
\hline
	${\rm EuCo_{1.94(2)}As_2}$					& 0.3607(1) 	& 3.9478(7) 	& 11.232(2) 	& 2.845(1) 	& 175.05(7) 		& 	52.71(2)			& 	\cite{Sangeetha2018}	\\	
	${\rm Eu(Co_{0.97(1)}Ni_{0.03(1)})_{1.92(1)}As_2}$   	& 0.3609(3)  	& 3.964(8) 	& 11.220(2)  	& 2.831(6) 	& 176.3(8) 		& 	53.1(2)			&	PW\\
	${\rm Eu(Co_{0.90(1)}Ni_{0.10(1)})_{1.92(2)}As_2}$     & 0.3608(2) 	& 3.978(5) 	& 11.07(1)		& 2.782(7) 	& 175.1(5) 		& 	52.7(2)			& 	PW\\
	${\rm Eu(Co_{0.80(1)}Ni_{0.20(1)})_{1.94(2)}As_2}$     & 0.3613(1) 	& 3.994(3)  	& 10.836(8) 	& 2.713(4)		& 172.8(3) 		& 	52.03(9)			&	PW\\
	${\rm Eu(Co_{0.68(1)}Ni_{0.32(1)})_{1.94(2)}As_2}$     & 0.3628(3) 	& 4.027(2)  	& 10.573(5) 	& 2.625(2)		& 171.5(2) 		& 	51.64(6)			&	PW \\
	${\rm Eu(Co_{0.58(1)}Ni_{0.42(1)})_{1.94(1)}As_2}$     & 0.36396(4)  	& 4.046(9) 	& 10.462(2) 	& 2.585(1)		& 171.2(8) 		& 	51.5(2)			&	PW\\
	${\rm Eu(Co_{0.46(1)}Ni_{0.54(1)})_{1.94(2)}As_2}$     & 0.36424(5)	& 4.062(2)		& 10.328(5) 	& 2.543(2) 	& 170.3(2)			& 	51.28(6)			& 	PW\\
	${\rm Eu(Co_{0.35(2)}Ni_{0.65(2)})_{1.94(2)}As_2}$     & 0.36476(5)	& 4.079(2) 	& 10.268(5) 	& 2.517(2) 	& 170.9(2) 		& 	51.46(6)			&	PW\\
	${\rm Eu(Co_{0.25(2)}Ni_{0.75(2)})_{1.94(2)}As_2}$     & 0.36531(5) 	& 4.091(2)  	& 10.217(5) 	& 2.497(2) 	& 171.0(2)			& 	51.49(6)			&	PW\\
	${\rm Eu(Co_{0.18(2)}Ni_{0.82(2)})_{1.94(2)}As_2}$     & 0.3657(1) 	&  4.100(1)	& 10.155(3) 	& 2.477(1) 	& 170.7(1) 		& 	51.40(3)			&	PW\\
	${\rm Eu(Co_{0.07(1)}Ni_{0.93(1)})_{1.94(2)}As_2}$     & 0.36627(4) 	&  4.099(3)	& 10.092(9) 	& 2.462(4) 	& 169.5(3) 		& 	51.04(9)			&	PW\\
	${\rm EuNi_{1.95(1)}As_2}$       					& 0.36653(8) 	&  4.1052(8)	& 10.027(2) 	& 2.442(1) 	& 168.99(7) 		& 	50.88(2)			&	\cite{Sangeetha2019a}	\\
\end{tabular}
\end{ruledtabular}
\end{table*}
\begin{figure}

\includegraphics[width=3in]{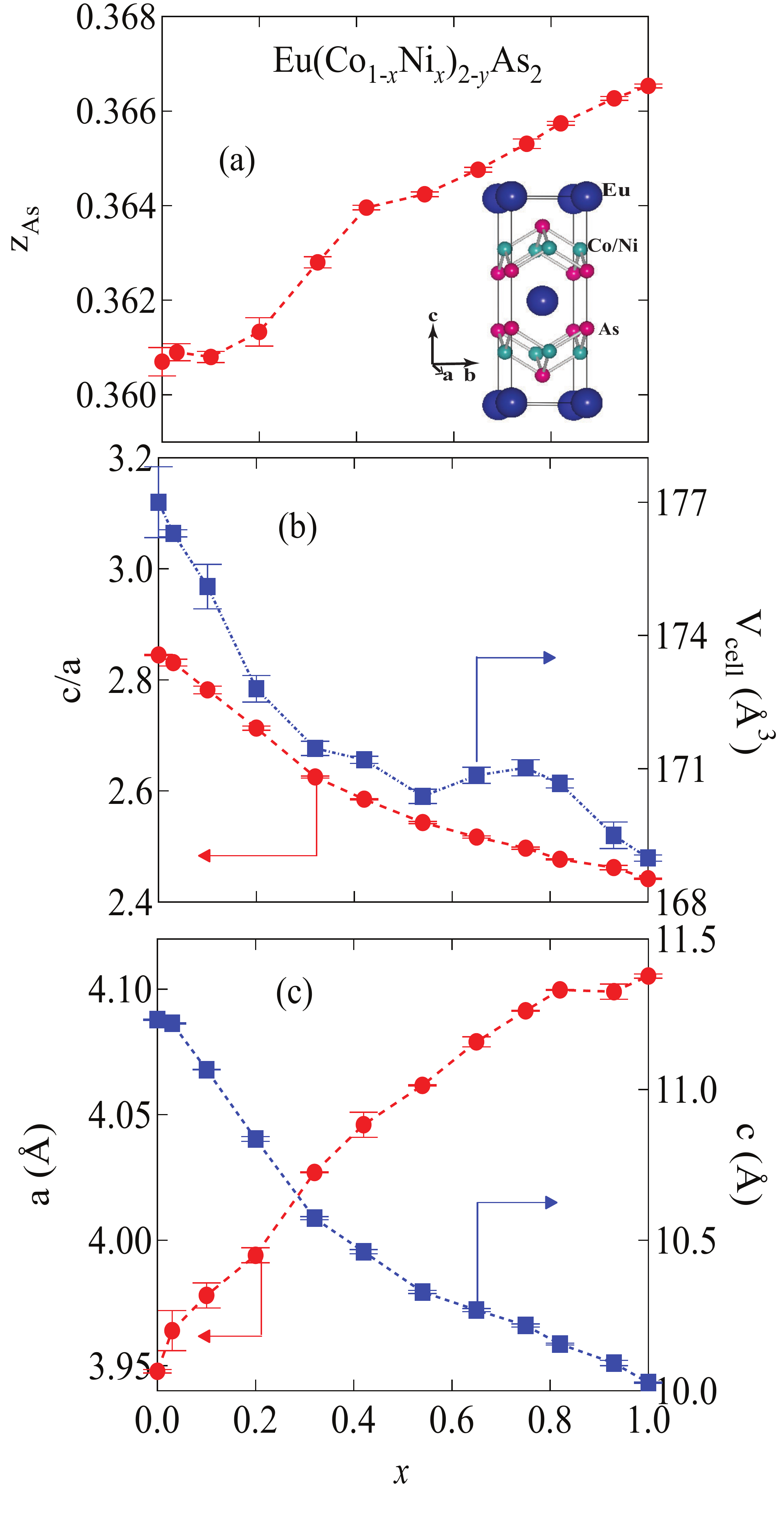}[ht]
\caption{Crystallographic parameters for \ecna\ single crystals versus composition $x$, including (a)~ the As $c$-axis parameter $z\rm_{As}$ (b) ~the $c/a$ ratio and unit-cell volume $V_{\rm cell}$, and (c)~the $a$ and~$c$ lattice parameters.  The dashed lines are guides to the eye. The inset of (a) shows a unit cell of \ecna.}
\label{latticeparameter}
\end{figure}

\section{\label{Sec:crystalstudy} Crystal structures}

\subsection{Single Crystals}

The chemical compositions and crystallographic data for the Ni-doped \ecaa\ single crystals  obtained from the single-crystal XRD and EDS measurements at room temperature are presented in Table~\ref{CrystalData}.  The  data confirm that \ecnaa\ with $x=0$ to~1 adopt the \tcs-type body-centered tetragonal structure with space group $I4/mmm$. We also confirm partial occupation of the Co/Ni site from both EDS and single-crystal XRD measurements, similar to \ecaa~\cite{Sangeetha2018}, with concentrations listed in Table~\ref{CrystalData}.

The lattice parameters $a$ and~$c$, the $c$-axis As position parameter $z_{\rm As}$, the ratio $c/a$, and the unit-cell volume $V_{\rm cell}=a^2c$ are plotted versus~$x$ in Fig.~\ref{latticeparameter}. The  $c$ lattice parameter decreases upon Ni doping, whereas the $a$ lattice parameter strongly increases with~$x$.  These behaviors give rise to the strongly nonlinear variation of $V_{\rm cell}$ with~$x$.  Furthermore, the $c/a$ ratio smoothly decreases with~$x$, whereas the $z_{\rm As}$ increases with $x$, suggesting important changes to the electronic properties. This smooth decrease in $c/a$ with $x$ indicates the occurrence of  a continuous structural crossover from the ucT structure at $x=0$ to the cT structure at $x=1$, similar to \scna~\cite{Sangeetha2019}.

The values of $c/a$ for $x=0.2$ (2.713) and $x=0.3$ (2.625) are in the vicinity of the crossover at $\approx 2.67$ between the ucT and cT structures~\cite{Anand2012}.  We therefore infer that a structural crossover occurs in \ecna\ at $x\approx 0.2$.  In the following sections we study how these structure changes due to electron doping of \ecaa\ correlate with the ground-state properties of the system.

\subsection{Polycrystalline samples}

\begin{figure}
\includegraphics[width = 3.3in]{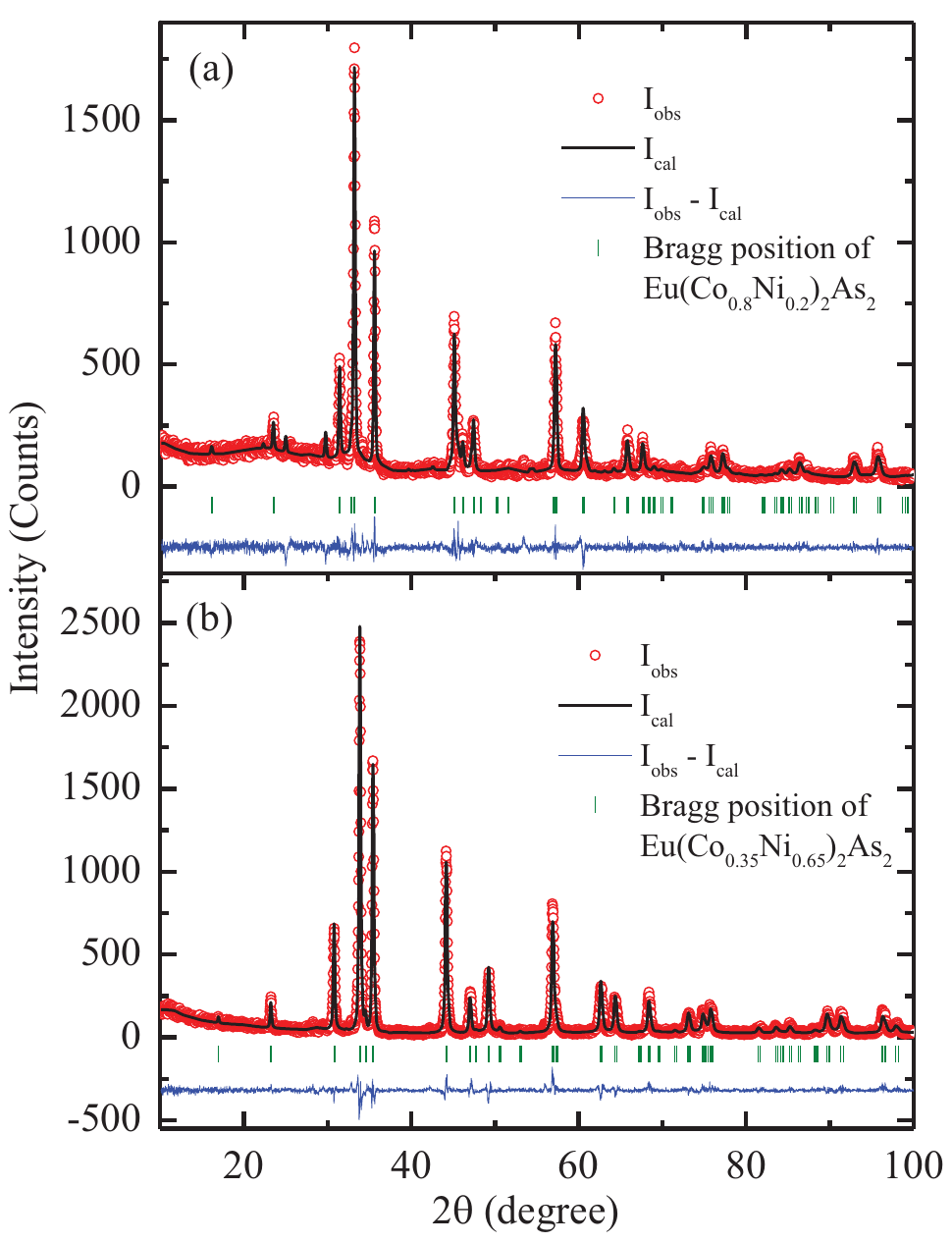}
\caption {Room temperature powder XRD patterns along with Rietveld refinement of \ecnaa\ polycrystalline compounds for (a)~$x = 0.2$ and (b)~$x = 0.65$. }
\label{Fig_XRD_polycrystal}
\end{figure}

Room-temperature powder XRD patterns of the polycrystalline \ecnaa\  samples with $x$ = 0.20 and 0.65 are shown in Figs.~\ref{Fig_XRD_polycrystal}(a) and \ref{Fig_XRD_polycrystal}(b), respectively. The Rietveld refinements confirm that both samples crystallize in the ThCr$_2$Si$_2$-type crystal structure with space group $I 4/mmm$. The lattice parameters are found to be $a = b = 4.0027(2)$ ~\AA\ and $c = 10.8636(8)$~\AA\ for $x = 0.20$, and $a = b = 4.0806(1)$\,\AA\ and  $c = 10.3261(4)$\,\AA\ for $x =0.65$, similar to the lattice parameters of the respective single crystals in Table~\ref{CrystalData}. The refinements yielded full occupancy of the Co/Ni site to within 1\% for both samples.

\section{\label{Sec:mag} Magnetic susceptibility measurements}

\begin{figure}
\includegraphics[width=3.5in]{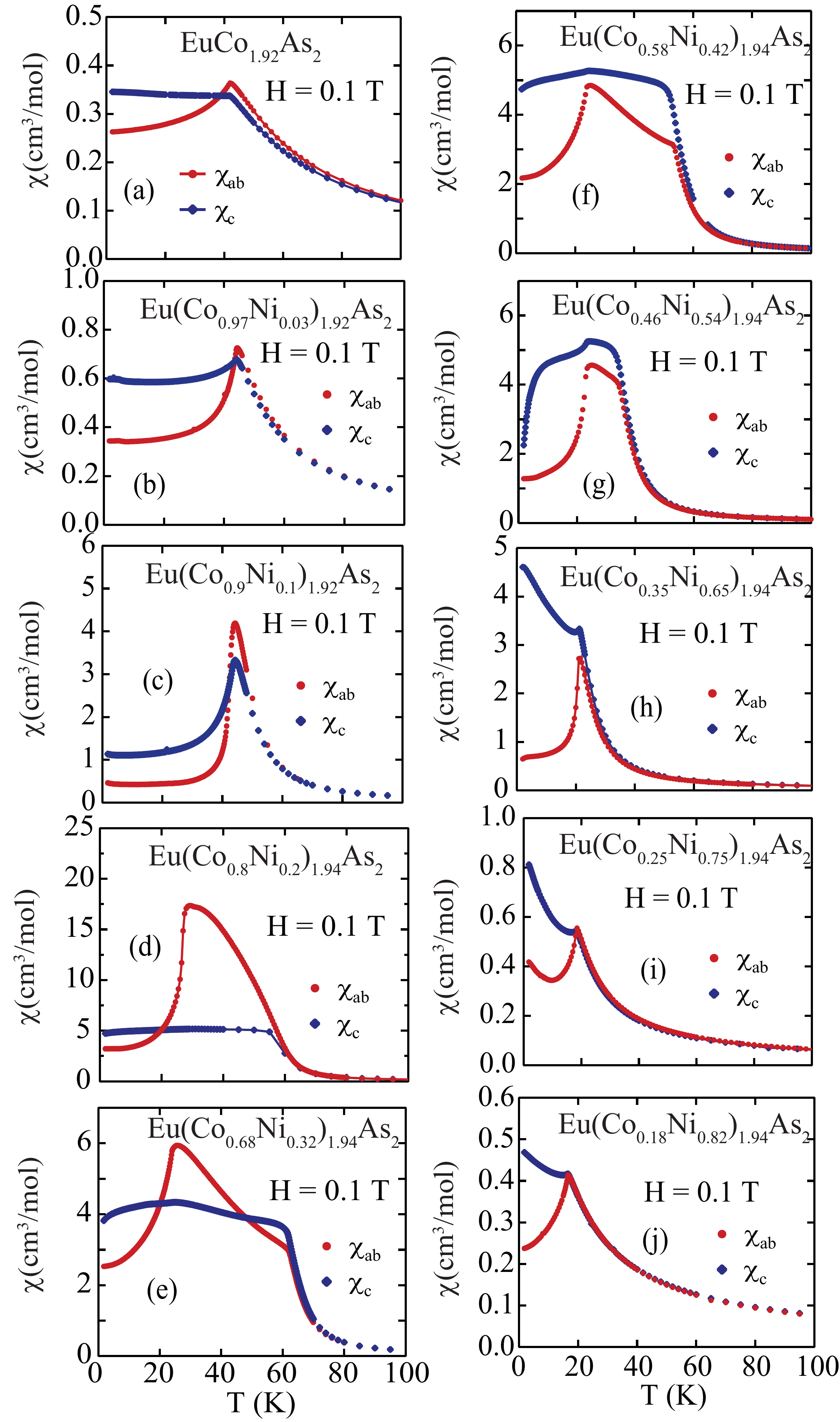}
\caption{ Zero-field-cooled (ZFC) magnetic susceptibility $\chi \equiv M/H$ of \ecna\ ($x =$ 0, 0.03, 0.10, 0.20, 0.32, 0.42, 0.54, 0.65, 0.75, and 0.82) single crystals as a function of temperature~$T$, from 1.8 to 300 K, measured in magnetic fields~$H = 0.1$~T applied in the $ab$~plane ($\chi_{ab}$, $H\parallel ab$) and along the $c$~axis ($\chi_c$, $H\parallel c$).  Note the factor of 50 increase in the ordinate scale between $x=0$ and $x = 0.2$ and the subsequent reduction by the same factor by $x=0.82$.}
\label{Fig:chi1kOe}
\end{figure}

\begin{table}
\caption{\label{Tab.Tn} Comparison of transition temperatures $T_{\rm N\,Eu}$ and $T_{\rm C\,Co/Ni}$ of \ecna\ ($x=0$--1) obtained from magnetic-susceptibility $\chi(T)$ data at an applied field $H=0.1$~T, and from electrical-resistivity $\rho(T)$ and heat-capacity $C_{\rm p}(T)$ data in $H=0$.}
\begin{ruledtabular}
\begin{tabular}{c|cc|cc|cc}
	
				&\multicolumn{2}{c|}{From $\chi(T)$} &\multicolumn{2}{c|} {From $\rho(T)$}  &\multicolumn{2}{c} {From $C_{\rm p}(T)$} \\
	$x$
	&$T_{\rm N\,Eu}$ 	& $T_{\rm C\,Co/Ni}$ 
	&$T_{\rm N\,Eu}$ 	& $T_{\rm C\,Co/Ni}$ 
	&$T_{\rm N\,Eu}$ 	& $T_{\rm C\,Co/Ni}$ \\
	& (K)				& (K)
	& (K)				& (K)
	& (K)				& (K)\\
\hline
	0	&45.1(8)	&		&45.0(4)&		&45.1(2)	\\
	0.03	&44(1)	&		&43.9(3)&		&43.6(2)	\\	
	0.1	&43(1)	&		&43.1(4)&		&42.8(1)	\\
	0.2	&33(1)	&60(2)	&			&58.2(4)	&29.8(4)	&57.8(4)\\
	0.32	&25(1)	&66(1)	&25.2(3)		&66.2(2)	&24.2(5)	&63.5(6)\\
	0.42	&25(1)	&58(1)	&24.4(3)		&58.2(2)	&23.9(1)	&54.3(2)\\
	0.54	&23(1)	&40(1)	&23.0(8)		&39.3(5)	&23.0(2)	&35.8(5)\\
	0.65	&21(1)	&25(1)	&&24.5(8)		&20.8(6)	\\
	0.75	&19(1)	&		&18.8(2)&		&17.8(3)	\\
	0.82	&16.6(3)	&		&16.5(2)&		&16.2(4)	\\
	1	&14.4(5)	&		&14.2(8)&		&14.4	\\
\end{tabular}
\end{ruledtabular}
\end{table}

The zero-field-cooled (ZFC) magnetic susceptibilities $\chi(H,T)\equiv M(T)/H$ of \ecna\ single crystals with $x=0$, 0.03, 0.10, 0.20, 0.32, 0.42, 0.54, 0.65, 0.75, and 0.82 measured in an applied field $H=0.1$~T aligned along the $c$~axis \mbox{($\chi_c,~H\parallel c$)} and in the $ab$~plane ($\chi_{ab},\ H\parallel ab$) are shown in Fig.~\ref{Fig:chi1kOe}. The Eu-spin $T_{\rm N}$ values of the crystals are estimated from the temperatures of the maxima of $d(\chi_{ab} T)/dT$ versus~$T$~\cite{Fisher1962} and are listed in Table~\ref{Tab.Tn}.   ${\rm EuCo_{1.92}As_2}$ exhibits $T_{\rm N} = 45.1$~K [Fig.~\ref{Fig:chi1kOe}(a)] as previously reported~\cite{Sangeetha2018}.   The $T_{\rm N}$ decreases monotonically  with increasing Ni doping $x$ and reaches 15~K at $x=1$, close to the value previously reported for \ena~\cite{Sangeetha2019a}. It is also apparent from Fig.~\ref{Fig:chi1kOe} that the magnitude of $\chi_{ab}(T)$ near $T_{\rm N}$ increases strongly with increasing Ni doping for $x=0.1$, 0.2, and 0.3.  In particular, the crystal with $x=0.20$ exhibits the strongest increase in $\chi_{ab}(T_{\rm N})$ compared with crystals with compositions on either side.  In contrast, for the crystals with $x=0.42$--0.65, the magnitude of $\chi_{c}(T)$ is greater than $\chi_{ab}(T)$ near $T_{\rm N}$.  Furthermore, the system \ecna\ exhibits an additional transition at $T_{\rm C\,Co/Ni}$ for compositions in the range $0.20\leq x \leq 0.54$ attributed to FM ordering associated with the Co/Ni atoms as shown in Figs.~\ref{Fig:chi1kOe}(d)--\ref{Fig:chi1kOe}(g) and listed in Table~\ref{Tab.Tn}. 

For the end-point compositions \ecaa\ and \enaa, neutron diffraction~\cite{Tan2016, Jin2019} and magnetic-susceptibility~\cite{Sangeetha2019, Sangeetha2019a} studies demonstrated that the AFM structure is a planar helix with the Eu ordered moments aligned in the $ab$ plane with the helix axis being the $c$ axis.  The data for $x=0.65$--0.82 suggest a $c$-axis helix magnetic structure as in the end-point compounds whereas the data for $x=0.03$ and~0.10 suggest the superposition of a $c$-axis helix and an $ab$-plane helix where the moments are aligned in an $ab$-$c$ plane perpendicular to the $ab$-plane helix axis.  However, the $\chi(T)$ data for intermediate compositions $x=0.2$--0.54 are more complicated because in this composition range two transitions are observed: a likely transition to a $c$-axis cone structure with a $c$-axis FM component  and a higher-temperature ferromagnetic transition associated with the Co/Ni atoms that is likely of itinerant origin (see below).  Coupled with these observations, the scale of the ordinates in Fig.~\ref{Fig:chi1kOe} increases by more than an order of magnitude between $x = 0$ and~0.2, remains larger than for $x = 0$ from \mbox{$x = 0.32$} to \mbox{$x= 0.75$}, then returns to the scale for $x=0$ at $x=1$.  In the following two sections we first discuss the helical orderings and then the ferromagnetically ordered component.

\subsection{Helical antiferromagnetic ordering}

\begin{figure}
\includegraphics[width=3.4in]{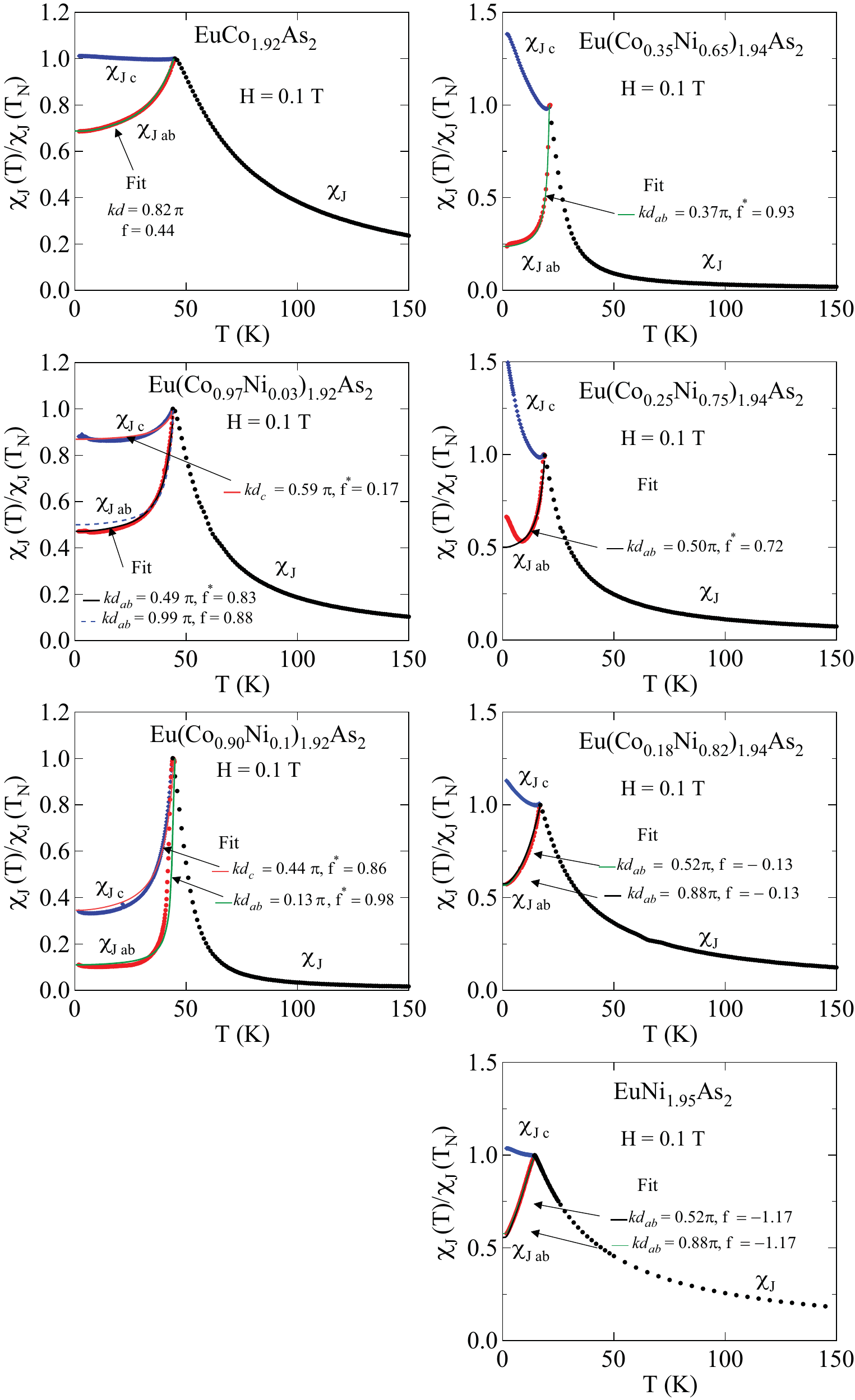}
\caption{Anisotropic magnetic susceptibility $\chi(T)$ of \ecna\ crystals with $x = 0$~\cite{Sangeetha2019}, 0.03, 0.10, 0.65, 0.75, 0.82 from Fig.~\ref{Fig:chi1kOe}, where the spherical average of the  $ab$-plane and $c$-axis data were obtained for $T>T_{\rm N}$, and then the $\chi_{ab}(T)$ and $\chi_c(T)$ data below $T_{\rm N}$ translated vertically to match the former data at $T_{\rm N}$.  The values $f=\theta_{\rm p\, ave}/T_{\rm N}$ are measured values, whereas the $f^\ast$ values are fitted.  }
\label{chiTTN}
\end{figure}

In this section we analyze the anisotropic $\chi(T)$ data for $x=0.03$, 0.10, and 0.65--0.82 in Fig.~\ref{Fig:chi1kOe} in terms of helical ordeing.  We also include the data and analyses of single-crystal $\chi(T)$ data for the $x=0$ and~1 end-point compounds~\cite{Sangeetha2019, Sangeetha2019a}.  As described in Refs.~\cite{Johnston2012, Johnston2015}, for  a Heisenberg spin system that has a helical AFM structure with the moments aligned in the $ab$~plane of a tetragonal crystal structure, $\chi_c$ is independent of $T$ and spin~$S$ below $T_{\rm N}$ whereas $\chi_{ab}$ monotonically decreases as the temperature decreases below $T_{\rm N}$ as shown for $x=0$ in Fig.~\ref{Fig:chi1kOe}(a).

When there is anisotropy above $T_{\rm N}$, for example due to anisotropic FM fluctuations that appear to be present in the \ecna\ system, we first take the spherical average of the data above $T_{\rm N}$ giving $\chi_J(T)$, thus eliminating the effects of anisotropy above $T_{\rm N}$.   Then we translate the \mbox{ $\chi_{ab}(T\leq T_{\rm N})$} and \mbox{$\chi_c(T\leq T_{\rm N})$} data vertically so that they meet the now isotropic PM $\chi_J(T\geq T_{\rm N})$ data at $T_{\rm N}$.  The results of this procedure for the crystals with $x=0$~\cite{Sangeetha2019}, 0.03, 0.10, and 0.65--0.82 are shown in Fig.~\ref{chiTTN}. 

\begin{figure}
\includegraphics[width=2.75in]{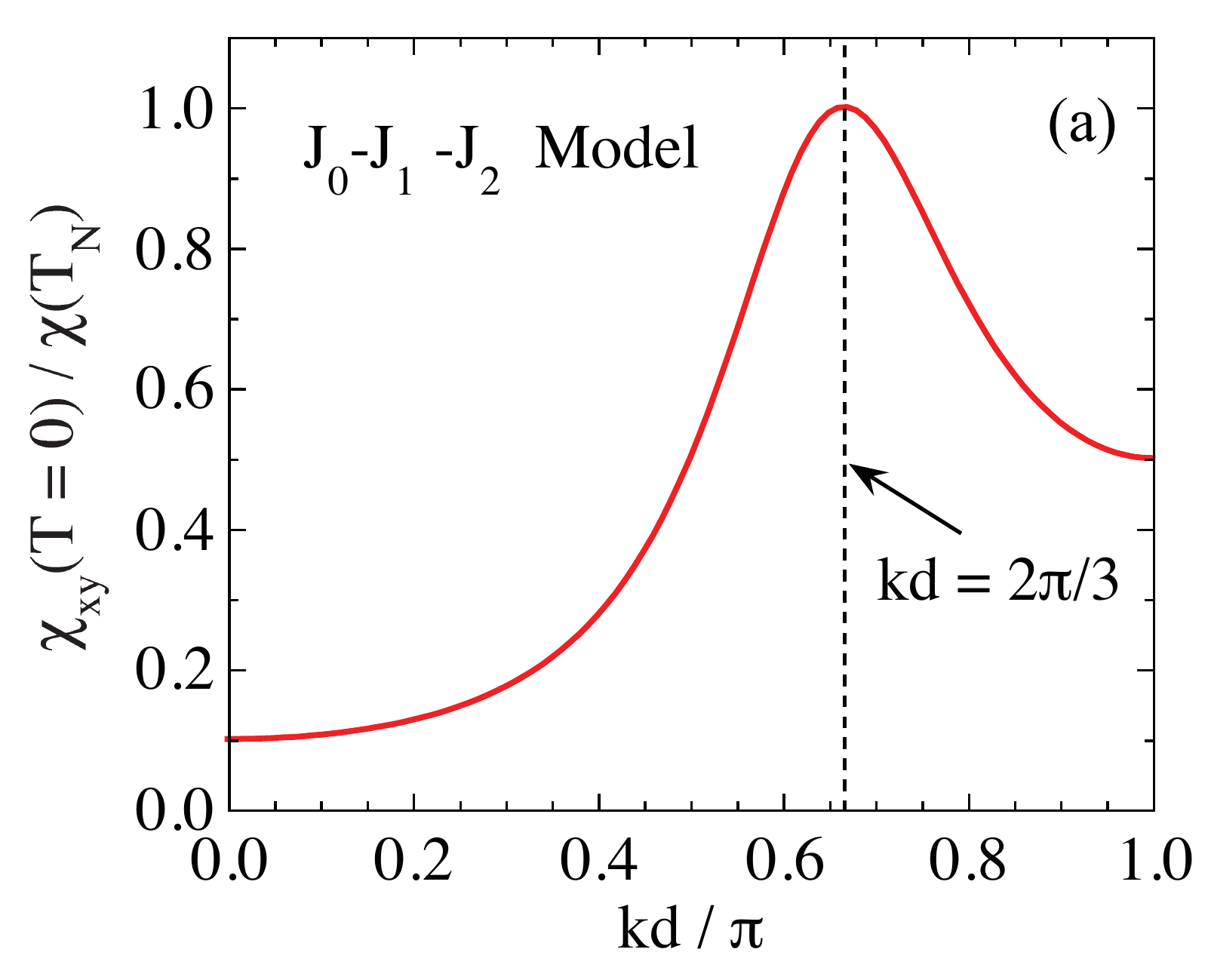}
\caption{The ratio $\chi_{xy}(T=0)/\chi(T_{\rm N})$ ($xy=ab$ here) versus the turn angle $kd/\pi$ between FM-aligned $ab$-plane layers for a $c$-axis Heisenberg helix~\cite{Johnston2012, Johnston2015}.  For $kd=2\pi/3$, $\chi_{xy} = \chi_z$ ($z=c$ here) is predicted to be independent of both the spin~$S$ and $T$ for $T\leq T_{\rm N}$, as observed~\cite{Johnston2012}.}
\label{chi0OnchiTNvskd}
\end{figure}

Within a one-dimensional $J_0$-$J_1$-$J_2$ MFT model for a $c$-axis Heisenberg helix, one obtains~\cite{Johnston2012, Johnston2015} 
\bea
\frac{\chi_{ab}(T\to0)}{\chi_{ab}(T_{\rm N})}= \frac{1}{2[1+\cos(kd)+\cos^2(2kd)]},
\label{chiabratio}
\eea
where $kd$ for a $c$-axis helix is the turn angle from magnetic-moment layer to layer along the $c$~axis between ferromagneticallyaligned moments aligned in the $ab$~plane.   A plot of this ratio versus $kd$ is shown in Fig.~\ref{chi0OnchiTNvskd}~\cite{Johnston2012, Johnston2015}.  It is evident that if the ratio is less than 1/2, there is only one solution to the turn angle~$kd$.  However, for ratios between 1/2 and 1, there are two possible solutions.  By comparison with neutron diffraction results if available, one can choose the appropriate value for the latter value.  If no such additional information is available, one can choose the more likely value from continuity of $kd$ with composition on approaching $x=0$ or $x=1$.  From the data for $x=0$ and $x=1$ in Fig.~\ref{chiTTN} and a comparison of the two calculated $kd$ values for each composition with the respective neutron-diffraction data, Eq.~(\ref{chiabratio}) gives $kd=0.85\pi$~rad for $x=0$~\cite{Sangeetha2018} and $kd = 0.88\pi$~rad for $x=1$~\cite{Sangeetha2019a}, where both values are in good agreement with the respective neutron-diffraction results~\cite{Tan2016, Jin2019}.

Within MFT, for $T\leq T_{\rm N}$ the perpendicular susceptibility $\chi_c$ of a $c$-axis helical Heisenberg AFM is predicted to be independent of~$T$ (and~$S$), in reasonable agreement with the data for $x=0$ and $x=1$ in Fig.~\ref{chiTTN}.  The normalized $T$-dependent ratio $\chi_{ab}(T \leq T_{\rm N})/\chi(T_{\rm N})$ within the $J_0$-$J_1$-$J_2$ Heisenberg model is given by~\cite{Johnston2012, Johnston2015}
\bse
\label{Eqs:Chixy}
\begin{equation}
\frac{\chi_{ab}(T \leq T_{\rm N})}{\chi(T_{\rm N})} =  \frac{(1+\tau^*+2f+4B^*)(1-f)/2}{(\tau^*+B^*)(1+B^*)-(f+B^*)^2},
\label{eq:Chi_plane}
\end{equation}
where $f=\theta_{\rm p\,ave}/T_{\rm N}$, $\theta_{\rm p\,ave}$ is the spherically averaged Weiss temperature obtained from the Curie-Weiss fit of the $\chi(T)$ data in the paramagnetic state as listed in Table~\ref{Tab.chidata} below,
\begin{equation}
B^*=  2(1-f)\cos(kd)\,[1+\cos(kd)] - f,
\label{eq:Bstar}
\end{equation}
\begin{equation}
t =\frac{T}{T_{\rm N}},\quad \tau^*(t) = \frac{(S+1)t}{3B'_S(y_0)}, \quad y_0 = \frac{3\bar{\mu}_0}{(S+1)t}, 
\label{Eq:ttauy0}
\end{equation}
$kd$ is the turn angle between the magnetic moments in adjacent planes  along the helix $c$~axis of $ab$-plane FM-aligned moment layers, and the ordered moment versus $T$ in $H=0$ is denoted by $\mu_0$.  The reduced ordered moment $\bar{\mu}_0 = \mu_0/\mu_{\rm sat}$ with the saturation moment $\mu_{\rm sat} = gS\mu_{\rm B}$ is determined by numerically solving the self-consistency equation
\begin{equation}
\bar{\mu}_0 = B_S(y_0),
\label{Eq:barmuSoln}
\end{equation}
\ese
 where $B_S(y)$ is the Brillouin function, and $B'_S(y_0) \equiv [dB_S(y)/dy]|_{y=y_0}$.   For $x=0$ and $x=1$, the data in Fig.~\ref{chiTTN} agree well with the MFT prediction for \mbox{$\chi_{ab}(T\leq T_{\rm N})/\chi(T_{\rm N})$},  following MFT behavior for a \mbox{$c$-axis} helix with $\mu_{\rm sat}= 7 \mu_{\rm B}$.  For $x=0$ with \mbox{$T_{\rm N} = 45.1$~K}, one has $f=0.44$ and $kd=0.82\pi$~rad~\cite{Sangeetha2018}, whereas for $x=1$,  $T_{\rm N} = 14.4$~K, $f=-1.17$, and \mbox{$kd =0.88\pi$~rad}~\cite{Sangeetha2019a}.
 
\begin{table*}
\caption{\label{Tab.kddata} Parameters obtained from fitting the magnetic susceptibility data below $T_{\rm N}$ of Eu  for \ecna.  Shown are the AFM transition temperature $T\rm_N$ and the Weiss temperature $\theta_{\rm p}$ taken from Table~\ref{Tab.chidata} below, the observed ratio $f=\theta_{\rm p}/T_{\rm N}$, the value $f^\ast$ obtained from a fit of $\chi_\alpha(T\leq T_{\rm N})$ by Eqs.~(\ref{Eqs:Chixy}), and the turn angle $kd$ between adjacent FM-aligned magnetic layers of the helix model obtained using Eq.~(\ref{chiabratio}).  For $\chi(T=0)/\chi(T_{\rm N}) > 1/2$ two values of $kd$ are possible.  The notation PW in the last column means the present work.}
\begin{ruledtabular}
\begin{tabular}{cccccccccc}
 		 & field	
		& $T\rm_N$	 					
		 & $\theta\rm_ {p}$ 
		& $f=\theta_{\rm p}/T_{\rm N}$	
		&$f^{*}$
		 & $kd$ 	
		& $kd$  
		& Ref.\\

	Compound &
	direction 	
	& (K)
	& (K) 
	&		
	& 
	& ($\pi$ radian) 	
	& (degree)
	&\\
\hline
EuCo$_{1.90(1)}$As$_2$							& $H\parallel ab$ 	&45.1(8)		& 19.76(9)		&0.44	&		& 0.82		& 147		&\cite{Sangeetha2018} \\	
Eu(Co$_{0.97(1)}$Ni$_{0.03(1)}$)$_{1.92}$As$_2$ 		& $H\parallel ab$ 	& 44(1) 		& 38.9(5) 		&0.88	& 0.83	& 0.49		& 88			& PW\\
											& $H\parallel c$ 	&		        	& 37.30(3)		& 0.84	& 0.17	& 0.59, 0.74	& 108, 133	&PW\\
Eu(Co$_{0.90(1)}$Ni$_{0.10(1)}$)$_{1.92}$As$_2$		& $H\parallel ab$ 	& 43(1) 		& 51.2(1)		&1.19 	& 0.98	& 0.13		& 23			&PW\\
											& $H\parallel c$ 	&			& 50.75(6) 	& 1.18	& 0.86	& 0.44		& 79			&PW\\
Eu(Co$_{0.35(2)}$Ni$_{0.65(2)}$)$_{1.94}$As$_2$ 		& $H\parallel ab$ 	& 21(1)	 	& 18.10(9) 	&0.86	& 0.93	& 0.37		& 67			&PW\\
Eu(Co$_{0.25(1)}$Ni$_{0.75(1)}$)$_{1.94}$As$_2$ 		& $H\parallel ab$ 	& 19(1)		& 6.74(6)		& 0.37	& 0.72	& 0.50, 0.92	& 91, 166		&PW\\
Eu(Co$_{0.18(1)}$Ni$_{0.82(1)}$)$_{1.94}$As$_2$ 		& $H\parallel ab$ 	& 16.6(3) 		& $-2.3$(1) 	&$-0.13$	& 	  	& 0.52, 0.88  	& 94, 159	 	&PW\\
\ena\											& $H\parallel ab$ 	&  14.4(5)		& $-15(1)$	 &$-1.17$	&		& 0.88		& 159		&\cite{Sangeetha2019a}\\
										
\end{tabular}
\end{ruledtabular}
\footnotetext{${^*}$ fitted parameter}
\end{table*}
 
The $\chi_\alpha(T< T_{\rm N})$ data for $x=0.03$ and especially for $x=0.1$ in Fig.~\ref{chiTTN} are quite interesting, because for these compositions not only does $\chi_{ab}$ decrease below $T_{\rm N}$, but so does $\chi_c$, an effect that is especially pronounced for $x=0.1$.   This behavior suggests that the AFM structure consists of a superposition of a $c$-axis helix with the moments aligned in the $ab$~plane and another helix with the helix axis and AFM propagation vector in the $ab$~plane and with the moments aligned in an $ab$-$c$ plane perpendicular to the in-plane AFM propagation vector, i.e., a so-called $2q$ structure.  Therefore, for $x=0.03$ and~0.1 we fitted both the $\chi_{ab}$ and $\chi_c$ data separately by the prediction in Eqs.~(\ref{Eqs:Chixy}) for the in-plane susceptibility for a helix, and good fits to both data sets were obtained as shown in Fig.~\ref{chiTTN}.   A summary of the quantities discussed above is given in Table~\ref{Tab.kddata}.

As discussed in the following section, a FM component to the ordering develops for $0.2 \leq x\leq 0.65$ which contains contributions from both the Eu spins and the Co/Ni sublattice.  The occurrence of the FM component for $x=0.65$ is correlated with the upturn in $\chi_c(T<T_{\rm N})$ in Fig.~\ref{chiTTN} for this composition, but the $\chi_{ab}(T<T_{\rm N})$ data  in Fig.~\ref{chiTTN} follow the same behavior as for $x=0.75$--1, so we include these data in the analysis of the $\chi_{ab}(T\leq T_{\rm N})$ data for crystals with compositions in the range $0.65\leq x\leq 1$.

\begin{figure}
\includegraphics[width=3in]{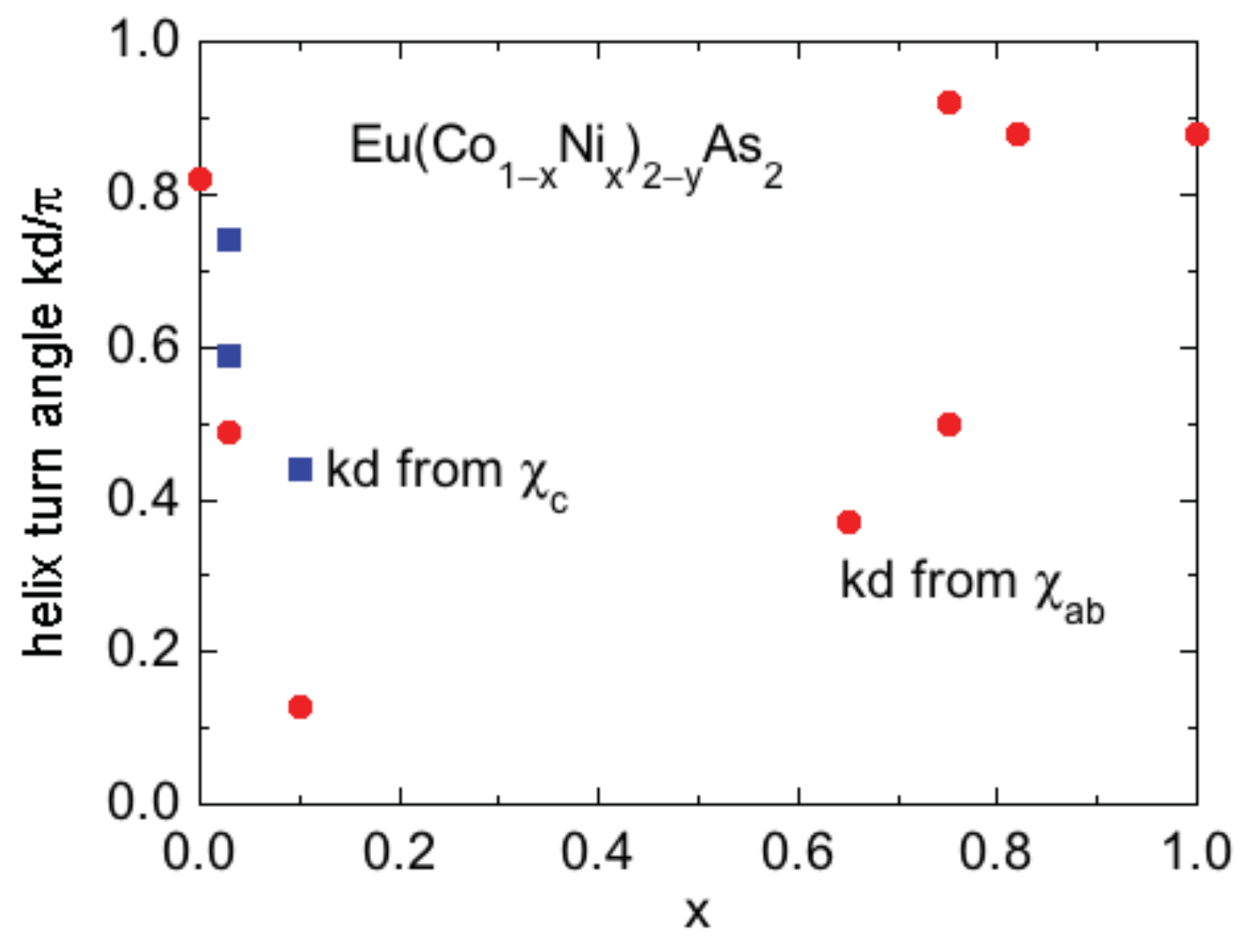}
\caption{Helix turn angle(s) versus composition obtained from the $ab$-plane susceptibility~$\chi_{ab}$ (filled red circles) and the \mbox{$c$-axis} susceptibility~$\chi_c$ (filled blue squares) at temperatures $T< T_{\rm N}$.  For $\chi_\alpha(0)/\chi(T_{\rm N}) > 1/2$, two values of $kd/\pi$ are possible according to Fig.~\ref{chi0OnchiTNvskd}, indicated by two values of $kd/\pi$ for the same composition~$x$ and for the particular field direction~$\alpha$ of $\chi_\alpha$.  The large gap between data points at $x=0.1$ and~$x=0.65$ is present because the Eu magnetic structure between these two compositions is not a simple helix (see text).}
\label{Eu(Co,Ni)2As2_kd_vs_x}
\end{figure}

Figure~\ref{Eu(Co,Ni)2As2_kd_vs_x} shows the helical angle $kd/\pi$ versus~$x$ for our crystals for which a helical magnetic structure can be identified from the magnetic susceptibility data. A turn angle $kd/\pi<1/2$ is indicative of a net ferromagnetic nearest-layer interaction, whereas a value $kd/\pi>1/2$ indicates a net antiferromagnetic interaction.  Therefore the data indicate that at intermediate compositions a net FM interaction develops between the Eu spins in adjacent layers, whereas near the ends of the composition range the net interaction is AFM.

\subsection{\label{Sec:chi} Ferromagnetic-ordering component}

\begin{figure}
\includegraphics[width=3.5in]{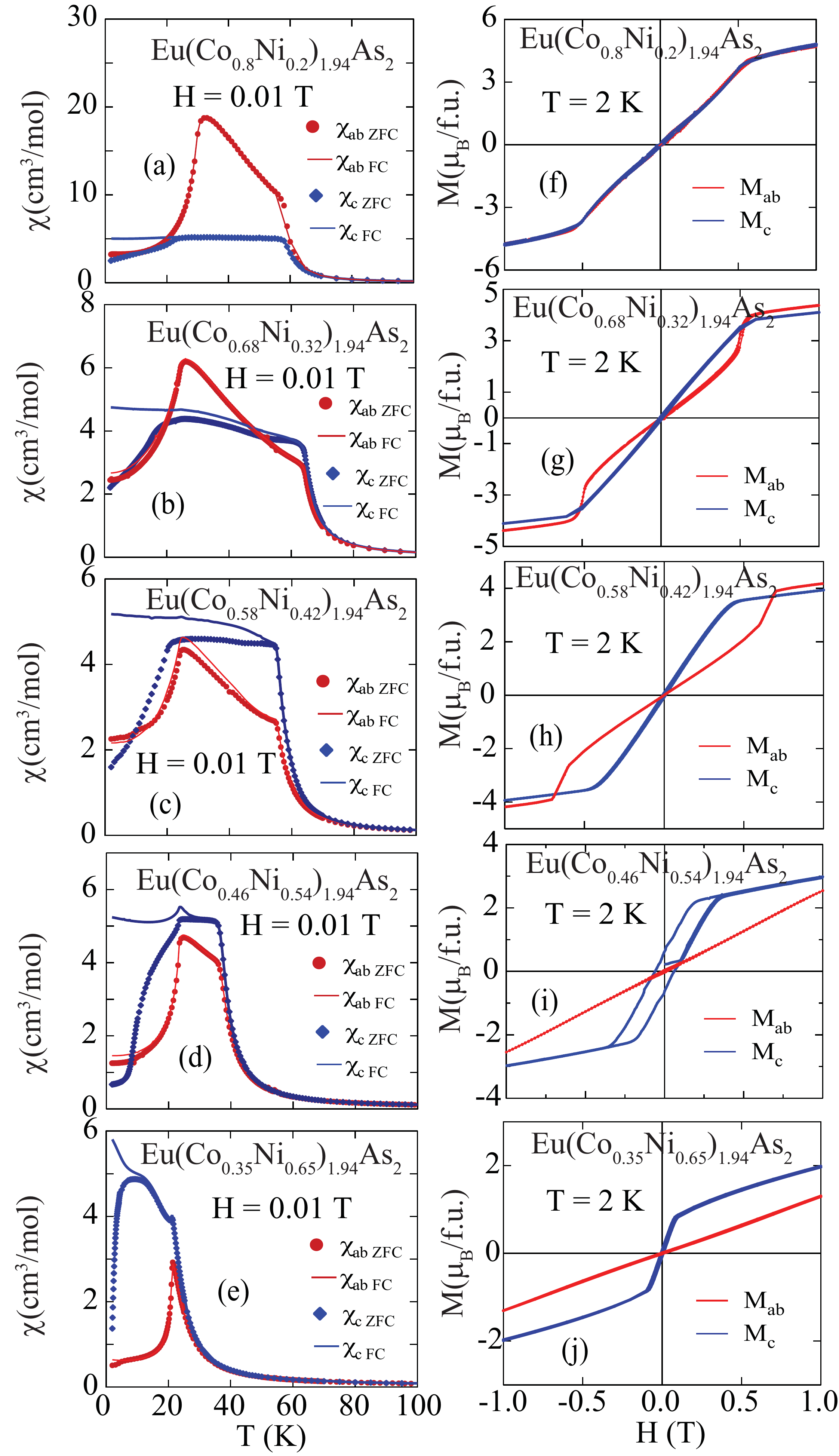}
\caption{(a)--(e) Zero-field-cooled (ZFC) and field-cooled (FC) magnetic susceptibility $\chi \equiv M/H$ of \ecna\ ($x =$ 0.20, 0.32, 0.42, 0.54, and 0.65) single crystals as a function of temperature~$T$, from 1.8 to 100 K, measured in magnetic fields~$H = 0.01$~T applied in the $ab$~plane ($\chi_{ab}$) and along the $c$~axis ($\chi_c$). (f)--(j)~Low-field isothermal magnetization $M$ of \ecna\ ($x =$ 0.20, 0.32, 0.42, 0.54, and 0.65) single crystals as a function of magnetic field~$H$\@.}
\label{Fig:MT100oe}
\end{figure}

\begin{figure}
\includegraphics[width=3.in]{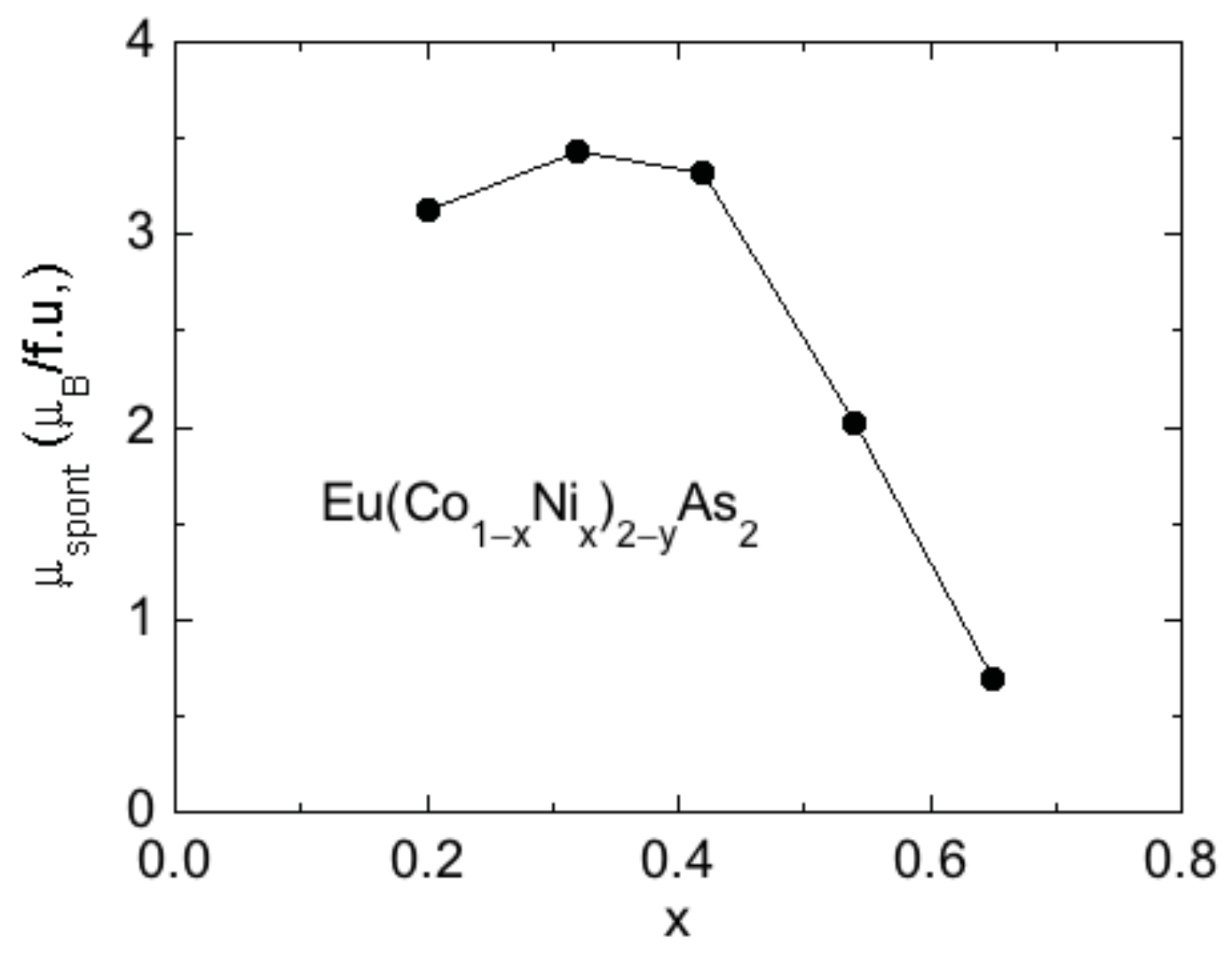}
\caption{Spontaneous moment $\mu_{\rm spont}$ at $T=2$~K versus $x$ for \ecna\ crystals with $x =$ 0.20, 0.32, 0.42, 0.54, and 0.65 obtained from the right-hand panels of Fig.~\ref{Fig:MT100oe}. }
\label{Fig:Eu(Co,Ni)2As2_spont_moment}
\end{figure}

To further investigate the magnetic behavior, we carried out zero-field-cooled (ZFC) measurements of $\chi(T)$ on warming and field-cooled (FC) measurements of $\chi(T)$ on cooling (FC)  versus~$T$ in the range 1.8 to 100~K with a small applied field $H=100$~Oe for the crystals with $x=0.20$, 0.32, 0.42, 0.54, and 0.65  for both $H\parallel ab$ and $H\parallel c$ which are plotted in Figs.~\ref{Fig:MT100oe}(a)--\ref{Fig:MT100oe}(e), respectively.  The ZFC data were taken after quenching the superconducting magnet to eliminate remanent fields in the magnet.  These data show virtually no hysteresis between the ZFC and FC measurements for $H\parallel ab$ but strong hysteresis below the high-$T$ transition for $H\parallel c$.  This indicates that there exists a high-$T$ FM transition for $x = 0.2$--0.54 in addition to the AFM ordering of the Eu spin sublattice.  Then from Fig.~\ref{Fig:chi1kOe}, we can assign the low-$T$ transition to an AFM transition involving the Eu spins designated as $T_{\rm N\,Eu}$ and the higher-$T$ transitions to FM ordering associated with the Co/Ni atoms at temperatures denoted by $T_{\rm C\,Co/Ni}$.  This interpretation is strongly supported by the $C_{\rm p}(T)$ measurements in Sec.~\ref{Sec:HC}  below.  The transition temperatures estimated from the $\chi(T)$ data are listed in Table~\ref{Tab.Tn}, where $T_{\rm N\,Eu}$ decreases monotonically upon doping whereas $T_{\rm C\,Co/Ni}$ first increases from 60~K for $x=0.2$ to 66~K for $x=0.32$,  then decreases for $0.32\leq x \leq 0.65$ and finally disappears for $x>0.65$. 

This interpretation is also supported by  low-field $M(H)$ measurements at $T =2$~K from $-1$ to $+1$~T as plotted in Figs.~\ref{Fig:MT100oe}(f)--\ref{Fig:MT100oe}(j) for the same compositions as in Figs.~\ref{Fig:MT100oe}(a)--\ref{Fig:MT100oe}(e), respectively.  An extrapolation of the linear data above the initial field dependence to $H = 0$ gives the spontaneous moment $\mu_{\rm spont}$ versus composition plotted in Fig.~\ref{Fig:Eu(Co,Ni)2As2_spont_moment}.  The variation of $\mu_{\rm spont}$ versus~$x$ is similar to that of the Co/Ni Curie temperature $T_{\rm C\,Co/Ni}$.  The large values of $\mu_{\rm spont}$ show that the magnetic structure has a large FM component that must arise from the Eu spins.  We infer that within the region in which ferromagnetism from the Co/Ni atoms occurs at higher temperatures, the Eu magnetic structure is a $c$-axis conical structure.  In addition, the $M(H)$ data show hysteresis with field for $x=0.54$ but the hysteresis is very small for the other compositions.

\subsection{\label{Eq:ChiPM} Magnetic susceptibility in the paramagnetic state}

We now turn our attention to $\chi_{\alpha}(T)$ ($\alpha=ab$ or~$c$) of the \ecna\ crystals in the PM region 100~K~$\leq T \leq 300$~K\@.  The data are analyzed within the local-moment Heisenberg model in terms of the modified Curie-Weiss law
\bea
\chi_{\alpha}(T) =\chi_{0\alpha} + \frac{C_{\alpha}}{T-\theta_{\rm p\alpha}},
\label{Eq.mcw}
\eea
where $\chi_{0\alpha}$ is a \mbox{$T$-independent} term, the Curie constant per mole of Eu spins is
\begin{subequations}
\begin{equation}
C_{\alpha}=\frac{N_{\rm A} g^2_{\alpha}S(S+1)\mu^2_{\rm B}}{3k_{\rm B}},
\label{Eq:cc}
\end{equation}
$N_{\rm A}$ is Avogadro's number, $g_\alpha$ is the possibly anisotropic $g$~factor, and $k_{\rm B}$ is Boltzmann's constant.  The effective moment $\mu_{{\rm eff}\alpha} = g_\alpha\sqrt{S(S+1)}$ of a spin in units of $\mu_{\rm B}$ is given by Eq.~(\ref{Eq:cc}) as
\begin{equation}
\mu_{\rm eff \alpha} = \sqrt{\frac{3k_{\rm B}C_{\alpha}}{N_{\rm A}{\rm\mu^2_B} }} ~ \approx \sqrt{8C_{\alpha}},
\label{Eq:mueff}
\end{equation}
\end{subequations}
where $C_\alpha$ is in cgs units of cm$^3$\,K/(mol Eu). 

\begin{figure}
\includegraphics[width=3.5in]{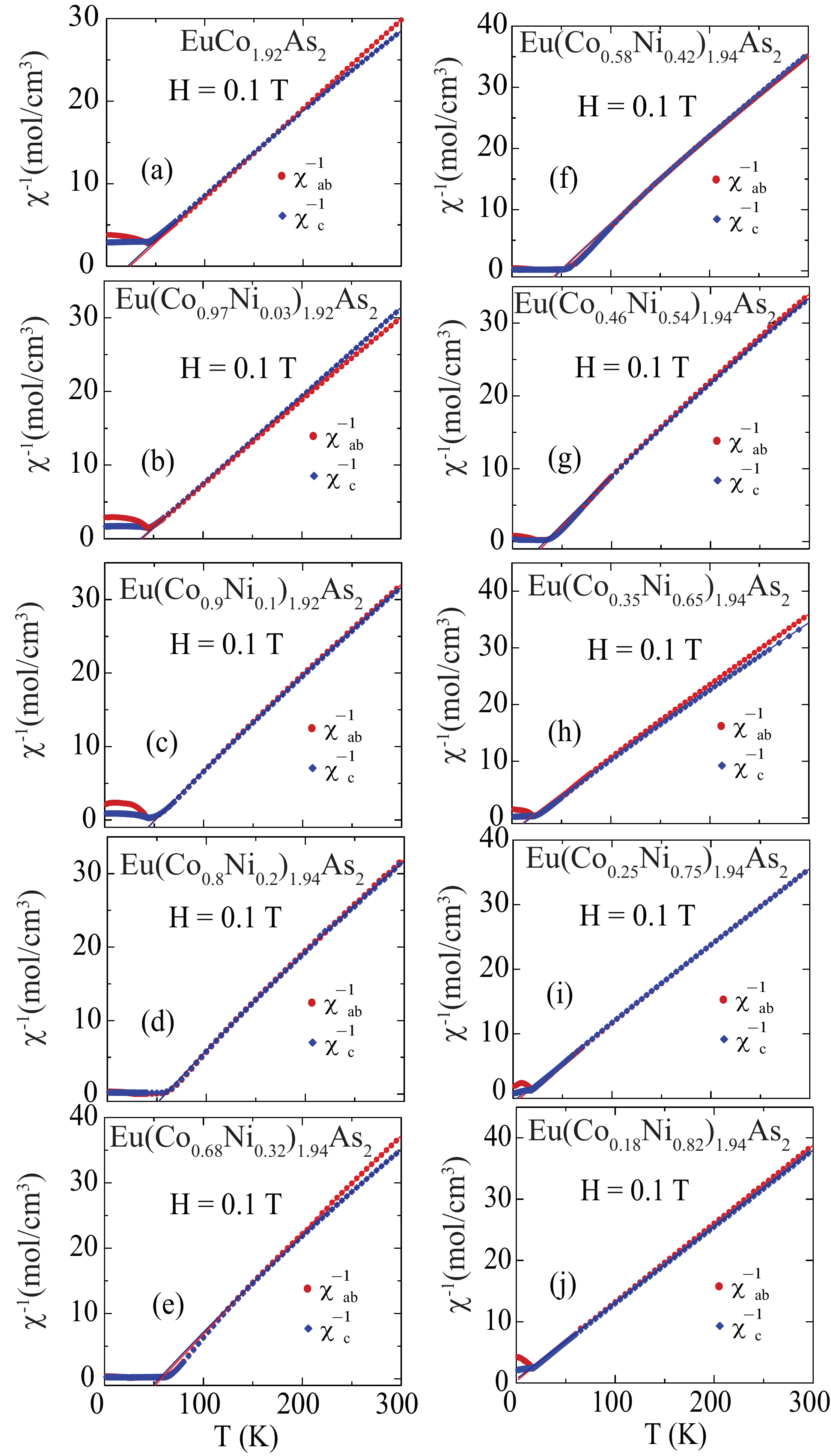}
\caption{ Inverse susceptibility $\chi^{-1}$ of \ecna\ ($x = 0$, 0.03, 0.10, 0.20, 0.32, 0.42, 0.54, 0.65, 0.75, and 0.82) single crystals as a function of temperature~$T$ from 1.8 to 300 K, measured in magnetic  fields~$H = 0.1$~T applied in the $ab$~plane ($\chi^{-1}_{ab}$, $H\parallel ab$) and along the $c$~axis ($\chi^{-1}_c$, $H\parallel c$). }
\label{Fig:Invchi}
\end{figure}

Figure~\ref{Fig:Invchi} shows the inverse susceptibility $\chi^{-1}$ with $H=0.1$~T applied along the $c$ axis ($\chi^{-1}_c$) and in the  $ab$~plane  ($\chi^{-1}_{ab}$) as a function of $T$ from 1.8 to 300~K\@. The fits in the PM regime between 100 and 300~K are shown as the solid lines and the fitted parameters are listed in Table~\ref{Tab.chidata}.

\begin{table*}
\caption{\label{Tab.chidata} Listed are the magnetic transition temperatures $T_{\rm N\,Eu}$ for \ecna, and $T_{\rm C\,Co,Ni}$ if present, together with the parameters obtained from modified Curie-Weiss fits of the magnetic susceptibility data between 100 and 300~K, including the $T$-independent contribution to the susceptibility $\chi_0$,  Curie constant per mole of formula units $C_\alpha$ with the applied field in the $\alpha = ab,\ c$ directions, and the Weiss temperature $\theta\rm_{p\alpha}$. Also listed are the effective moments for Eu$^{+2}$ with $S=7/2$ and $g=2$ given by $\mu_{\rm eff\,\alpha} = \sqrt{8 C_\alpha}$ and the  angle-averaged effective moment $\mu_{\rm eff\,ave}=(2 \mu_{{\rm eff} ab} +\mu_{ {\rm eff}c})/3$.  }
\begin{ruledtabular}
\begin{tabular}{ccccccccccc}
 		& 	
		& $T_{\rm N\,Eu},\ T_{\rm C\,Co/Ni}$	 
		& $\chi_0$  							
		& $C_\alpha$ 	
		& $\mu\rm_{eff\alpha}$	
		& $\theta\rm_ {p \alpha}$ 	
		& $\mu\rm_{eff,ave}$  
		& $\theta_{\rm p ave}$
		& Ref.\\

	Compound &	
	& (K) 		
	& $\rm{\left(10^{-3}~\frac{cm^3}{mol}\right)}$	 
	& $\rm{\left(\frac{cm^3 K}{mol}\right)}$	
	& $\rm{\left(\frac{\mu_B}{f.u.}\right)}$	 
	& (K) 	
	&$\rm{\left(\frac{\mu_B}{f.u.}\right)}$ 
	& (K) \\
\hline
 EuCo$_{1.90(1)}$As$_2$\footnotemark[1] 			& $H\parallel ab$ 	&45.1(8),\ ---	& $-1.4(2)$		&  8.98(1)			& 8.476(4)		& 19.76(9)		& 8.47(1)	& 19.07(7)			&\cite{Sangeetha2018} \\		
											& $H\parallel c$ 	&			& $-1.2(1)$		&  8.970(5)		& 8.471(2) 	& 17.70(5)				& \\ 
 Eu(Co$_{0.97(1)}$Ni$_{0.03(1)}$)$_{1.92}$As$_2$ 		& $H\parallel ab$ 	& 44(1),\ --- 	& 1.6(7)			& 8.27(5) 			& 8.13(2)		& 38.9(5)		& 8.16(2)	& 38.3(3)			& PW\\
											& $H\parallel c$ 	&		        & $-0.2.4(1)$ 		& 8.45(1) 			& 8.221(5)		& 37.30(3)				&\\
Eu(Co$_{0.90(1)}$Ni$_{0.10(1)}$)$_{1.92}$As$_2$		& $H\parallel ab$ 	& 43(1),\ --- 	& 2.44(5) 			& 7.15(1) 			& 7.563(5)		& 51.2(1)		& 7.58(5)	& 51.05(8)			&PW\\
											& $H\parallel c$ 	&			& 2.33(2)			& 7.282(7) 		& 7.632(4)		& 50.75(6)				&\\
Eu(Co$_{0.80(1)}$Ni$_{0.20(1)}$)$_{1.94}$As$_2$ 		& $H\parallel ab$ 	& 33(1), 60(2) 	& 1.7(1) 			& 7.25(4) 			& 7.62(2)		& 54.3(4)		& 7.47(1)	& 52.7(3)			&PW\\
											& $H\parallel c$ 	&			& 2.35(3)			& 6.45(1) 			& 7.183 (5)	& 49.6(1)				&\\
Eu(Co$_{0.68(1)}$Ni$_{0.32(1)}$)$_{1.94}$As$_2$ 		& $H\parallel ab$ 	& 25(1), 66(1) 	& 0.8(2)			& 6.35(7) 			& 7.13(4)		& 55.3(8)		& 7.09(2)	& 55.2(6)			&PW\\
											& $H\parallel c$ 	&			& 3.4(4)			& 6.15(1) 			& 7.01 (2)		& 55.0(2)				&\\
Eu(Co$_{0.58(1)}$Ni$_{0.42(1)}$)$_{1.94}$As$_2$ 		& $H\parallel ab$ 	& 25(1), 58(1) 	& 2.45(3) 			& 6.52(1) 			& 7.22(1)		& 48.3(1)		& 7.38(1)	& 48.58(9)			&PW\\
											& $H\parallel c$ 	&			& 1.93(2)			& 7.424(7) 		& 7.707 (3)	& 49.15(7)				&\\
Eu(Co$_{0.46(1)}$Ni$_{0.54(1)}$)$_{1.94}$As$_2$		& $H\parallel ab$ 	& 23(1), 40(1)	& 2.83(9)	 		& 7.07(3) 			& 7.52(1)		& 33.6(5)		& 7.55(1)	& 33.3(4)			&PW\\
											& $H\parallel c$ 	&			& 2.86(5) 			& 7.23(2) 			& 7.60(1)		& 32.9(2)				&\\
Eu(Co$_{0.35(2)}$Ni$_{0.65(2)}$)$_{1.94}$As$_2$ 		& $H\parallel ab$ 	& 21(1), 25(1) 	& 1.21(2) 			& 7.50(7) 			& 7.75(3)		&18.10(9)		& 7.79(2)	& 18.43(9)			&PW\\
											& $H\parallel c$ 	&			& 1.55(2) 			& 7.736(8) 		& 7.867(4)		&19.1(1)				&\\
Eu(Co$_{0.25(1)}$Ni$_{0.75(1)}$)$_{1.94}$As$_2$ 		& $H\parallel ab$ 	& 19(1),\ ---	& 0.124(1)			& 7.88(6)			& 7.940(3)		& 6.74(6)		& 7.97(1)	& 5.76(7)			&PW\\
											& $H\parallel c$ 	&			& 0.080(3) 		& 8.09(1) 			& 8.04(5)		& 3.8(1)				&\\
Eu(Co$_{0.18(1)}$Ni$_{0.82(1)}$)$_{1.94}$As$_2$ 		& $H\parallel ab$ 	& 16.6(3),\ --- 	& 0.252(3)		 	& 7.72(1) 			& 7.859(5)		& $-2.3$(1)	& 7.92(1)	& $-3.2(1)$		&PW\\
											& $H\parallel c$	&			& $-0.196(2)$ 		& 8.09(1) 			& 8.044(5)		& $-5.1$(2)			&\\
\ena\											& $H\parallel ab$ 	&  14.4(5)	,\ ---	&1.01(2)			& 7.8(1) 			& 7.90(5)		& $-15(1)$	& 7.95(3)	& $-15.2(8)$		&\cite{Sangeetha2019a}\\
											& $H\parallel c$ 	&			&1.09(8)			& 8.13(3) 			& 8.06(1)		& $-15.6(5)$			&\\
\end{tabular}
\end{ruledtabular}
\footnotetext[1]{Grown in Sn flux}
\end{table*}

The theoretical values of $C\rm_{ave}$ and $\mu\rm_{eff, ave}$  for ${\rm Eu^{+2}}$ spins with $S=7/2$ and $g=2$  are $7.878~{\rm cm^3\,K/mol~Eu}$ and $7.937~\mu{\rm_B/Eu}$, respectively. From Table~\ref{Tab.chidata},  the $\mu\rm_{eff, ave}$ values significantly decrease gradually from 8.47 to 7.09~$\mu{\rm_B/Eu}$ for $0\leq x \leq 0.32$, then increase for $x\geq0.32$ and finally reach the theoretical value 7.94~$\mu{\rm_B/Eu}$ for $x=1$. It is also of note that the $\mu\rm_{eff, ave}$ values  for $0.10\leq x \leq 0.32$ are much smaller than the theoretical value. The latter observation suggests the possibilities of either an Eu valence increase and/or coexistence of both local moment and itinerant magnetism in \ecna. Other types of measurements are required to understand this strange magnetic behavior. 

Also apparent in Table~\ref{Tab.chidata} is a small difference of $\approx 1$~K between the Weiss  temperatures for fields along the $c$~axis and in the $ab$~plane. This anisotropy may arise from the anisotropic magnetic dipole interactions between the Eu spins~\cite{Johnston2016}, from single-ion uniaxial anisotropy~\cite{Johnston2017}, and/or from anisotropy in the Ruderman-Kittel-Kasuya-Yosida (RKKY) interactions between Eu spins.
 The positive $\theta_{\rm p, ave}$ values  for  $0\leq x \leq 0.75$  indicate predominantly FM exchange interactions between the Eu$^{+2}$ spins, whereas for $x=$ 0.82 and 1, the $\theta_{\rm p, ave}$ values are negative, indicating predominantly AFM exchange interactions.

\begin{figure}
\includegraphics[width=3.3in]{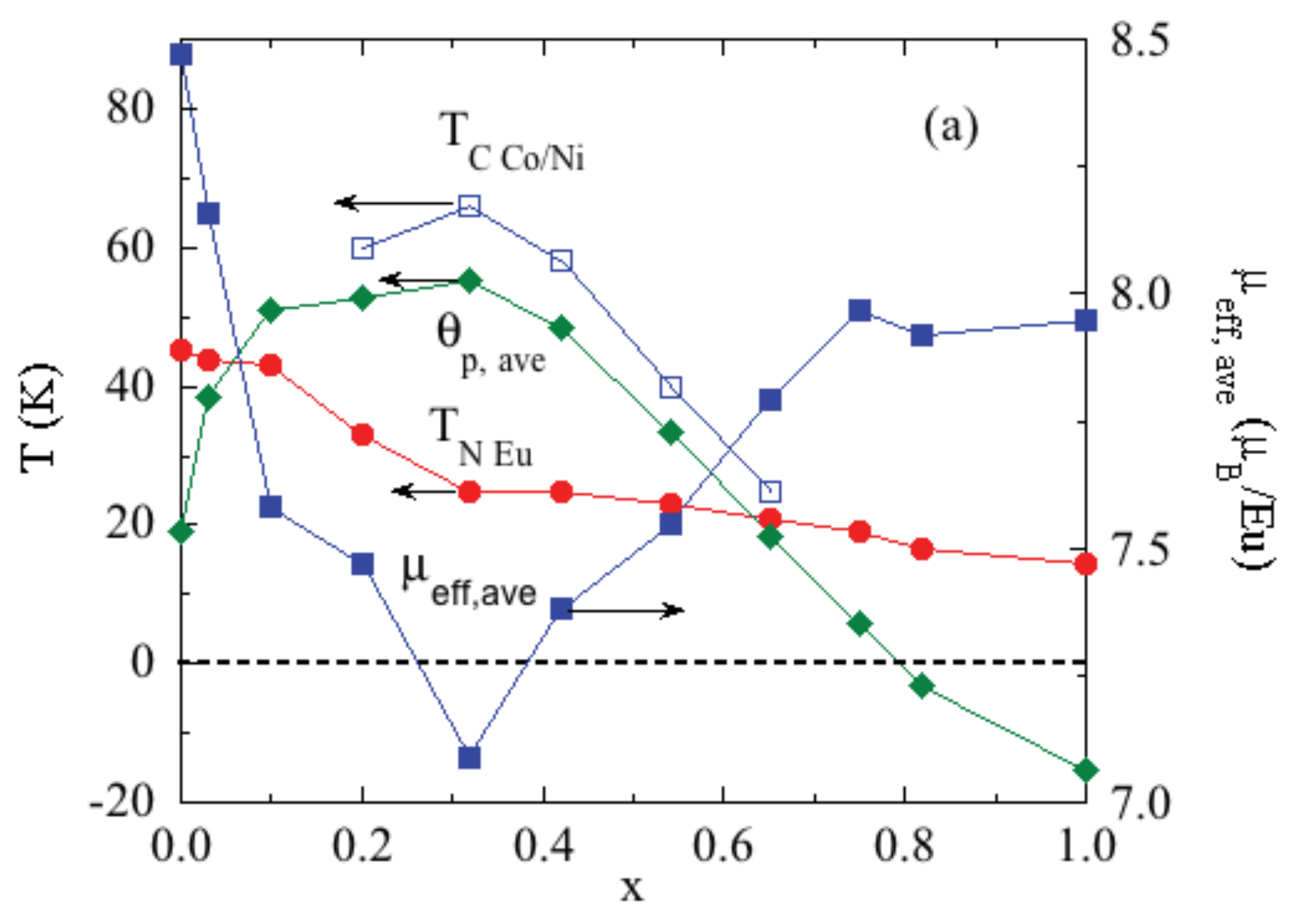}
\includegraphics[width=3.3in]{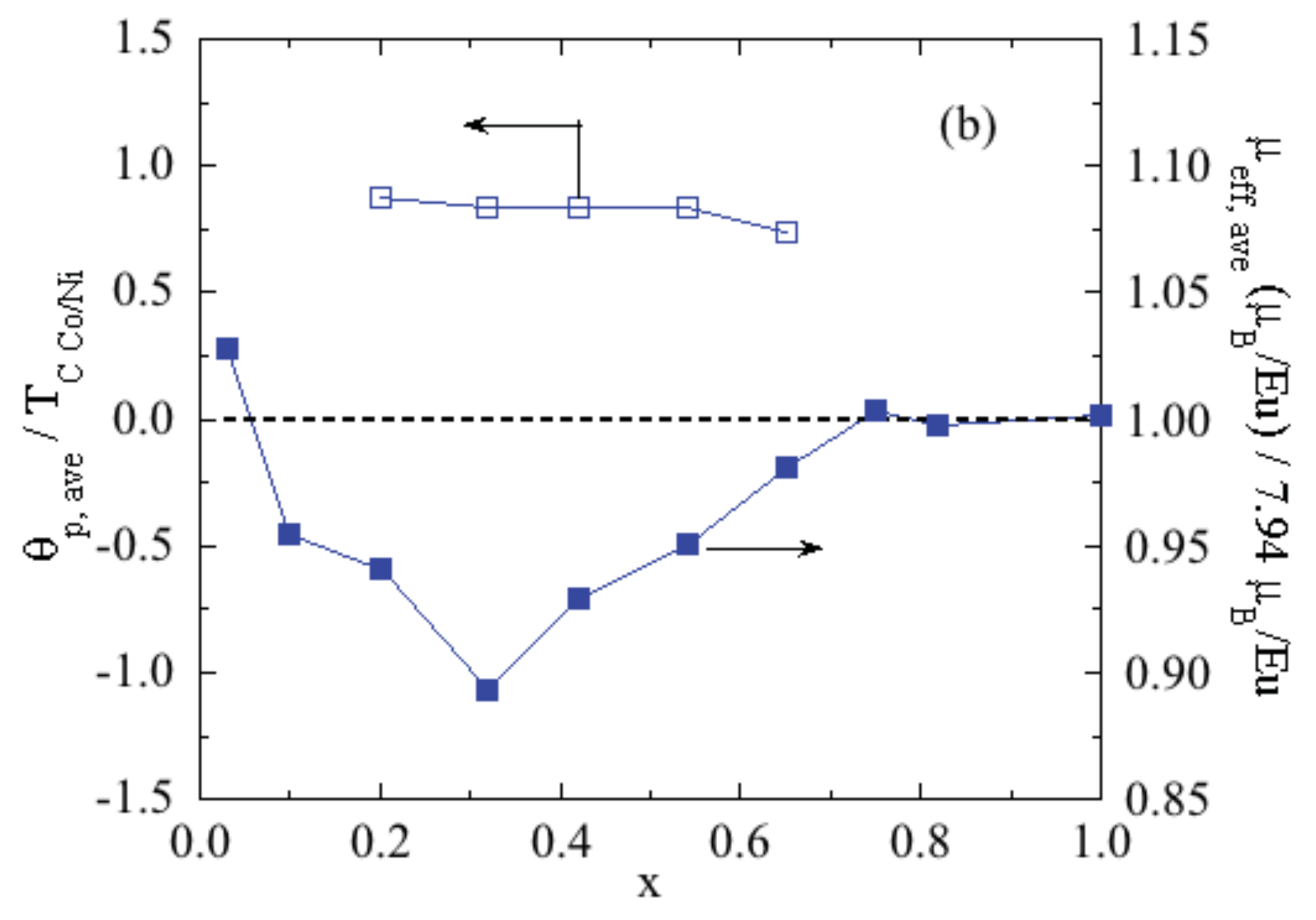}
\caption{(a)~N\'eel temperature  $T_{\rm N\,Eu}$ (filled red circles), spherically averaged Weiss temperature $\theta_{\rm p\,ave}$ (filled green diamonds), average effective moment $\mu_{\rm eff,\,ave}$ in $\mu_{\rm B}$/Eu (filled blue squares), and the second transition temperature $T_{\rm C\,Co/Ni}$ (open blue squares) versus~$x$ for \ecna\ crystals.   (b)~The ratios $f\equiv\theta_{\rm p,\,ave}/T_{\rm C\, Co/Ni}$ (empty blue squares) and $\mu_{\rm eff\,ave}/(7.94~\mu_{\rm B}$/Eu) (filled blue squares) versus~$x$.  The value $\mu_{\rm eff} = 7.94~\mu_{\rm B}$/Eu is the value expected for Eu$^{+2}$ spins with $S=7/2$ and $g=2$.  The data were obtained from Table~\ref{Tab.chidata}.  The lines are guides to the eye.}
\label{Eu(Co,Ni)2As2_magnetic_data}
\end{figure}

Figure~\ref{Eu(Co,Ni)2As2_magnetic_data}(a) shows plots of $T_{\rm N\,Eu}$, $T_{\rm C\,Co/Ni}$, $\theta_{\rm p\,ave}$, and $\mu_{\rm eff,\,ave}$ in units of $\mu_{\rm B}$/Eu from $x=0$ to~1.  It is informative to plot the ratio $\theta_{\rm p\,ave}/T_{\rm C\,Co/Ni}$ versus~$x$ and the ratio of $\mu_{\rm eff,\,ave}$ to the value $7.94~\mu_{\rm B}$/Eu versus~$x$ that the Eu spins would have for $S = 7/2$ and $g=2$.  These ratios are plotted in Fig.~\ref{Eu(Co,Ni)2As2_magnetic_data}(b).  Several important features are evident from Fig.~\ref{Eu(Co,Ni)2As2_magnetic_data}(b).  First, relative to the expected value of $7.94~\mu_{\rm B}$/Eu expected for Eu$^{+2}$ with $S=7/2$ and $g=2$, $\mu_{\rm eff}$ is enhanced for $x=0$ as previously reported~\cite{Sangeetha2018}.  However, on Ni doping this ratio decreases rapidly and reaches about 89\% of the expected value at $x=0.3$, then reverses course and increases, reaching the expected value of unity at $x=0.7$ and remaining at that value up to full substitution of Ni for Co at $x=1$.  The deviations of this ratio from unity may reflect the composition-dependent polarization of the Co $3d$ bands by the Eu moments.  Alternatively, the negative deviation of this ratio from unity may reflect the presence of an intermediate-valent state for the Eu atoms, where Eu$^{+3}$ does not carry a local moment.  However, the latter possibility is ruled out for $x=0.2$ and~0.65 by the $^{151}$Eu M\"ossbauer measurements in Sec.~\ref{Sec:Mossbauer} below.

The second important feature of Fig.~\ref{Eu(Co,Ni)2As2_magnetic_data}(b) is that the ratio $f = \theta_{\rm p,\,ave}$/$T_{\rm C\,Co/Ni}$ is nearly independent of composition with a value close to unity.  In a local-moment picture, $f\approx1$ is only found for a ferromagnet or a material which is nearly ferromagnetic~\cite{Johnston2015}.  This result is therefore consistent with the conclusion in Sec.~\ref{Sec:chi} that between $x=0.2$ and 0.65 the system has a FM component to the ordering which we show from the heat capacity measurements in Sec.~\ref{Sec:HC} is associated with the Co/Ni sublattice.

\section{\label{Sec:HighFieldM(H)} HIgh-field Magnetization versus applied magnetic field isotherms}

\begin{figure}
\includegraphics[width=3.5in]{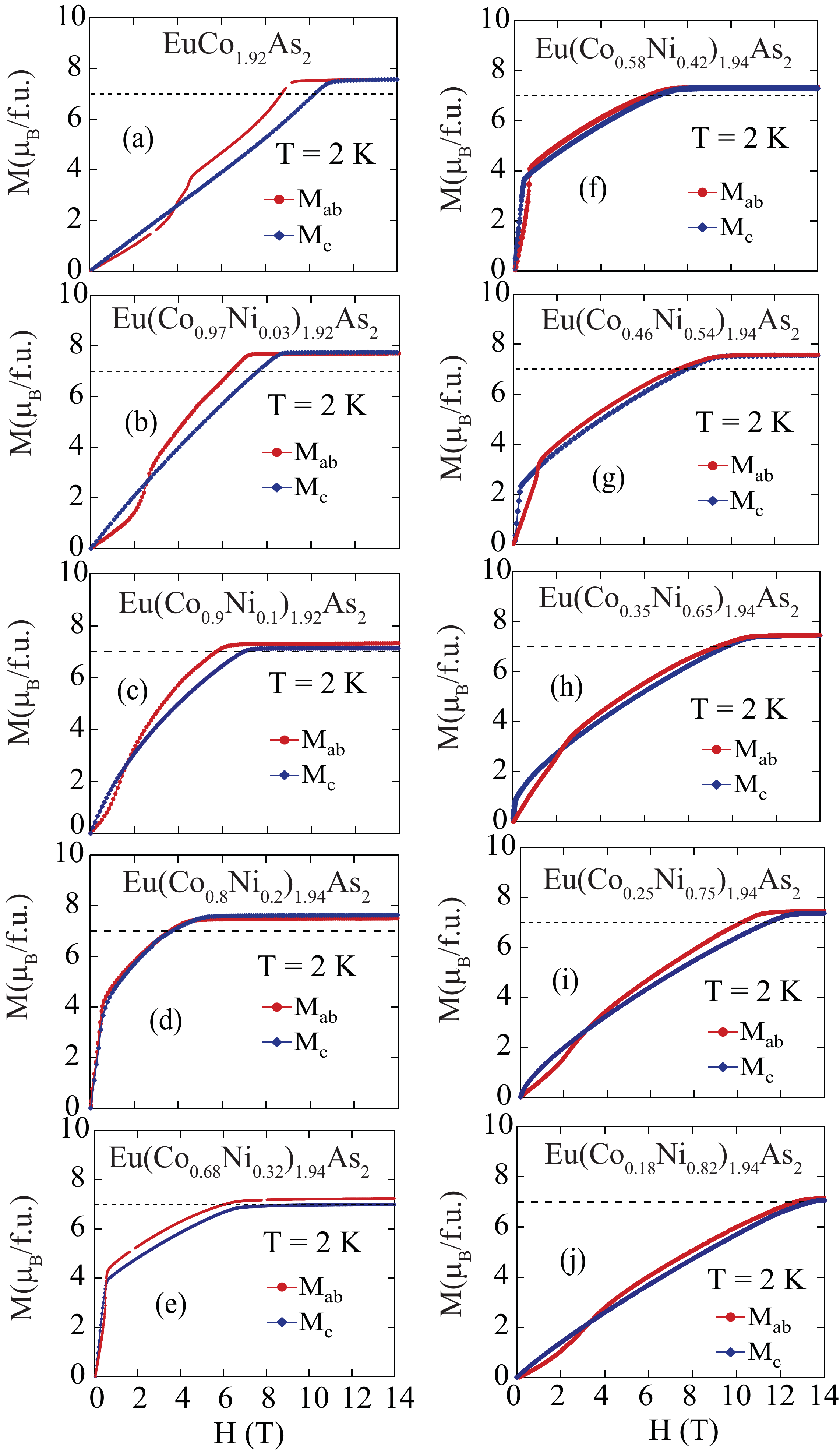}
\caption{Isothermal magnetization $M$ of \ecna\ ($x =$ 0, 0.03, 0.10, 0.20, 0.32, 0.42, 0.54, 0.65, 0.75, and 0.82) single crystals as a function of magnetic field~$H$ from 0 to 14~T measured at $T=2$~K with $H$ applied in the $ab$ plane ($M_{ab}$, $H\parallel ab$) and along the $c$ axis ($M_c$, $H\parallel c$).  The horizontal dashed lines indicate the magnetization $M = gS\mu_{\rm B}$/Eu = $7\mu_{\rm B}$/Eu assuming $S=7/2$ and $g=2$.}
\label{Fig:MH}
\end{figure}

\begin{figure}
\includegraphics[width=3.5in]{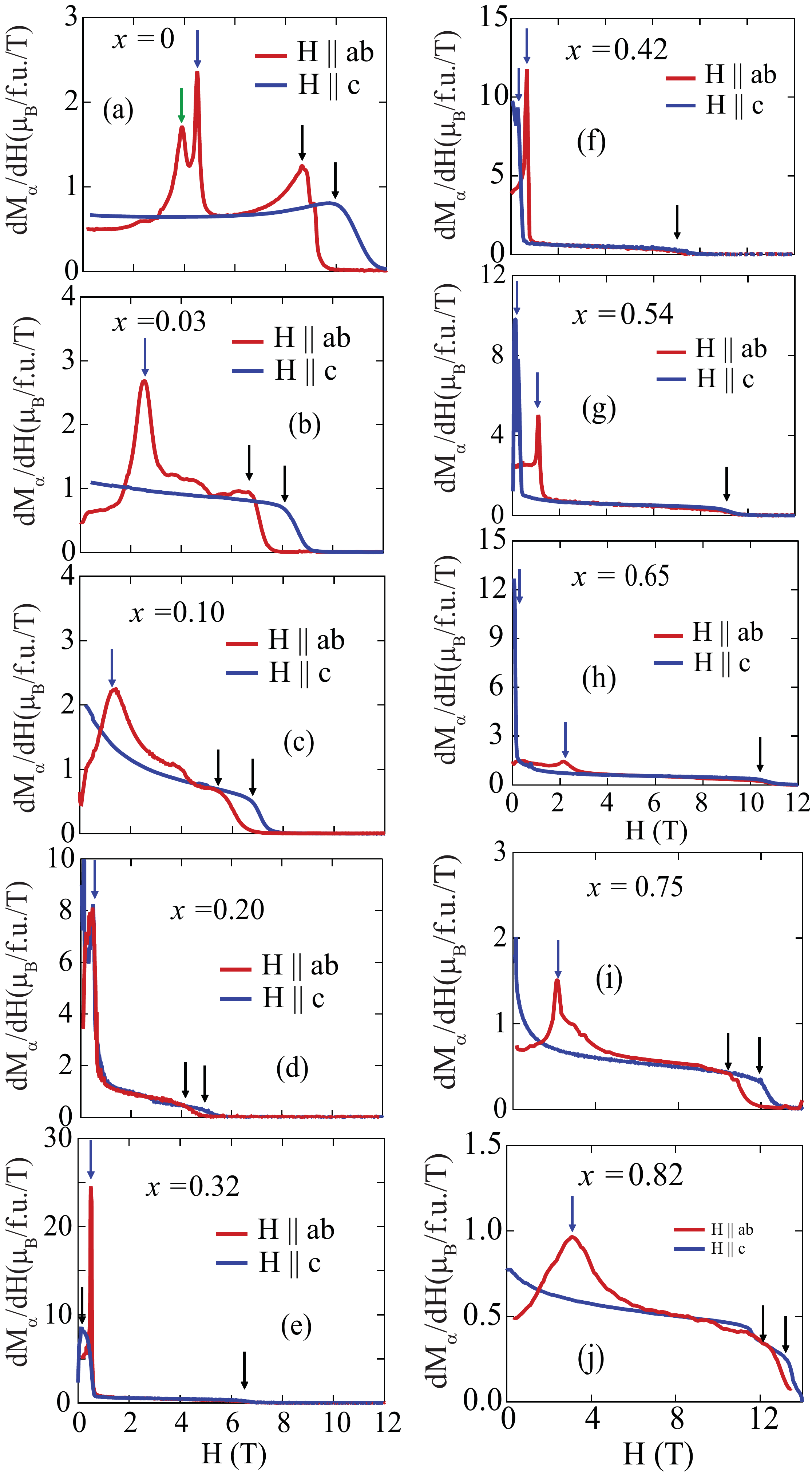}
\caption{Derivatives $dM_\alpha/dH$ of the isothermal magnetization $M_\alpha(H)$ of \ecna\ single crystals ($x =$ 0, 0.03, 0.10, 0.20, 0.32, 0.42, 0.54, 0.65, 0.75, and 0.82) from $H=0$ to 14~T measured at $T=2$~K with $H$ applied in the $ab$ plane ($M_{ab}$, $H\parallel ab$) and along the $c$ axis ($M_c$, $H\parallel c$). The transition fields $H_{\rm mm1}$ and $H_{\rm mm2}$ are marked by green and blue arrows, respectively. The critical fields $H_{\rm c\alpha}$ are marked by black arrows.}
\label{Fig:dMH}
\end{figure}

Isothermal magnetization versus applied magnetic field $M(H)$ isotherms for \ecna\  ($x =$ 0, 0.03, 0.10, 0.20, 0.32, 0.42, 0.54, 0.65, 0.75, and 0.82) measured at $T=2$~K with $0\leq H \leq 14$~T applied in the $ab$ plane ($M_{ab}, H\parallel ab$) and along the $c$~axis ($M_c, H \parallel c$) are shown in Fig.~\ref{Fig:MH}.  The $M_c(H)$ data are nearly linear in field for $x=0$ and 0.03 as predicted by MFT for a $c$-axis helix with the field applied along the helix axis~\cite{Johnston2015}. For the crystal with $x=0.1$, the $M_c(H)$ data exhibit negative curvature and the crystals  eventually show a spontaneous magnetization at low field upon doping for $0.2\leq x \leq 0.65$ signifying a metamagnetic transition at a field $H_{\rm mm}^c$. The $H_{\rm mm}^c$ and the critical field $H_{\rm c\perp}$ which separates the AFM from the PM phases versus~$T$ are taken to be the fields at which $dM/dH$ versus $H$ exhibit a peak or discontinuity as appear in Fig.~\ref{Fig:dMH}. The estimated values of $H_{\rm mm}^c$ along the $c$ axis are given in Table~\ref{Tab:MH}.  The $H_{\rm mm}^c$ first appears at 0.48~T for $x=0.2$, then decreases upon doping and finally disappears for $x>0.65$.  For the crystals with higher Ni doping $x=0.75$ and~0.82, the $M_c(H)$ data are again nearly linear in field in Figs.~\ref{Fig:MH}(i) and~\ref{Fig:MH}(j) as expected for a $c$-axis helical antiferromagnet.  The saturation moment $M_{\rm sat}$ and the critical field $H_{c\perp}$ are given in Table~\ref{Tab:MH}.  The symbol $\perp$ in $H_{\rm c\perp}$ refers to the critical field with $H$ applied parallel to the $c$ axis, i.e., with $H$ applied perpendicular to the zero-field plane of the ordered moments which is the $ab$ plane in this case.  The critical field $H_{\rm c\perp}$ initially shifts to lower field with Ni doping $0 \leq x \leq 0.2$ and then shifts to higher field upon Ni doping with $0.32 \leq x \leq 1$.   It is notable that the estimated value of $M_{\rm sat}$ at $H=14$~T is larger with Ni doping than the theoretical value $\mu_{\rm sat} = gS\mu_{\rm B}=7~\mu_{\rm B}$/Eu for $S=7/2$ and $g=2$, except for $x=0.32$ and 0.82 where $\mu_{\rm sat} = 7~\mu_{\rm B}$/Eu.

For $H\parallel ab$, the $M_{ab}(H)$ data in Fig.~\ref{Fig:MH}(a) for $x=0$ show a metamagnetic transition at $H\rm_{mm1} \approx 3.9$~T and another at $H_{\rm mm2}^{ab} \approx 4.5$~T which appears to be second order as reported in Ref.~\cite{Sangeetha2018}. The transition fields $H^{ab}_{\rm mm1}$ and  $H_{\rm mm2}^{ab}$ and the critical field $H_{\rm c\parallel}$ which separates the AFM from the paramagnetic (PM) phases  are taken to be the fields at which $dM/dH$ versus $H$ exhibits a peak or a discontinuity (shown in Fig.~\ref{Fig:dMH}). The results are given in Table~\ref{Tab:MH}. One sees that $H^{ab}_{\rm mm1}$ disappears with a trace amount of Ni doping whereas $H_{\rm mm2}^{ab}$ initially decreases continuously upon Ni doping from 4.5~T at $x=0$ to 0.5~T at $x=0.2$ and then increases for $x>0.2$ and reaches 9.51~T at $x=1$. 

According to Figs.~\ref{Fig:MH}(d)--\ref{Fig:MH}(f), both $M_{ab}(H)$ and $M_c(H)$  are almost isotropic for $x=0.20$, 0.32, and~0.42.  As discussed above in Sec.~\ref{Sec:chi}, the \ecna\ system exhibits two magnetic transitions for $0.2 \leq x \leq 0.65$. 

\begin{table}
\caption{\label{Tab:MH} Metamagnetic fields $H^{ab}_{\rm mm1}$ and $H^{ab}_{\rm mm2}$, and the critical fields with the field parallel to the $ab$~plane $H_{\rm c\parallel}$ and parallel to the $c$~axis ($H\parallel c,~H_{\rm c\perp})$ of \ecna\ single crystals at $T=2$~K, determined from the isothermal magnetization $M_\alpha(H)$ data in Figs.~\ref{Fig:MT100oe}(f)--\ref{Fig:MT100oe}(j), \ref{Fig:MH}, and~\ref{Fig:dMH}.  Also shown are the respective saturation moments $\mu_{\rm sat}$ and the spherical average $\mu_{\rm sat\,ave}$, all in units of $\mu_{\rm B}$/f.u. }
\begin{ruledtabular}
\begin{tabular}{c|cccc|ccc|c}
				&\multicolumn{4}{c|}{$H\parallel ab$} & \multicolumn{3}{c|} {$H\parallel c$}  \\ 
$x$				&  $H^{ab}_{\rm mm1}$  & $H_{\rm mm2}^{ab}$ &  $H_{\rm c\parallel}$ & $\mu_{\rm sat}$ &$H_{\rm mm}^c$& $H_{\rm c\perp}$  & $\mu_{\rm sat}$ &  $\mu_{\rm sat,\,ave}$\\
  	        	 &    (T)           	 & (T)     	& (T)  	&	$(\mu_{\rm B}$)	& (T) & (T) & $(\mu_{\rm B}$) & $(\mu_{\rm B}$)\\
\hline
0 	  	& 3.9		&4.5		& 8.79		&7.59	&		&  9.9	& 7.57 		&7.58\\
0.03	 &			&2.6		& 6.8		&7.70	&		&8.02	&7.76		&7.72\\
0.10		& 	 		&1.3		& 5.5		&7.32	&		&6.7	&7.13		&7.25\\
0.20		 &			&0.5		&3.6		&7.50	& 0.48	& 4.6	&7.62		&7.54\\
0.32	 &			&0.53		&6.25		&7.23	&0.35	&6.4	&7.01		&7.15\\
0.42	&			&0.77		&7.4		&7.3	&0.28	&7.3	&7.3		&7.3\\
0.54	&			&1.09		&9.2		&7.58	&0.25	&9.3	&7.54		&7.56\\
0.65	&			&2.2		&10.2		&7.44	&0.15	&10.4	&7.44		&7.44\\
0.75	&			&2.1		&10.8		&7.46	&		&12.1	&7.33		&7.42\\	
0.82	& 			&3.07		& 12.9		&7.19	&		&13.4	&7.04		&7.14\\
1		&3.15		&9.51		& $>14$	&		& 		&$>14$ &			&\\
\end{tabular}
\end{ruledtabular}
\end{table}

\section{\label{Sec:Mossbauer} M\"ossbauer spectroscopy of polycrystalline samples}

\begin{figure}
\includegraphics[width = 3.3in]{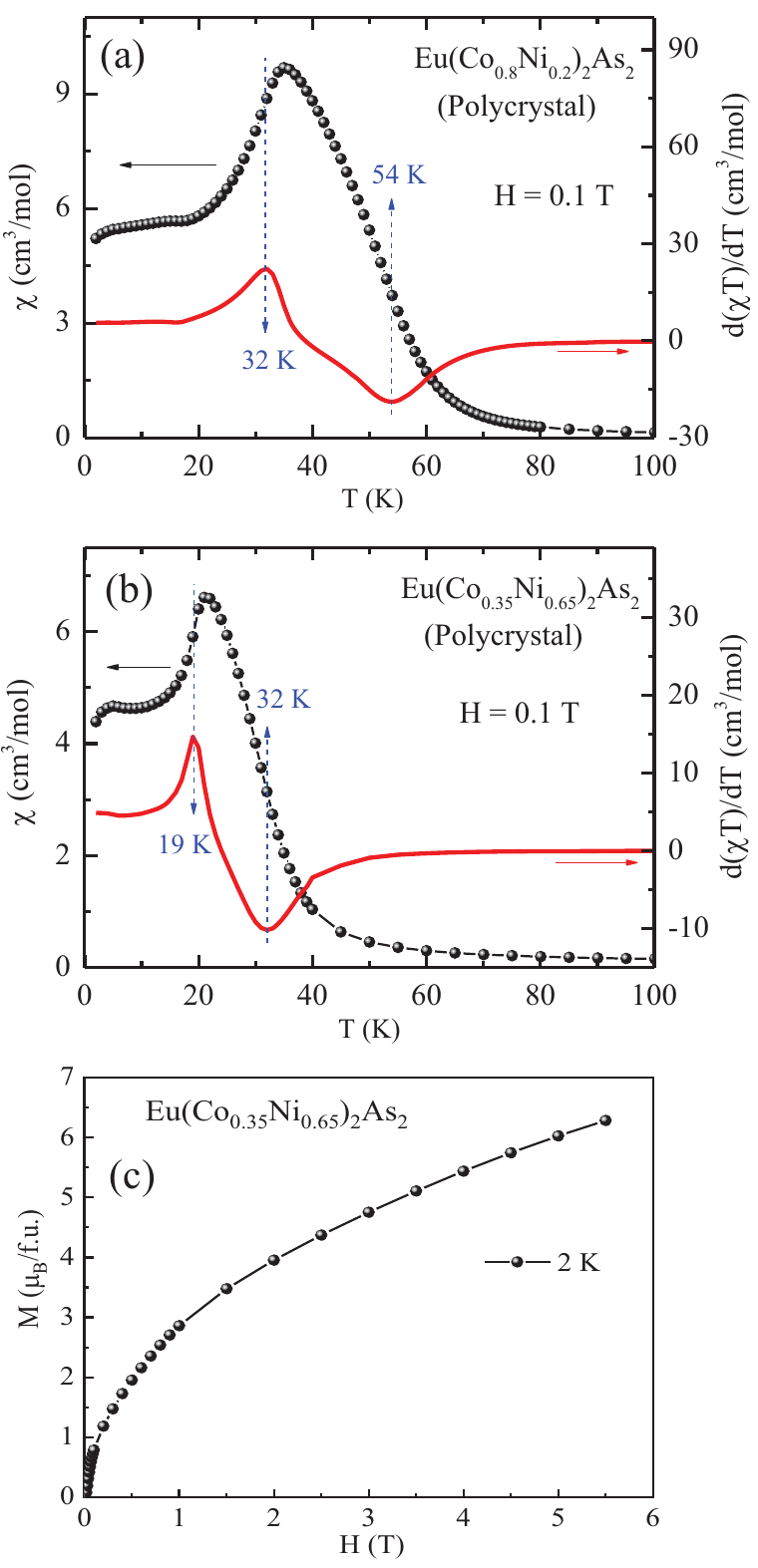}
\caption {(Left ordinate) Temperature dependence of the magnetic susceptibility ($\chi = M/H$) in a magnetic field $H =0.1$~T of \ecnaa\ polycrystalline samples with \mbox{(a)~$x = 0.20$}, and (b)~$x = 0.65$. The right ordinate of (a) and~(b) is the temperature derivative $d(\chi T)/dT$ versus~$T$ obtained from the respective $\chi(T)$ data. (c)~Magnetization~$M$ versus applied magnetic field~$H$ for $x=0.65$ at $T=2$~K\@.}
\label{Fig_polycrystal_magnetic}
\end{figure}

Magnetic susceptibility data were obtained for the polycrystalline $x=0.2$ and 0.65 samples used in the M\"ossbauer experiments.  Figures~\ref{Fig_polycrystal_magnetic}(a) and~\ref{Fig_polycrystal_magnetic}(b) (left-hand ordinates) depict $\chi(T)$ of these \ecnaa\  samples measured in $H = 0.1$~T under zero-field-cooled (ZFC) conditions, respectively. The magnetic transition temperatures obtained from the temperatures of the peaks in $d(\chi T)/dT$ versus~$T$ plots (right-hand ordinates) are $\sim 54$~K and $\sim 32$~K for $x = 0.2$ and $\sim 32$~K and $\sim 19$~K for $x = 0.65$.  The higher-temperature troughs in Figs.~\ref{Fig_polycrystal_magnetic}(a) and~\ref{Fig_polycrystal_magnetic}(b) are at the FM ordering temperatures $T_{\rm C\,Co/Ni}0$ associated with the Co/Ni atoms and the lower-temperature peaks are at the N\'eel temperatures $T_{\rm N\,Eu}$ of the Eu atoms.  The $M(H)$ isotherm at $T=2$~K for $x=0.65$ is shown in Fig.~\ref{Fig_polycrystal_magnetic}(c), where the data at low fields suggest a FM component to the magnetic ordering and at higher fields the data tend towards saturation of the Eu spins to the expected value $\approx 7\,\mu_{\rm B}$/Eu.

\begin{figure}
\includegraphics[width=2.5in]{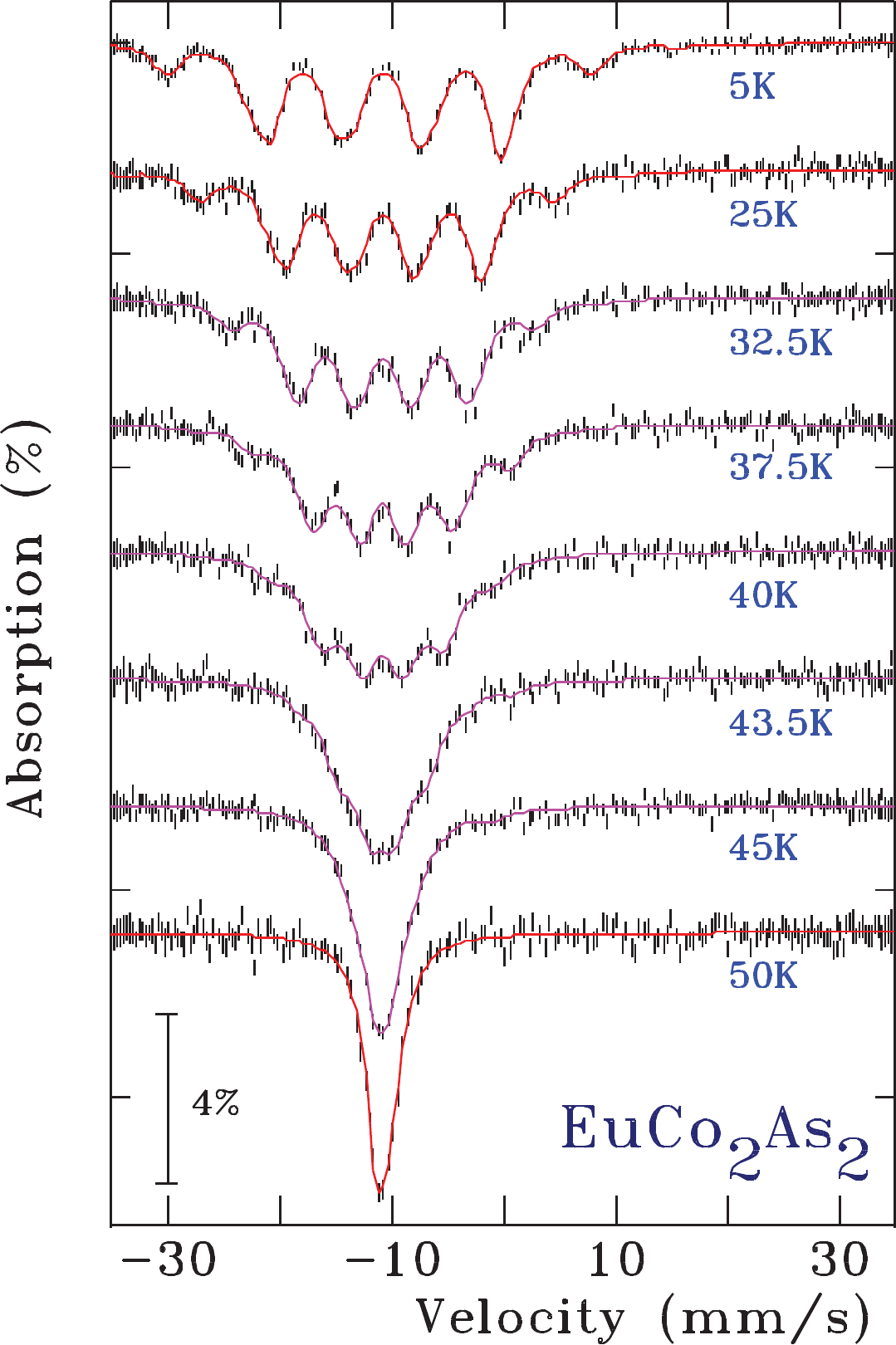}
\caption{$^{151}$Eu M\"ossbauer spectra of $\rm EuCo_{2-y}As_2$ showing the
evolution of the spectra with temperature.  The solid lines are fits derived 
from either a full Hamiltonian solution (red lines: $T\leq30$~K and $T=50$~K) or
from the modulated model (magenta lines: 32.5~K $\leq T \leq50$~K).
See text for details.  }
\label{fig:Ni00_spectra}
\end{figure}

The $^{151}$Eu M\"ossbauer spectra of \ecaa\ shown in
Fig.~\ref{fig:Ni00_spectra} confirm that the europium is fully divalent [isomer
shift of $-10.72(2)$~mm/s at 5.4~K] and the sharp lines [HWHM = 1.11(2)~mm/s] show
that the Eu$^{+2}$ moments are in a single magnetic environment with a hyperfine
field ($B_{\rm hf}$) of 26.21(5)~T\@. There is a small asymmetry apparent in the
spectrum (the four strongest lines increase in intensity going from negative to
positive velocity) due to the presence of an electric quadrupole contribution.  If we assume that the moments are aligned in the $ab$~plane~\cite{Tan2016, Sangeetha2018} and are
thus perpendicular to $V_{zz}$ ($\theta=90^{\circ}$), then we obtain a
quadrupole contribution of $-3.2(4)$~mm/s. The contribution is too small for a
free fit of both $\theta$ and $V_{zz}$ to yield a unique solution.

%%%%%%%%%%%%%%%%%%%
%%%%%%%%%%%%%%%%%%%

In all three samples that were studied using $^{151}$Eu M\"ossbauer
spectroscopy, we observed significant line broadening at intermediate
temperatures. This broadening is too severe to be consistent with a varying
canting angle between the $c$~axis and the moments, and its form is inconsistent
with the presence of slow paramagnetic relaxation. We conclude that the
broadening indicates that time-averaged moments on the europium atoms do not
take a single value throughout the sample over the temperature range studied. The form of this broadening is more consistent with the development of
a static incommensurate modulated contribution to the magnetic order in which
the magnitude of the europium moments varies from site to site within the
material. The intermediate-temperature spectra were therefore fitted using a
model that derives a distribution of hyperfine fields from an (assumed)
incommensurate sinusoidally-modulated magnetic structure~\cite{bonville349,
maurya216001}.

If we assume that the moment modulation along the
direction of the propagation vector {\bf k} can be written in terms of its
Fourier components, and further assume that the observed hyperfine field is a
linear function of the magnitude of the Eu moment at any given site, then the
variation of $B_{\rm hf}$ with distance $x$ along the propagation vector {\bf k} can
be written~\cite{bonville349}
\begin{equation}
B_{\rm hf}(kx) = Bk_0 + \sum^n_{l=0} Bk_{2l+1} \sin [ (2l+1)kx ],
\label{eqn:fourier}
\end{equation}
where the $Bk_n$ are the odd Fourier coefficients of the field (moment) modulation. Since
$+B_{\rm hf}$ and $-B_{\rm hf}$ are indistinguishable, $kx$ only needs to run over
half the modulation period, and in this case, a square-wave modulated structure
can be modeled either as a sum over a very large number of Fourier
coefficients, or by simply using the $Bk_0$ term with all of the other $Bk_n$
set to zero. We found the fits to be far more stable with the $Bk_0$ term
included rather than using a large set of $Bk_n$; however, the two approaches are
effectively equivalent. Variations of this model have also been used to fit spectra of \mbox{EuPdSb}~\cite{bonville349} and Eu$_4$PdMg~\cite{ryan17d108}. 

%%%%%%%%%%%%%%%%%%%
\begin{figure}
\includegraphics[width=2.75in]{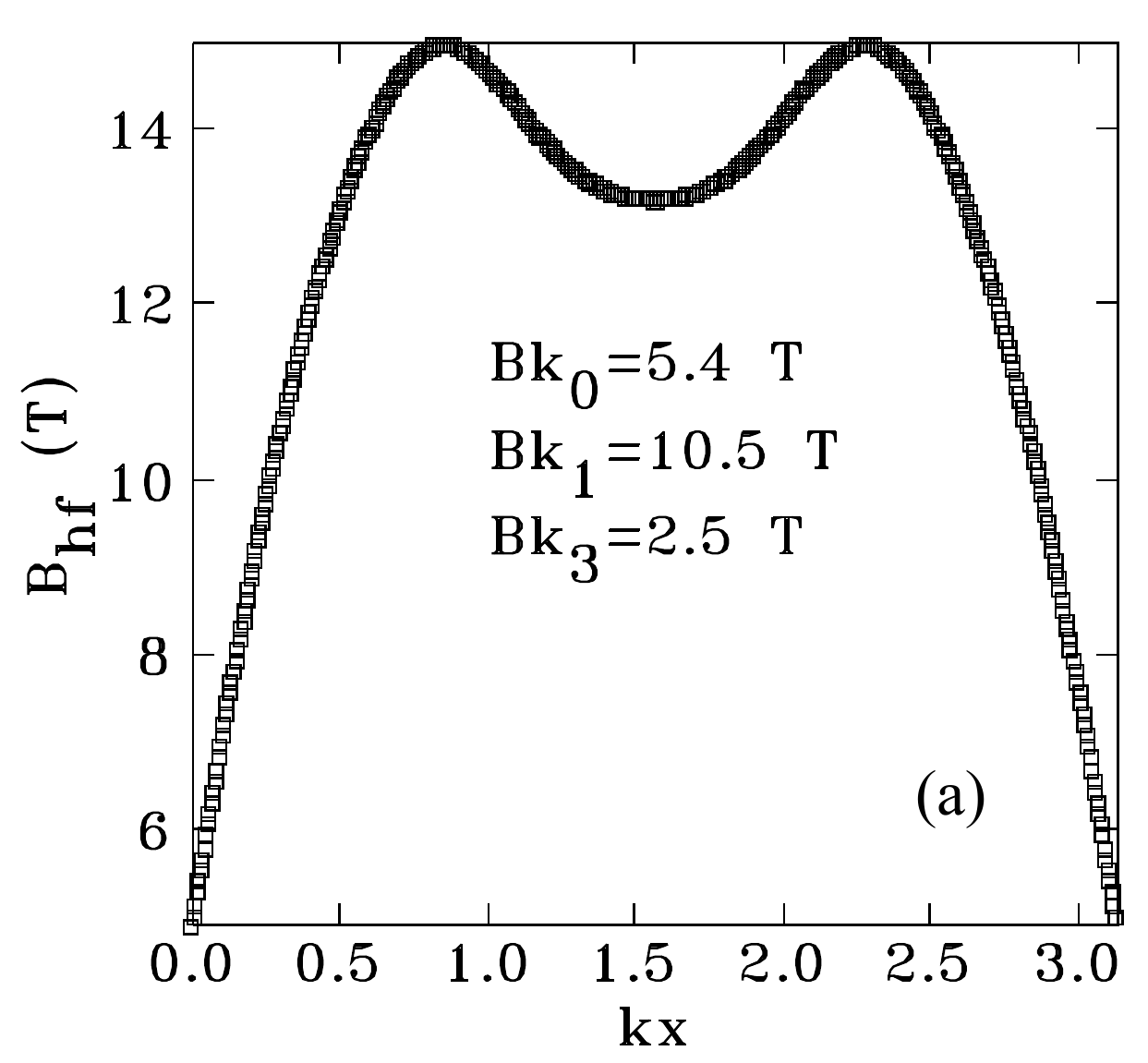}
\includegraphics[width=2.75in]{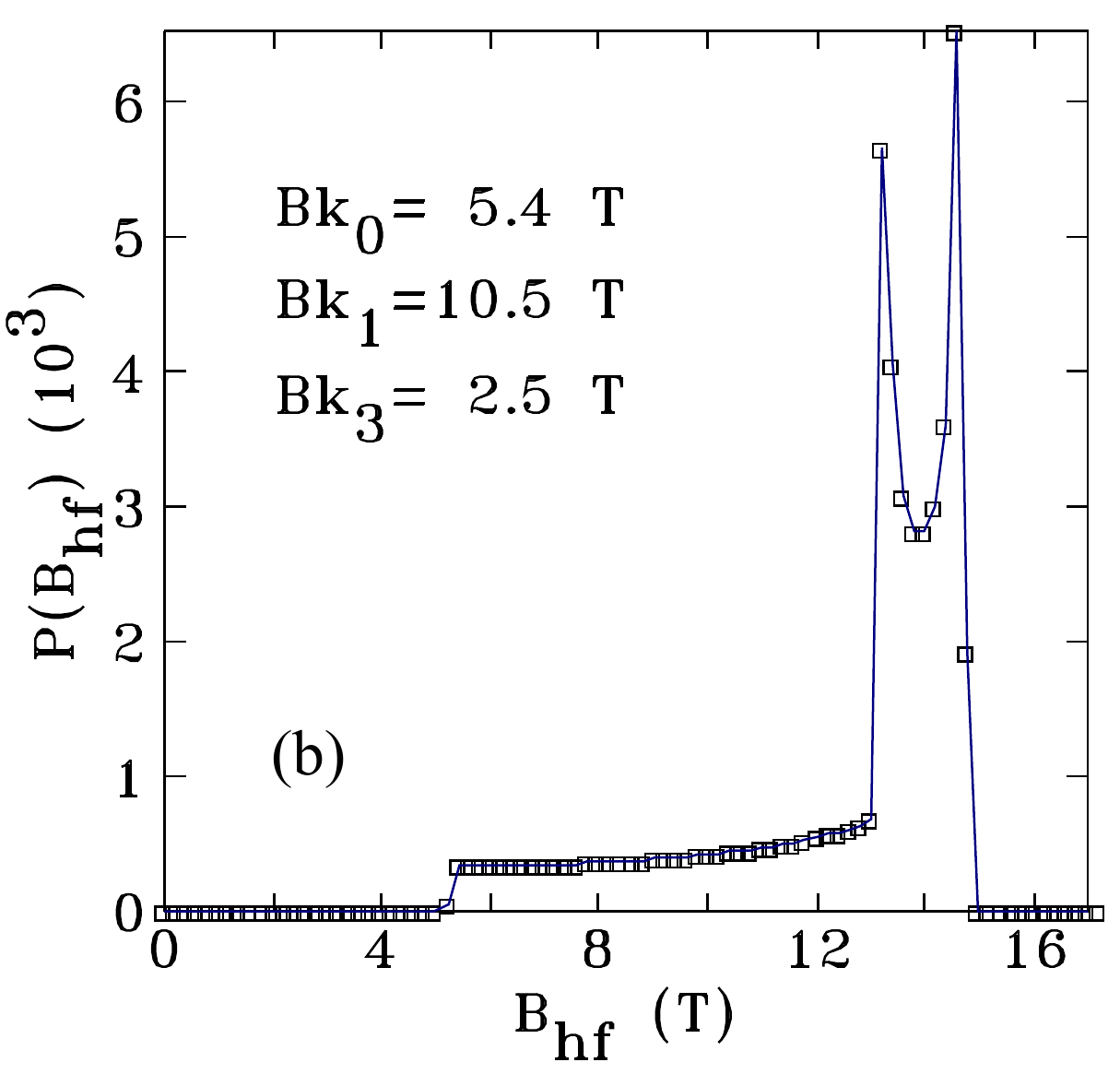}
\caption{(a)~Variation of the hyperfine field $B_{\rm hf}$ versus $kx$ at $T= 40$~K obtained from the Fourier components of the 40~K data in Fig.~\ref{fig:Ni00_Bhf} and Eq.~(\ref{eqn:fourier}). The third harmonic $Bk_3$, not plotted in Fig.~{\ref{fig:Ni00_Bhf}}, is responsible for the dip at $kx=\pi/2$.  (b)~Unnormalized probability distribution $P(B_{\rm hf})$ versus $B_{\rm hf}$.}
\label{B_vs_kx}
\end{figure}
%%%%%%%%%%%%%%%%%%%

The sequence for fitting the spectra  is as follows.  Starting with an initial set of $Bk_n$, the
evolution of $B_{\rm hf}$ along the propagation vector~$x$ can be evaluated [a
typical example for EuCo$_{2-y}$As$_2$ at 40~K is shown in Fig.~\ref{B_vs_kx}(a)]. From
$B_{\rm hf}$ versus $kx$, a histogram containing the probability $P(B_{\rm hf}$) versus $B_{\rm hf}$ can be constructed [Fig.~\ref{B_vs_kx}(b)], and the calculated $^{151}$Eu M\"ossbauer spectrum is
then obtained through a sum of magnetic patterns weighted according to the histogram.
A conventional nonlinear least-squares minimization routine is then used to
adjust the $Bk_n$ values (and a number of other general parameters including a uniform
baseline, overall scale factor, and isomer shift) to minimize $\chi ^2$.

\begin{figure}
\includegraphics[width=2.75in]{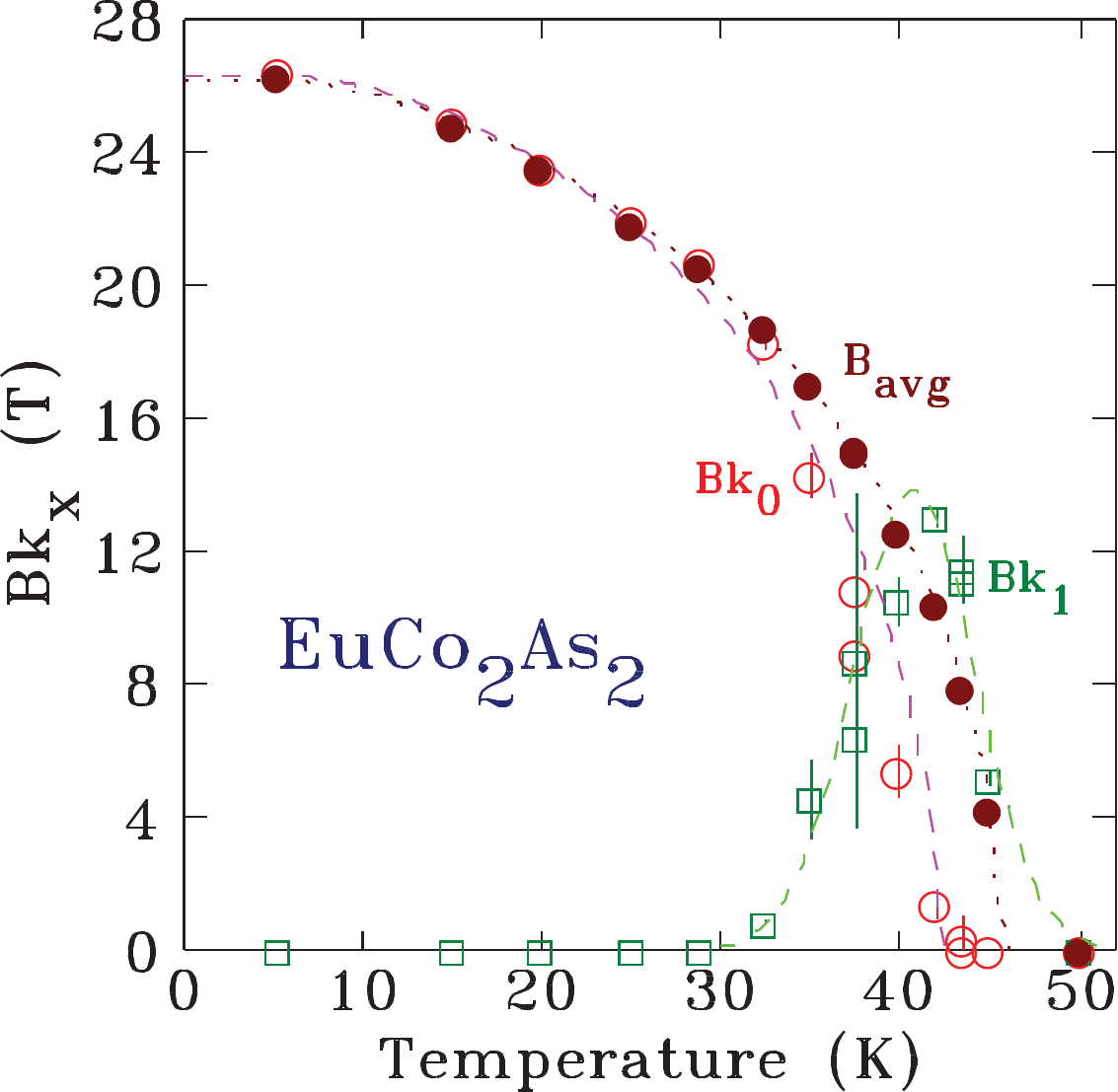}
\caption{Temperature dependence of the fitted hyperfine field contributions 
in $\rm EuCo_2As_2$ derived from the modulated model. The average hyperfine field
($B_{\rm avg}$) is plotted as solid round symbols with a dotted line showing a fit (see text)
yielding a transition temperature of 45.6(1)~K, while the amplitude of the
uniform term ($Bk_0$) is plotted as open round symbols with a dashed line fit
giving a transition temperature of 42.0(1)~K\@. The behavior of the first Fourier
component ($Bk_1$) is plotted as open square symbols and shows a peak at 40.8(3)~K\@. The third harmonic $Bk_3(T)$ is not plotted.}
\label{fig:Ni00_Bhf}
\end{figure}

On increasing the temperature, Fig.~\ref{fig:Ni00_Bhf} shows a conventional reduction in the average hyperfine field $B_{\rm avg}(T)$ where $B_{\rm avg}$ was calculated from the fitted hyperfine field distribution in Eq.~(\ref{eqn:fourier}) using only the positive half of the period.  Fitting $B_{\rm avg}(T)$ by a $S=7/2$ Brillouin function yields a transition temperature of 45.6(1)~K as shown by the black dashed curve in Fig.~\ref{fig:Ni00_Bhf}, close to the $T_{\rm N\,Eu}$ for $x=0$ in Table~\ref{Tab.chidata}.  However, closer inspection of the spectra in Fig.~\ref{fig:Ni00_spectra}, particularly the 37.5~K spectrum, reveals a characteristic ``settling'' towards the center of the pattern consistent with the development of an incommensurate modulation of the magnitude of the moments. The modulated contribution starts to appear by 32.5~K and rapidly comes to dominate the spectrum with increasing temperature. Figure~\ref{fig:Ni00_Bhf} shows that the uniform term ($Bk_0$) vanishes on heating through 42~K as the first harmonic ($Bk_1$) peaks.  As we will show, this ordering sequence (uniform $\rightarrow$ modulated $\rightarrow$ paramagnetic) is observed on warming in all three of the samples studied here using $^{151}$Eu M\"ossbauer spectroscopy. 

%%%%%%%%%%%%%%%%%%%
%%%%%%%%%%%%%%%%%%%

From the Fourier components in Fig.~\ref{fig:Ni00_Bhf}, one can calculate the hyperfine field $B_{\rm hf}$ seen by the Eu spins versus position along the (unspecified) $x$~axis using Eq.~(\ref{eqn:fourier}).  Shown in Fig.~\ref{B_vs_kx}(a) is a plot of $B_{\rm hf}$ versus $kx$, where $k$ is the propagation vector of the Eu-spin ordering in the $x$~direction which is presumably the $c$-axis direction for a $c$-axis helix.   Because the $T=0$ hyperfine field from Fig.~\ref{fig:Ni00_Bhf} is about 26~T, the modulation of the hyperfine field at 40~K in Fig.~\ref{B_vs_kx}(a) corresponds to a modulation of the Eu moment of $\approx 0.58(7\mu_{\rm B}) = 4.0\mu_{\rm B}$, a rather large modulation.  The unnormalized probability distribution $P(B_{\rm hf})$ versus $B_{\rm hf}$ is plotted in Fig.~\ref{B_vs_kx}(b).  These data were obtained as the inverse of the slope of the plot in Fig.~\ref{B_vs_kx}(a) versus $B_{\rm hf}$ after binning the data to avoid divergences at the peaks.

It is essential to note that since the magnetic-moment modulation is explicitly assumed to be
incommensurate with the crystal lattice, positions along $kx$ in Fig.~\ref{B_vs_kx}(a) do
not correspond in any way to specific positions in the chemical cell. Rather, 
they represent how $B_{\rm hf}$, which by assumption is proportional to $\mu _{\rm Eu}$, is sampled by the europium atoms. As the periodicity of the modulation and the crystal cell are
not related by a simple rational fraction, neighboring points along the $kx$
axis, in reciprocal space, will be far apart in real space.

Within MFT, if an amplitude-modulated magnetic structure occurs on cooling below $T_{\rm N\,Eu}$ as inferred from Figs.~\ref{fig:Ni00_spectra}--\ref{fig:Ni00_Bhf}, the heat-capacity jump on cooling below $T_{\rm N\,Eu}$ is expected to be significantly smaller~\cite{Blanko1991, Rotter2001} than the MFT value of 21.4~J/mol~spins for $S=7/2$ ~\cite{Sangeetha2018}.  However, for an \ecaa\ crystal the measured magnetic contribution $C_{\rm mag}(T)$ to the heat capacity exhibits a MFT heat capacity jump close to the theoretical value~\cite{Sangeetha2018}.

\begin{figure}
\includegraphics[width=2.5in]{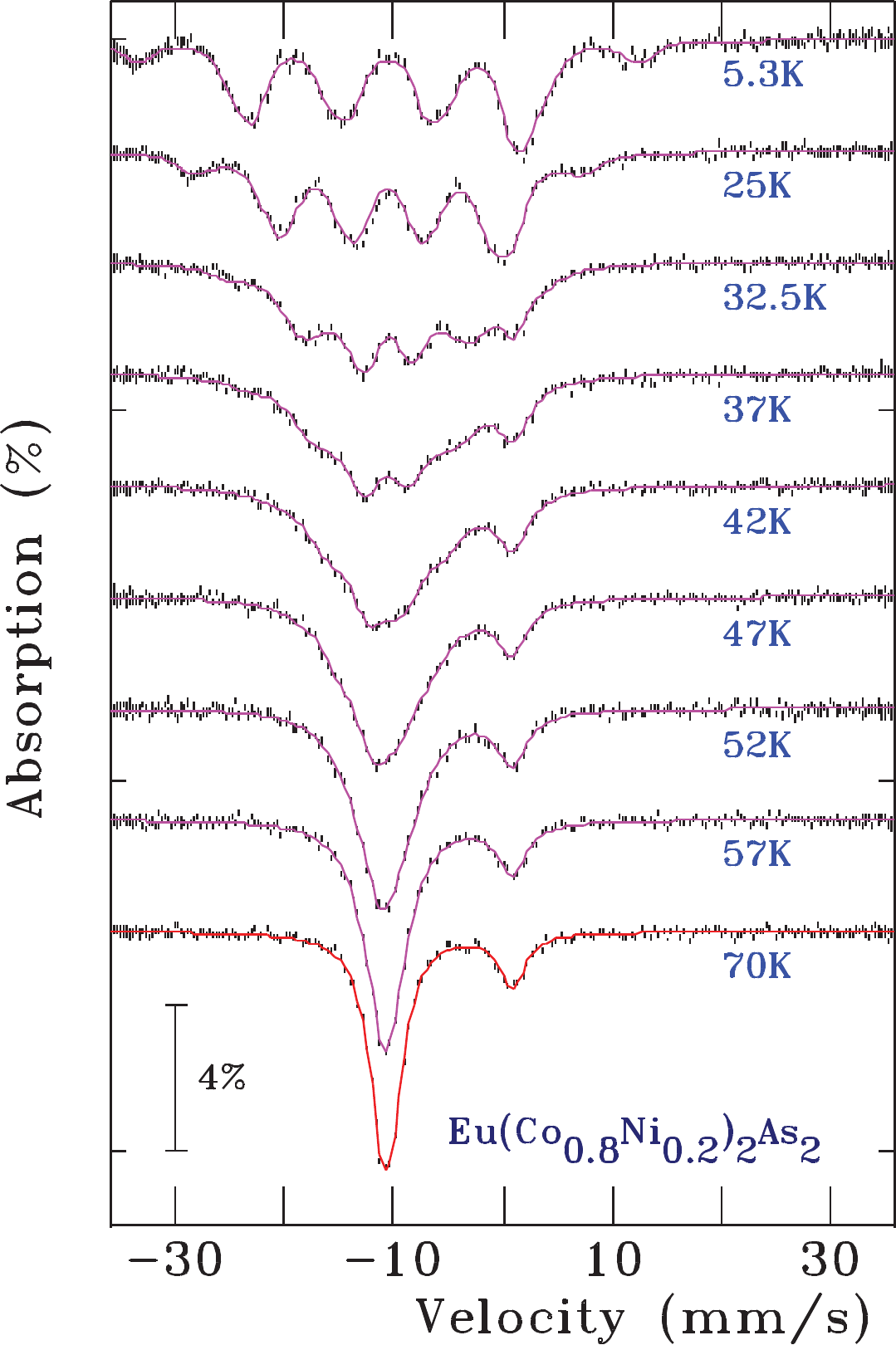}
\caption{$^{151}$Eu M\"ossbauer spectra of $\rm Eu(Co_{0.8}Ni_{0.2})_2As_2$ showing the
evolution of the spectra with temperature.
The solid lines are fits derived
from either a full Hamiltonian solution (red line: $T=70$~K) or from the modulated
model (magenta lines: $T<70$~K). See text for details. The additional feature
near 0~mm/s is due to a nonmagnetic Eu$^{+3}$ impurity, possibly introduced during handling,
that accounts for 7.6(4)\% of the total spectral area at 5.3~K. }
\label{fig:Ni20_spectra}
\end{figure}

\begin{figure}
\includegraphics[width=2.5in]{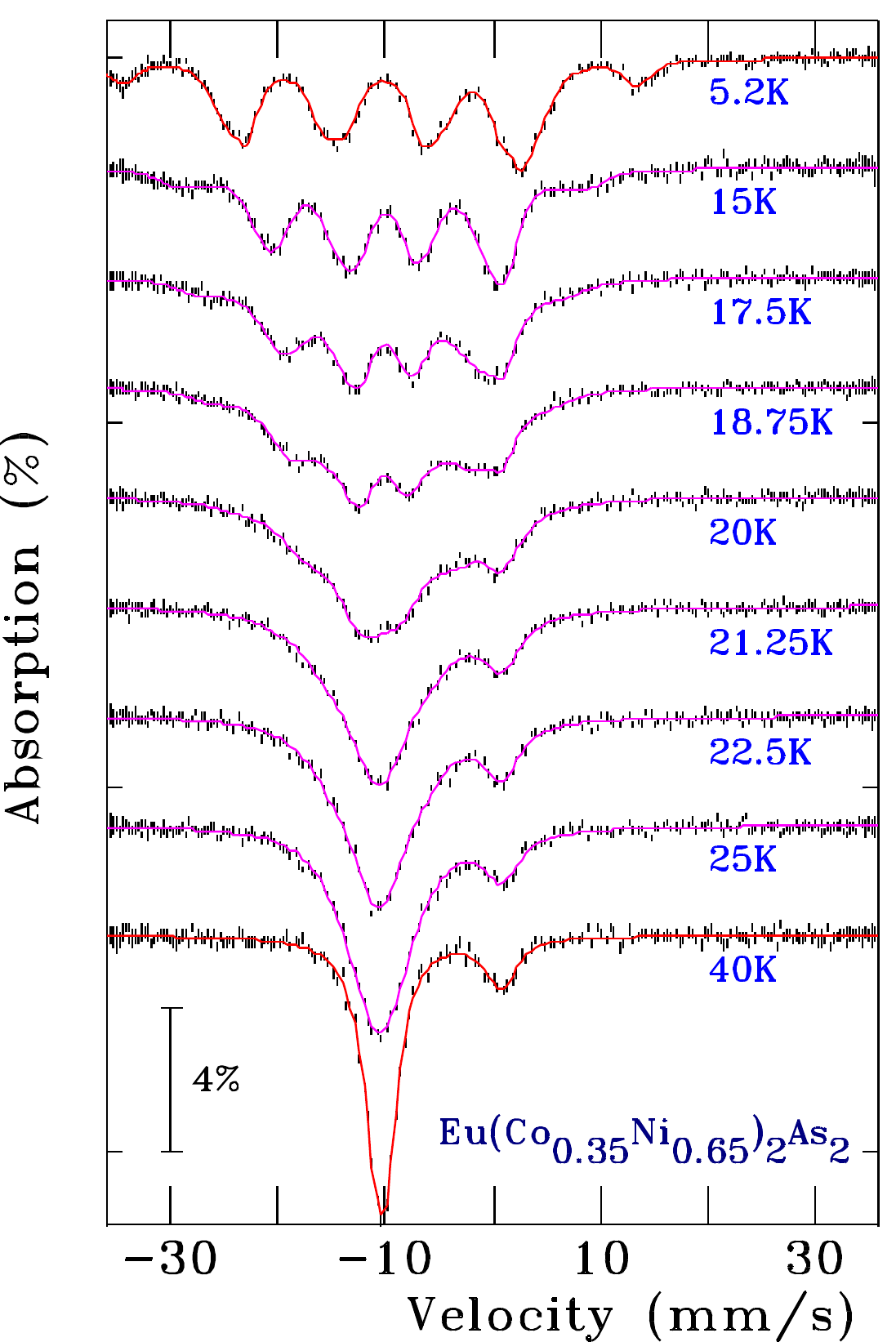}
\caption{$^{151}$Eu M\"ossbauer spectra of $\rm Eu(Co_{0.35}Ni_{0.65})_2As_2$
showing the evolution of the spectra with temperature.  The solid lines are fits
derived from either a full Hamiltonian solution (red lines: $T=5.2$ and 40~K) or
from the modulated model (magenta lines: 15~K~$\leq T \leq 25$~K). See text for
details. The additional feature near 0~mm/s is due to a nonmagnetic Eu$^{+3}$
impurity, possibly introduced during handling, that accounts for 7.4(3)\% of the
total spectral area at 5.2~K.}
\label{fig:Ni65_spectra}
\end{figure}

The $^{151}$Eu M\"ossbauer spectra of $\rm Eu(Co_{0.8}Ni_{0.2})_2As_2$ shown in
Fig.~\ref{fig:Ni20_spectra} and of $\rm Eu(Co_{0.35}Ni_{0.65})_2As_2$ shown in
Fig.~\ref{fig:Ni65_spectra} exhibit the same progression as seen for $\rm
EuCo_2As_2$ with one clear difference. There is an additional feature at
$\sim0$~mm/s that accounts for about 7\% of the total area. We attribute this to
a nonmagnetic Eu$^{+3}$ impurity, possibly introduced during handling. As
this impurity contribution is well defined and consistent in its behavior it is
easily included in the analysis without distorting the results.

%%%%%%%%%%%%%%%%%%%
%%%%%%%%%%%%%%%%%%%

%%%%%%%%%%%%%%%%%%%
%%%%%%%%%%%%%%%%%%%

\begin{figure}
\includegraphics[width=2.75in]{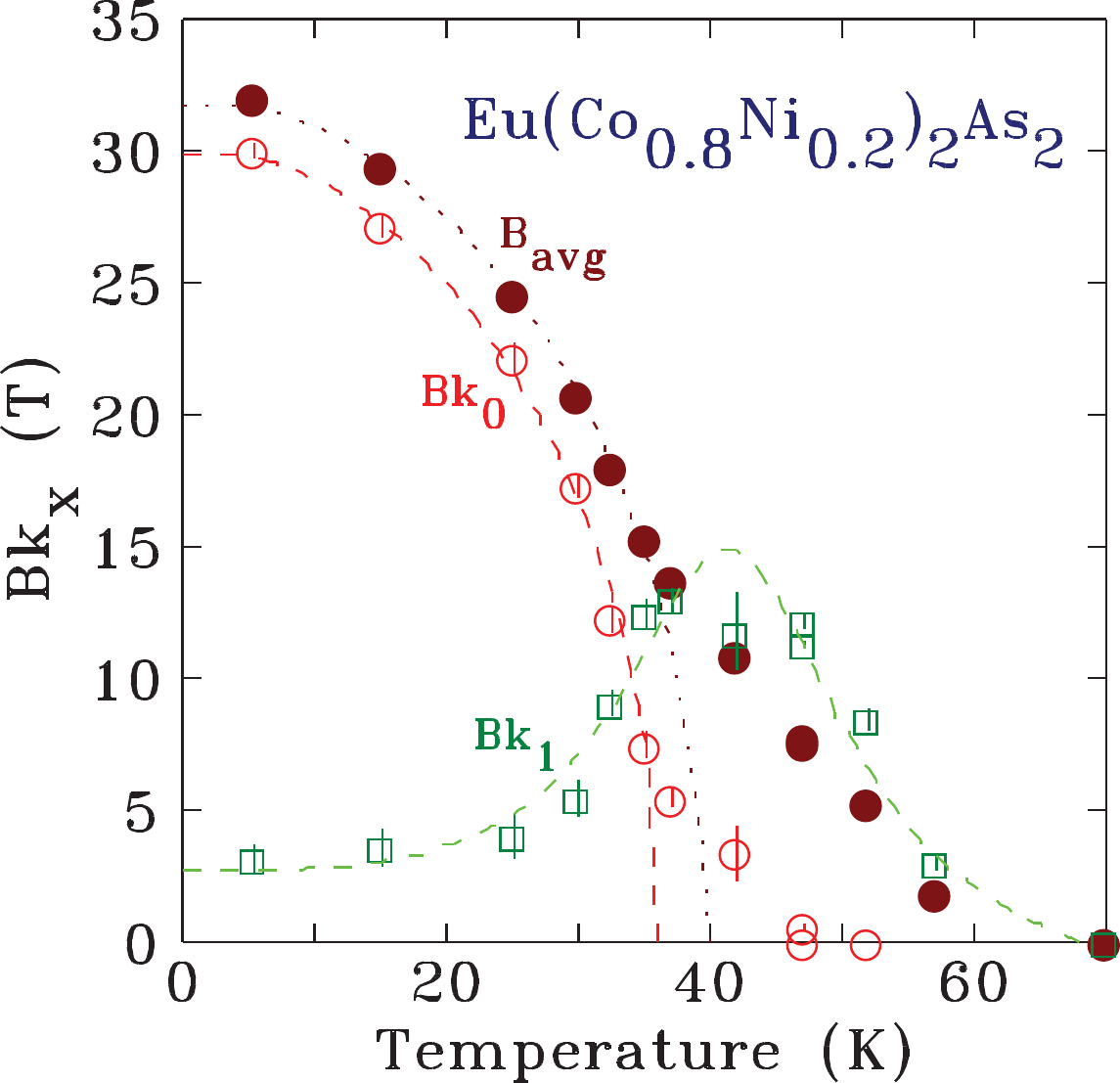}
\caption{Temperature dependence of the fitted hyperfine-field contributions in
$\rm Eu(Co_{0.8}Ni_{0.2})_2As_2$ derived from the modulated model. The average
hyperfine field ($B_{\rm avg}$) is plotted as solid round symbols with a dotted line
showing a fit (see text) yielding a transition temperature of 40.1(3)~K, while
the amplitude of the uniform term ($Bk_0$) is plotted as open circles with
a dashed line fit giving a transition temperature of 36.1(2)~K\@. The behavior of
the first Fourier component ($Bk_1$) is plotted as green open squares and shows a
broad peak centered at 41.5(6)~K\@. Note: Some magnetic broadening is present at
least up to 57~K.}
\label{fig:Ni20_Bhf}
\end{figure}

\begin{figure}
\includegraphics[width=2.75in]{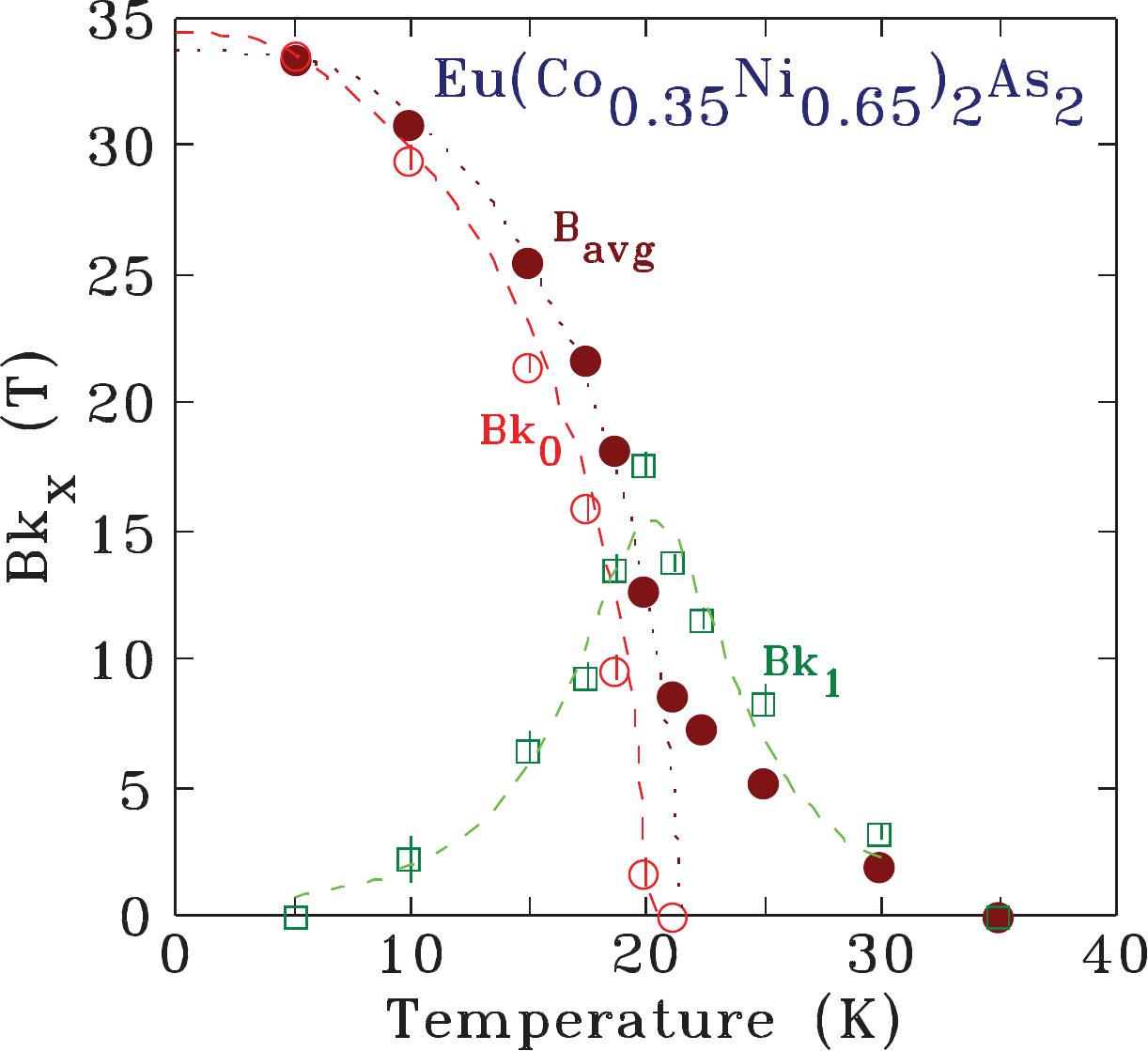}
\caption{Temperature dependence of the fitted hyperfine field contributions in
$\rm Eu(Co_{0.35}Ni_{0.65})_2As_2$ derived from the modulated model. The average
hyperfine field ($B_{\rm avg}$) is plotted as solid round symbols with a dotted line
showing a fit (see text) yielding a transition temperature of 21.5(1)~K, while
the amplitude of the uniform term ($Bk_0$) is plotted as open round symbols with
a dashed line fit giving a transition temperature of 20.0(1)~K\@. The behavior of
the first Fourier component (Bk$_1$) is plotted as square symbols and shows a
cusp-like peak centered at 20.3(5)~K\@. Note: Some magnetic broadening is present
until about 35~K\@.}
\label{fig:Ni65_Bhf}
\end{figure}

It is clear from Figs.~\ref{fig:Ni20_Bhf} and \ref{fig:Ni65_Bhf} that the
temperature range over which the modulated order is a significant component of
the total order is much wider in
$\rm Eu(Co_{0.8}Ni_{0.2})_2As_2$ and $\rm Eu(Co_{0.35}Ni_{0.65})_2As_2$ than it is in
\ecaa. For $x=0.2$ and~65, the uniform term ($Bk_0$) is lost
well before the overall magnetic contribution goes to zero.  This leads to the systems exhibiting two
distinct transitions.  In $\rm Eu(Co_{0.8}Ni_{0.2})_2As_2$, fitting $Bk_0$ versus 
temperature yields $T_{\rm c}=36.1(2)$~K while the spectra continue to exhibit
magnetic broadening up to $\sim 60$~K\@. For $\rm Eu(Co_{0.35}Ni_{0.65})_2As_2$,
the corresponding temperatures are 20.0(1)~K and $\sim35$~K\@. These upper temperatures are the same as the FM ordering temperatures associated with the Co/Ni sublattice determined above.  By contrast, for $x=0$, $Bk_0$ goes to zero at 42~K while the magnetic broadening is gone by 46~K
and the two transitions are not well resolved.  According to the magnetic susceptibility and magnetization versus field measurements, there is no FM component to the ordering for $x=0$, so the magnetic broadening  up to 46~K may be due to dynamic short-range magnetic ordering above $T_{\rm N\,Eu}$ associated with the Eu spins as revealed in the magnetic contribution $C_{\rm mag}(T)$ to the heat capacity above $T_{\rm N\,Eu}$ discussed in Sec.~\ref{Sec:HC} below.

%%%%%%%%%%%%%%%%%%%
%%%%%%%%%%%%%%%%%%%

%%%%%%%%%%%%%%%%%%%

\begin{table}[h]
\begin{tabular}{|c|c|c|c|c|c|c|}
\hline
$x$ & $\delta$ & $B_{\rm hf}$ & $T_{\rm c}(Bk_0$) & $T_{\rm c}(B_{\rm avg}$) & $T_{\rm pk}(Bk_1$) & end \\
 & (mm/s) & (T) & (K) & (K) & (K) & (K) \\
\hline
& & & & & & \\
0 & $-$10.72(2) & 26.21(5) & 42.0(1) & 45.6(1) & 40.8(3) & 46 \\
& & & & & & \\
0.20 & $-$10.53(3) & 31.80(8) & 36.1(2) & 40.1(3) & 41.5(6) & 60 \\
& & & & & & \\
0.65 & $-$10.30(2) & 33.29(7) & 20.0(1) & 21.5(1) & 20.3(5) & 32 \\
& & & & & & \\
\hline

\end{tabular}
\caption{Results derived from the analysis of the $^{151}$Eu M\"ossbauer spectra
of Eu(Co$_{1-x}$Ni$_x)_2$As$_2$. The isomer shift ($\delta$) and hyperfine field
($B_{\rm hf}$) are taken from the 5~K spectra. Transition temperatures for fits to
the average hyperfine field ($B_{\rm avg}$) and the uniform term in the Fourier
expansion ($Bk_0$) are given along with the location in the peak of the first
harmonic ($Bk_1$).  The final column lists the approximate temperature above which magnetic broadening of the spectrum disappears. For $x=0.2$ and~0.65 these correspond approximately to the FM transition temperature $T_{\rm C\,Co/Ni}$  in Table~\ref{Tab.Tn} associated with the Co/Ni atoms as determined from other measurements.}
\label{tab:MSsummary}
\end{table}

\subsection*{Discussion}

As M\"ossbauer spectroscopy is fundamentally a short-ranged probe of magnetic
order, it provides very little direct information about the detailed nature of
the long-range magnetic order in this system. However, some key statements can
be made.

There are no significant changes in the isomer shifts of the Eu$^{+2}$ component
of the spectra in any of the samples, nor is there any sign of a new Eu$^{+3}$
component (beyond the previously-noted impurity) developing with changing
temperature. We therefore conclude that the europium in all three samples
remains fully divalent from $\sim5$ to $\sim 295$~K\@.

At the lowest temperatures in all three samples ($x=0$, 0.20, 0.65) the europium
is in a single magnetic environment (modulo rotations about the $c$~axis, to which
the technique is insensitive due to the 4/$mmm$ point symmetry of the Eu site).

In all three compositions, the initial ordering of the Eu spins that develops on cooling from the
high-temperature paramagnetic state appears to include an incommensurate modulated component in which the magnitudes of the europium moments vary,  so that the $^{151}$Eu nuclei experience a broad distribution of
magnetic environments that can be modeled using a sum of Fourier components.
This modulated regime is quite narrow for $x=0$ but is much broader for $x=0.20$ and
0.65.

The magnetic transitions in EuCo$_{2-y}$As$_2$ are only separated by a few degrees
(see Table~\ref{tab:MSsummary}) and do not appear to exhibit separate signatures
in our other measurements.  However, in $\rm Eu(Co_{0.8}Ni_{0.2})_{2}As_2$ the two
transitions defined by extrapolation of $Bk_0$~versus~$T$ and by the ultimate loss of
magnetic splitting in the spectrum (``end'' column) are about 20~K apart (see
Table~\ref{tab:MSsummary}) and are clearly resolved in both $\chi(T)$ and
$C_{\rm p}(T)$.  We also see two transitions separated by about 10~K in $\rm
Eu(Co_{0.35}Ni_{0.65})_2As_2$; however, only the lower one is consistently seen by the 
bulk techniques.

There are two basic possibilities for the nature of the modulated phase. Given
that helical order exists at the lowest temperatures, it is possible that
in addition to the rotation about the $c$~axis the magnitude of the moments
also varies from layer to layer as the upper transition is approached.
Alternatively, the long-range order could break down into a short-range
correlated state before the long-range order collapses entirely. The former would appear
consistent with the behavior of the $x=0$ and 0.20 samples where the transitions, when
resolved, are seen by multiple techniques.  The latter form may be more
consistent with the behavior of the $x=0.65$ sample where the upper loss of order
is not consistently seen and the decay of $B_{\rm avg}(T)$ is quite soft, perhaps
reflecting a gradual breakdown of magnetic correlations.  Only direct measurement
by either neutron diffraction or resonant magnetic x-ray diffraction can
distinguish these two possibilities. 

For $\rm Eu(Co_{0.8}Ni_{0.2})_2As_2$ (Fig.~\ref{fig:Ni20_Bhf}) and 
$\rm Eu(Co_{0.35}Ni_{0.65})_2As_2$ (Fig.~\ref{fig:Ni65_Bhf}), $B_{\rm avg}$ exhibits a clear tail
to high temperatures, well beyond where the extrapolation of the lower
temperature behavior would predict that it would go to zero. The observed
hyperfine fields are too large to simply be the result of a direct transferred field from
the itinerant FM Co/Ni order; however, exchange interactions between the Co/Ni
order and the europium moments leading to an effective applied magnetic field
supporting the ordering of the europium moments could certainly account for the
observed fields. The $^{151}$Eu M\"ossbauer spectra therefore provide further
evidence of the Co/Ni order for intermediate nickel doping.

\begin{figure}
\includegraphics[width=3.3in]{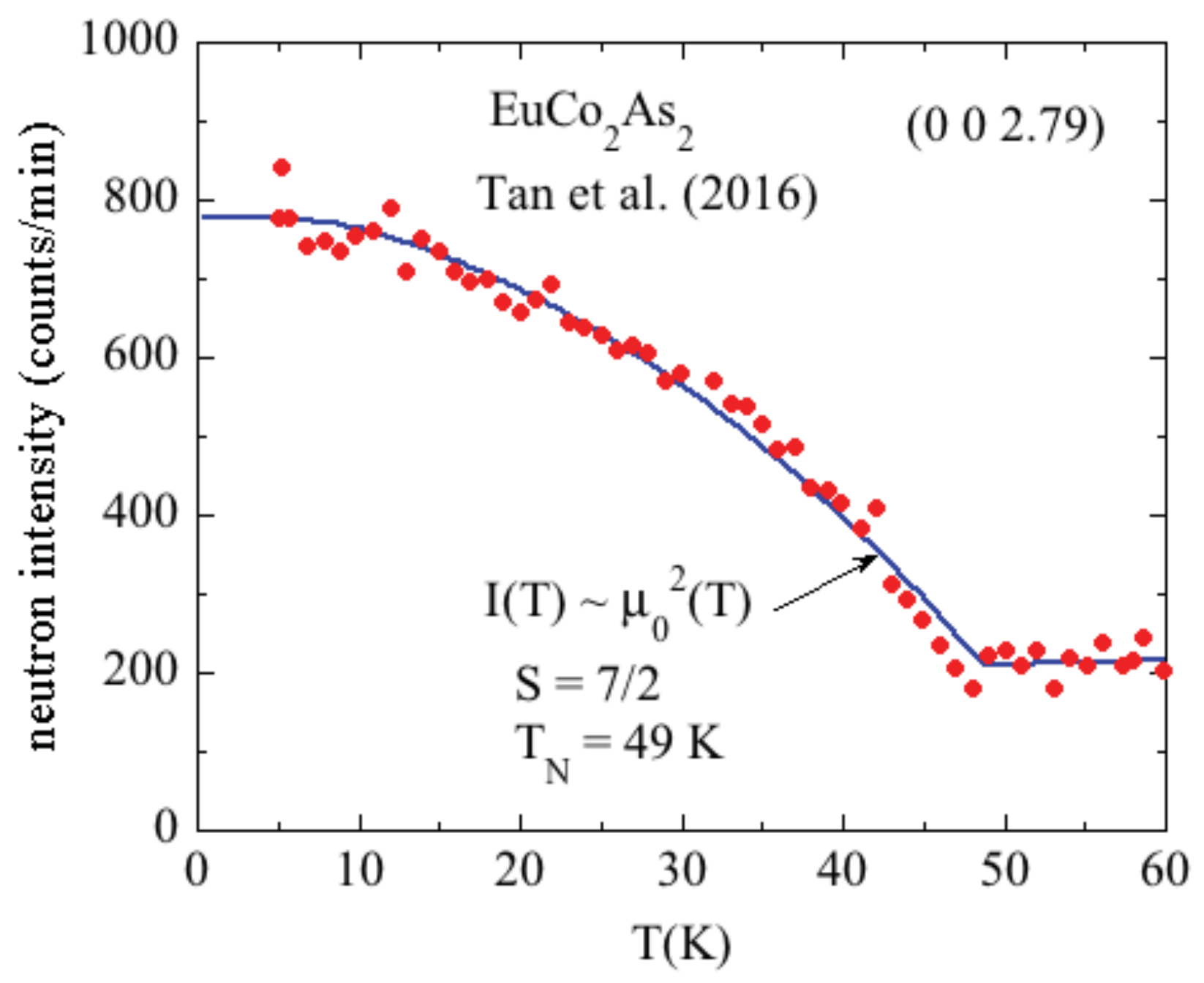}
\caption{Neutron intensity $I(T)$ of the helical (0 0 2.79) magnetic peak of \ecaa\ reported in Ref.~\cite{Tan2016} (filled red circles).  Our least-squares fit by molecular-field theory for spin $S=7/2$ [see Eq.~(\ref{Eq:barmuSoln})] is shown by the solid blue curve.}
\label{Tan_Fig2c_Fit}
\end{figure}

We fitted the neutron-diffraction intensity $I(T)$ of the helical \mbox{(0\ 0\ 2.79)} magnetic peak for \ecaa\ versus temperature that was presented in Ref.~\cite{Tan2016}.  Because the spin $S=7/2$ of Eu$^{+2}$ is large, the MFT prediction should be quite accurate.  We indeed obtained a good fit to the data by MFT for $S=7/2$ as shown by the solid blue curve representing the square of the ordered moment versus temperature in Fig.~\ref{Tan_Fig2c_Fit}, consistent with the good fit of $B_{\rm avg}(T)$ in Fig.~\ref{fig:Ni00_Bhf}, apart from a possible small systematic deviation of the data from the fitted curve in Fig.~\ref{Tan_Fig2c_Fit} on approaching $T_{\rm N\,Eu}$ from below.

\section{\label{Sec:HC}  Heat capacity}

\begin{figure}
\includegraphics[width=3.5in]{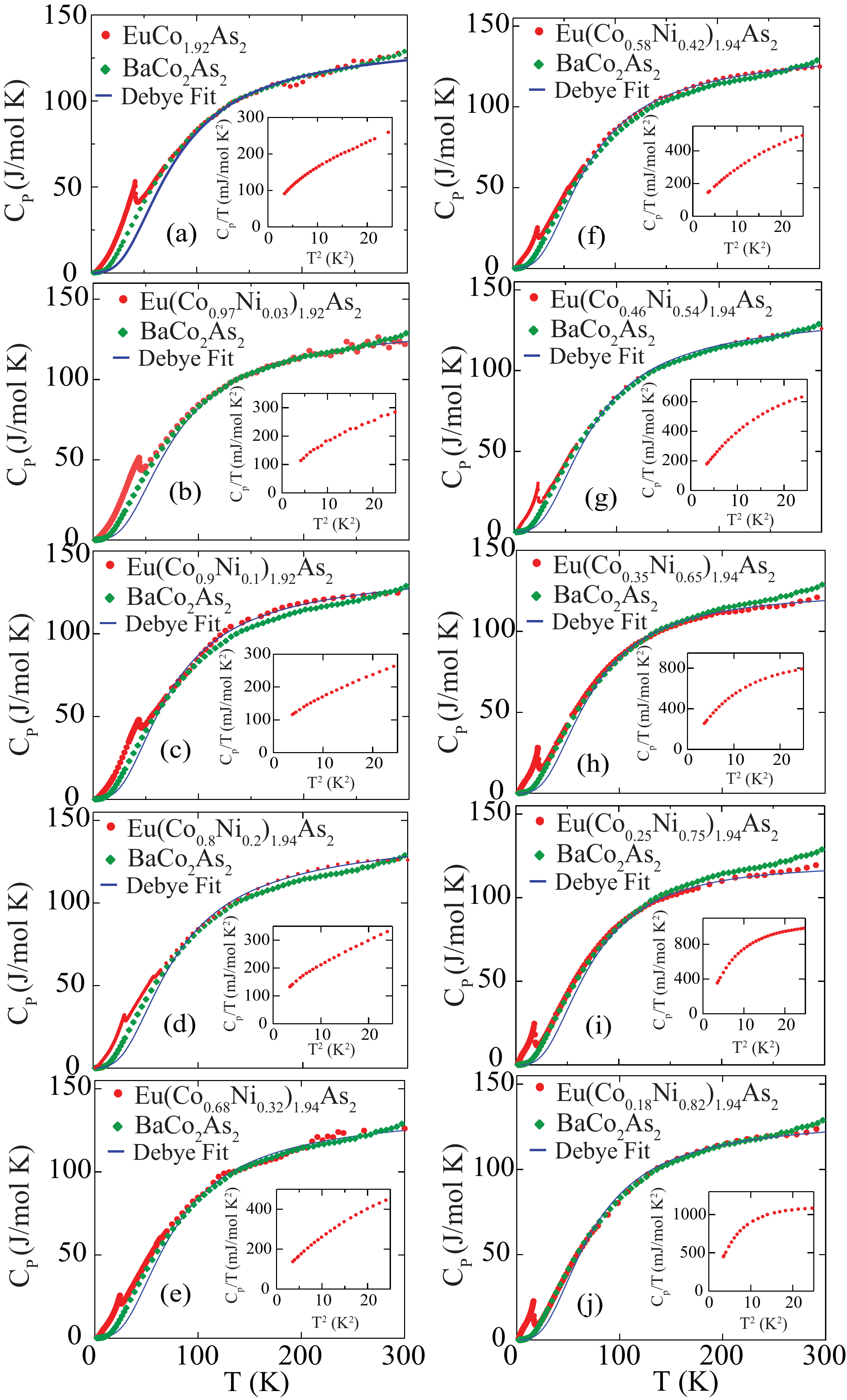}
\caption{The temperature $T$ dependence of the heat capacity $C_{\rm p}$ for \ecna\  with $x = 0$, 0.03, 0.10, 0.20, 0.32, 0.42, 0.54, 0.65, 0.75, and 0.82 and \bca\ single crystals in $H=0$~T\@. The solid black curves are fits of the data between 70 and 300~K by the Debye lattice heat capacity model in Eqs.~(\ref{Eq:Cpdebye}). Insets: Expanded plots of $C_{\rm p}(T)/T$ vs $T^2$ from 1.8 to 5~K. }
\label{Fig:HC1}
\end{figure}

\begin{figure}
\includegraphics[width=3.5in]{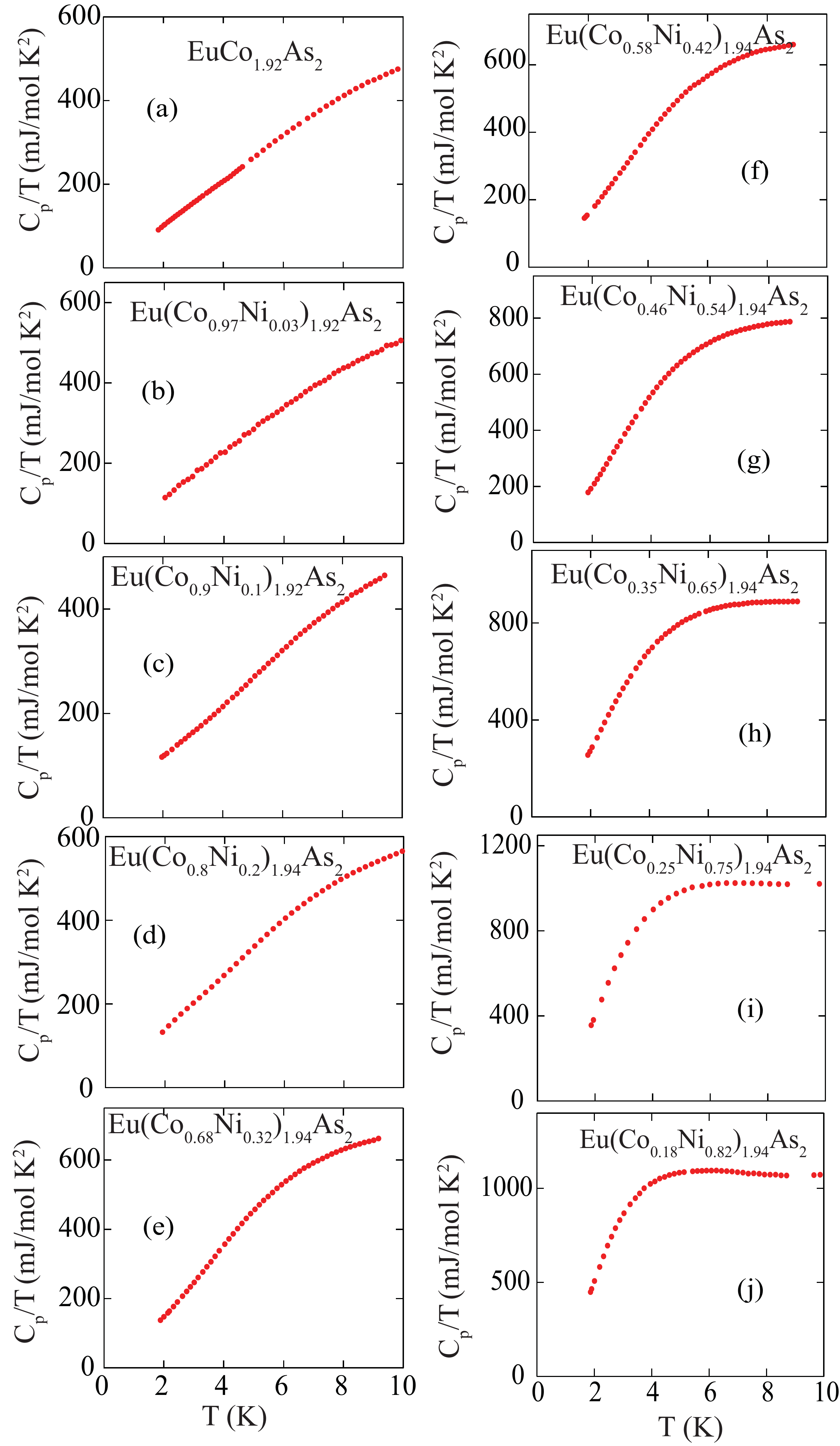}
\caption{Heat capacity $C_{\rm p}/T$ vs $T$ at zero field from 1.8 to 10~K for each \ecna\ crystal ($x =$ 0, 0.03, 0.10, 0.20, 0.32, 0.42, 0.54, 0.65, 0.75, and 0.82).}
\label{Fig:HC2}
\end{figure}

Figure~\ref{Fig:HC1} shows the  heat capacities $C_{\rm p}(T)$ for \ecna\ single crystals with compositions $x=0$, 0.03, 0.1, 0.2, 0.32, 0.42, 0.54, 0.65, 0.75, and 0.82, and for the nonmagnetic reference compound ${\rm BaCo_2As_2}$~\cite{Sangeetha2018} measured in the temperature range from 1.8 to 300~K in zero applied magnetic field. For each composition, the $C_{\rm p}(T=300$~K) of \ecna\ attains a value of $\approx 123.2$~J/mol K which is close to the classical Dulong-Petit high-$T$ limit  $C_{\rm V}=3nR=123.2$~J/mol~K, where $n=4.94$ is the number of atoms per formula unit and $R$ is the molar gas constant.

The insets in Fig.~\ref{Fig:HC1} show plots of $C_{\rm p}/T$ versus $T^2$  from 1.8 to 5~K for each crystal.  The purpose of these plots is to show that the low-temperature data cannot be fitted by the conventional expression $C_{\rm p}(T)/T = \gamma + \beta T^2$, where $\gamma$ is the Sommerfeld coefficient associated with the conduction electrons and $\beta$ is the coefficient of the $T^3$ lattice term. 

The $C_{\rm p}/T$ versus $T$ data for the \ecna\ crystals in the temperature range from 1.8 to 10~K are plotted in Figure~\ref{Fig:HC2}. One sees that each crystal shows approximately linear behavior over a certain $T$ range; for example, for $x=0$ to 0.32 this range is from 2 to 6~K, whereas  for $x\geq0.54$ the range is below 4~K, which means that $C_{\rm p}$ has an approximately $T^2$ contribution over the respective $T$ range. This behavior may arise from the temperature-dependent heat capacity of AFM spin waves.

The $C_{\rm p}(T)$ data for \ecna\ in the PM regime \mbox{$100~{\rm K} \leq T \leq 300~{\rm K}$} are analyzed using the sum of an electronic $\gamma T$ term and the lattice term given by the Debye model according to
\begin{subequations}
\label{Eq:Cpdebye}
\begin{equation}
C_{\rm p}(T) = \gamma T+ nC_{\rm V \,Debye}(T),
\end{equation}
where $\gamma$ is again the Sommerfeld electronic heat capacity coefficient, $C_{\rm V \,Debye}(T)$ is the Debye lattice heat capacity given by
\begin{equation}
C_{\rm V \,Debye}(T) = 9R\left(\frac{T}{\Theta_{\rm D}}\right)^3\int_{0}^{\Theta_{\rm D}/T}\frac{x^4 dx}{(e^x-1)^2}dx,
\end{equation}
\end{subequations}
and $n=4.94$ is again the number of atoms per formula unit.  The black solid curves in Fig.~\ref{Fig:HC1} represent the fits of the $C_{\rm p}(T)$ data for ${\rm 100~K\leq {\it T} \leq 300~K}$ by Eqs.~(\ref{Eq:Cpdebye}) obtained using the accurate analytic Pad\'{e} approximant function for $C_{\rm V \,Debye}$ versus  $T/\Theta_R$ given in Ref.~\cite{Goetsch2012}. The fitted values of $\gamma$ and $\Theta\rm_D$ for \ecna\ are listed in Table~\ref{Tab:HC}.

The density of degenerate conduction carrier states at the Fermi energy $E\rm_F$ for both spin directions $D_\gamma(E\rm_F$) is obtained from $\gamma$ according to
\bse
\label{Eqs:Dgamma}
\bea
{\cal D}_\gamma(E\rm_F)=\frac{3\gamma}{\pi^2k_B^2},
\label{Eq:DOS1}
\eea
\rm{which gives}
\bea
{\cal D}_\gamma(E\rm_F)\left[\frac{states}{eV~f.u.}\right] = \frac{1}{2.359}~\gamma\left[\frac{mJ}{mol~ K^2}\right],
\label{Eq:DOS} 
\eea
\ese
where ``mol'' refers to a mole of formula units (f.u.).  The ${\cal D}_\gamma$($E\rm_F$) values calculated for our \ecna\ crystals from their $\gamma$ values using Eq.~(\ref{Eq:DOS}) are listed in Table~\ref{Tab:HC}, where the fitted values are close to each other for $x=0$ to 0.82.

\begin{table}
\caption{\label{Tab:HC} The parameters $\gamma$, $\theta_{\rm D}$ and density of states at the Fermi energy $D_\gamma(E_{\rm F})$ obtained from fits to the heat capacity by Eqs.~(\ref{Eq:Cpdebye}) in the temperature range 100~K~$\leq T\leq300$~K.}
\begin{ruledtabular}
\begin{tabular}{cccc}				
	$x$ 
	&$\gamma$  
	& $\theta_{\rm D}$  
	& ${\cal D}_\gamma({\rm E_F})$ \\

	& (mJ/mol\,K$^2$)     
	 & (K)				 
	&($\rm{\frac{states}{eV. f.u}}$)\\
\hline
	0		&15(2)				&308(3)		&6.3(8)\\
	0.03	&18(2)				&310(3)		&7.6(8)\\
	0.1		&25(1)				&286(2)		&10.7(4)\\
	0.2		&33(1)				&306(1)		&13.9(4)\\
	0.32	&25(3)				&313(5)		&11(1)\\
	0.42	&23(1)				&298(1)		&9.7(4)\\
	0.54	&23(2)				&311(4)		&9.7(8)\\
	0.65  	&23(2)				&294(2)		&9.7(8)\\
	0.75	&22(1)				&298(2)		&9.3(4)\\
	0.82	&21.3(8)			&303(2)		&9.0(3)\\

\end{tabular}
\end{ruledtabular}
\end{table}

\begin{figure}
\includegraphics[width=3.4in]{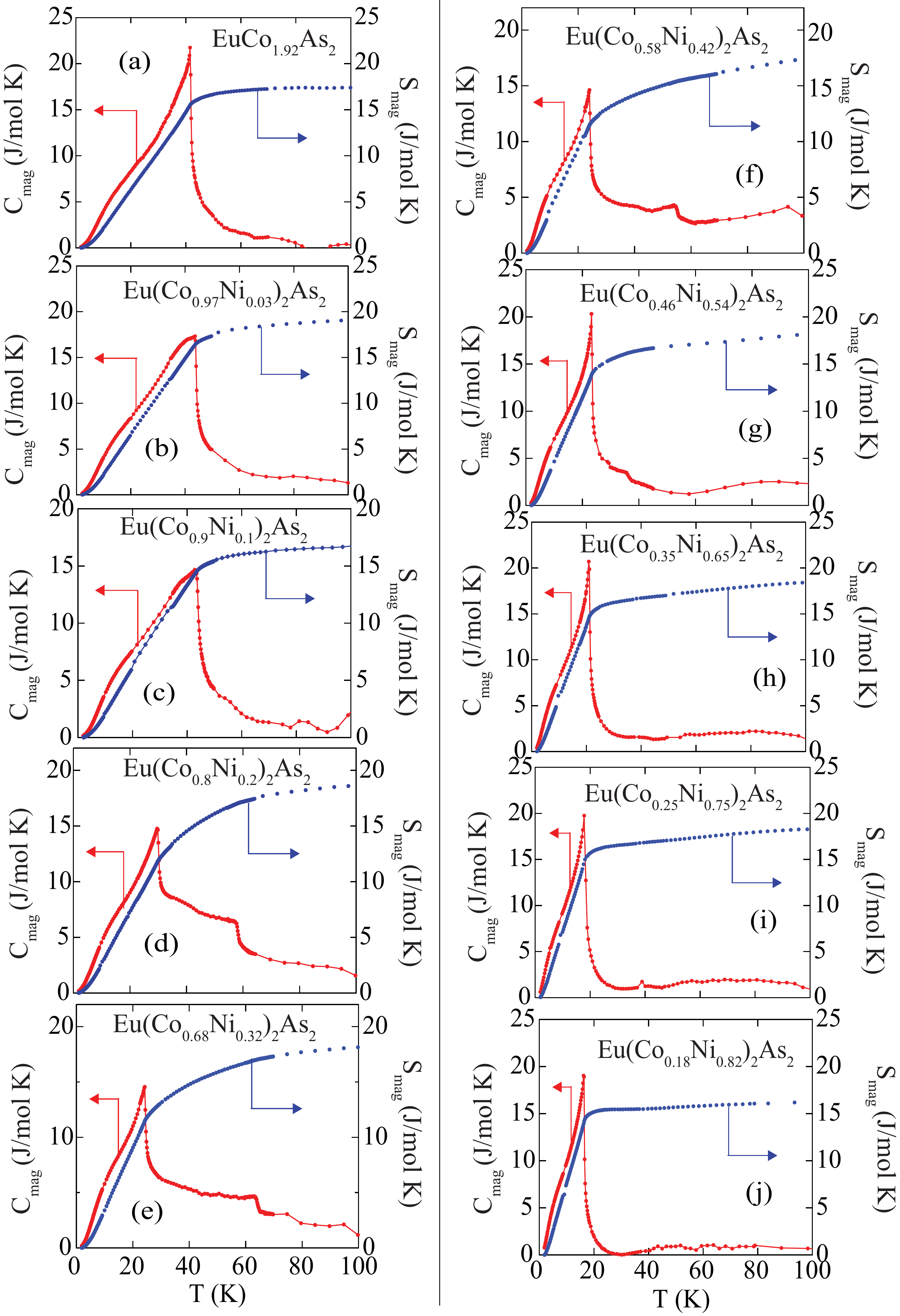}
\caption{Temperature~$T$-dependent magnetic contribution $C{\rm_{mag}}(T)$ obtained by subtracting the nonmagnetic contribution $C_{\rm_p}(T)$ of \bca\ (left-hand ordinates) and the magnetic entropy $S_{\rm mag}(T)$ calculated from the experimental $C_{\rm mag}(T)/T$ data using Eq.~(\ref{Eq:SmagCalc}) for \ecna\ (right-hand ordinates) for $x = 0$, 0.03, 0.10, 0.20, 0.32, 0.42, 0.54, 0.65, 0.75, and 0.82 from 1.8 to 100~K\@.  The sharp $\lambda$-like transitions are associated with AFM ordering of the Eu spins and the steplike features for $x=0.2$--0.54 are associated with itinerant FM ordering associated with the Co/Ni sublattice.}
\label{Fig:HC3}
\end{figure}

\begin{figure}
\includegraphics[width=3.5in]{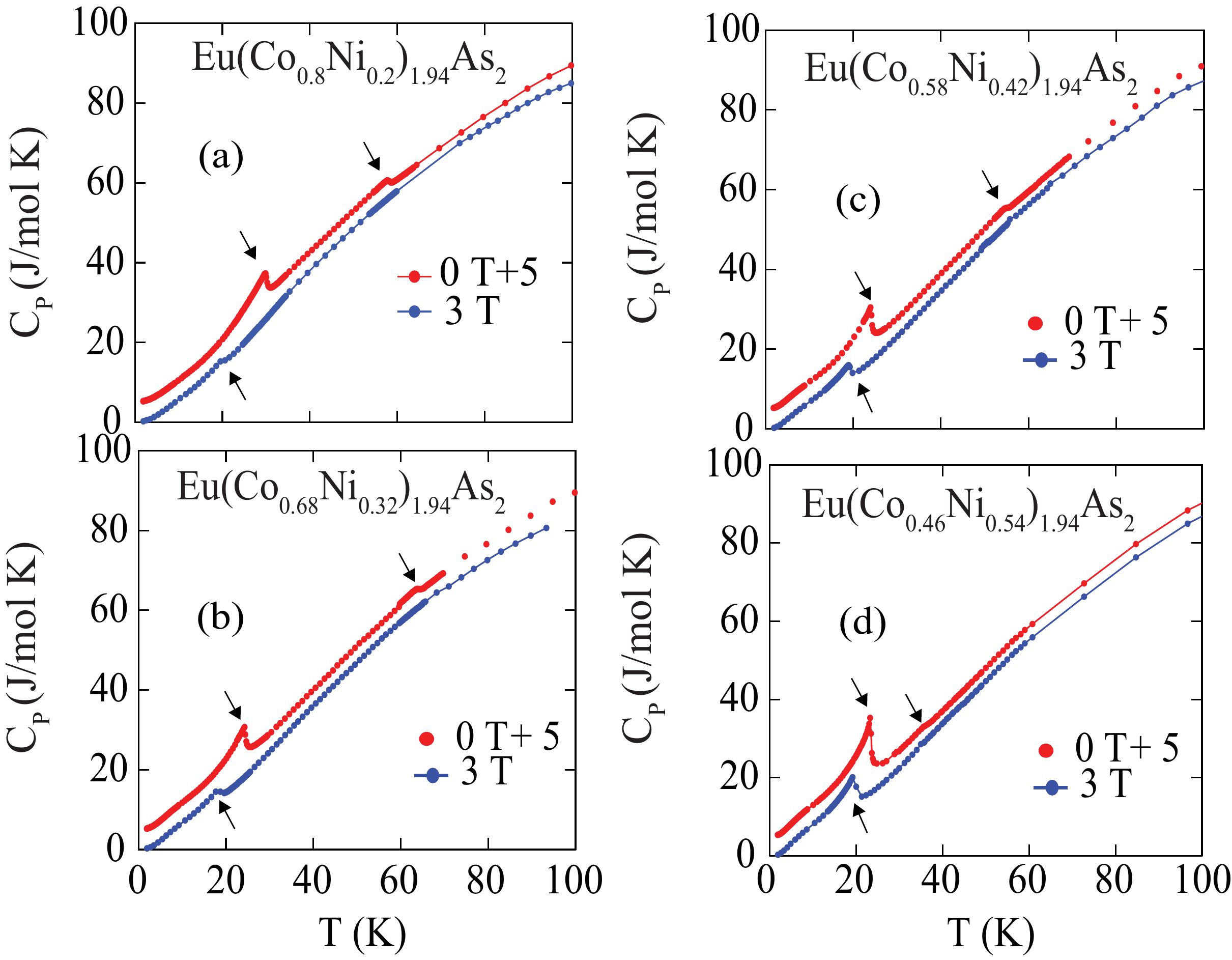}
\caption{Heat capacity $C\rm_{p}$ versus temperature~$T$ of \ecna\  with $x =$ 0.20, 0.32, 0.42, and 0.54 in magnetic fields $H=0$ and 3~T applied parallel to the $c$~axis, $H\parallel c$. For clarity, the data for $H=0$~T are offset by 5~J/mol~K as indicated.  Phase transitions are marked by arrows.}
\label{Fig:HC4}
\end{figure}

The magnetic contribution $C_{\rm mag}(T)$ to $C_{\rm p}(T)$ of \ecna\ obtained after subtracting the lattice contribution, taken to be $C_{\rm p}(T)$ of ${\rm BaCo_2As_2}$~\cite{Sangeetha2018}, is shown in Fig.~\ref{Fig:HC3}.  Within MFT the discontinuity in $C_{\rm mag}$ at $T=T_{\rm N}$ is given by~\cite{Johnston2015}
\begin{eqnarray}
\Delta C{\rm_{mag}}=R\frac{5S(1+S)}{1+2S+2S^2} = 20.14~ \rm {J/mol~K},
\label{Eq:deltaCp}
\end{eqnarray}
where the second equality is calculated for $S = 7/2$.  The experimental heat capacity jump for $x=0$ in  Fig.~\ref{Fig:HC3}(a) at $T_{\rm N} =45$~K is $\approx 21.74$~J/mol~K, somewhat larger than the theoretical prediction of MFT\@. With doping, the value of $\Delta C{\rm_{mag}}$ decreases to values smaller than the theoretical prediction up to $x=0.42$, and then increases again.  The $C_{\rm mag}(T)$ is nonzero for $T_{\rm N}<T\leq 40$~K, indicating the presence of dynamic short-range AFM ordering of the Eu spins above $T_{\rm N\,Eu}$, thus accounting at least in part for the decrease in $\Delta C_{\rm mag}$ from the theoretical value.

From Fig.~\ref{Fig:HC3}, we find two magnetic transition peaks for $x=0.2,0.32, 0.42$, and 0.54, consistent with the above studies.  It is clear that the magnetic ordering of the Eu spins for all compositions is associated with the $\lambda$-type anomaly that decreases monotonically from a temperature $T_{\rm N\,Eu} = 45.1(2)$~K to 16.2(4)~K at $x=0.84$, as listed above in Table~\ref{Tab.Tn}.  We therefore ascribe the upper transition temperature manifested by a step discontinuity in $C_{\rm mag}(T)$ in Fig.~\ref{Fig:HC3} to magnetic ordering of the Co/Ni atoms, which from the above magnetic measurements is FM\@.  Therefore, the latter transition temperatures are denoted by $T_{\rm C\,Co/Ni}$ in Table~\ref{Tab.Tn}.

The magnetic entropy is calculated from the $C_{\rm mag}(T)$ data for \ecna\ in Fig.~\ref{Fig:HC3} using 
\begin{equation}
S_{\rm mag}(T) = \int_{0}^{T}\frac{C_{\rm mag}(T^\prime)}{T^\prime}dT^\prime,
\label{Eq:SmagCalc}
\end{equation}
 and the results are shown in Fig.~\ref{Fig:HC3} (right-hand ordinates). The theoretical high-$T$ limit is $S_{\rm mag}(T) = R~{\rm ln}(2S+1) = 17.29$~J/mol~K for $S=7/2$. The entropy reaches almost 90$\%$ of $R{\rm ln}(8)$ at $T_{\rm N\,Eu}$\@.  The high-$T$ limit of the data is slightly different from the theoretical value, which may be due to a small error in estimating the lattice contribution to $C_{\rm p}(T)$.

The $C_{\rm p}(H,T)$ data for \ecna\ crystals with compositions $x=0.2, 0.32, 0.42$, and~0.54 showing two magnetic transitions  measured in magnetic fields $H=0$ and $H=3$~T applied along the $c$~axis are shown in Fig.~\ref{Fig:HC4}. It is evident that both transitions are intrinsic and magnetic because $T_{\rm N\,Eu}$ shifts to lower temperatures marked by an arrow and the heat capacity jump at $T_{\rm N\,Eu}$ decreases with applied magnetic field, both as predicted from MFT for a field parallel to the helix axis \cite{Johnston2015}.  The second transition at $T_{\rm C\,Co/Ni}$ as listed in Table~\ref{Tab.Tn} is seen to be suppressed by an applied field of 3~T\@. We take $T_{\rm N\,Eu}$ at each field to be the temperature of the peak in $C_{\rm p}$ versus~$T$ instead of the temperature at half-height of the transition, because the latter is ambiguous to estimate due to the significant contribution of short-range AFM ordering to $C_{\rm p}$ above the respective $T_{\rm N\,Eu}$.  Table~\ref{Tab.Tn} lists the $x$-dependent $T_{\rm N\,Eu}$ and $T_{\rm C\,Co/Ni}$ values obtained from the $\chi(T)$ and $C_{\rm p}(T)$ measurements and from the $\rho(T)$ measurements discussed in the following section. The values obtained from the different measurements are quite close to each other. The table shows that the $C_{\rm p}(T)$ measurements do not reveal two transitions for $x=0.65$ even though the two transitions are evident in both the $\chi(T)$ and $\rho(T)$ (see below) measurements.

\section{\label{Sec:Res} Electrical Resistivity}

\begin{figure}
\includegraphics[width=3.5in]{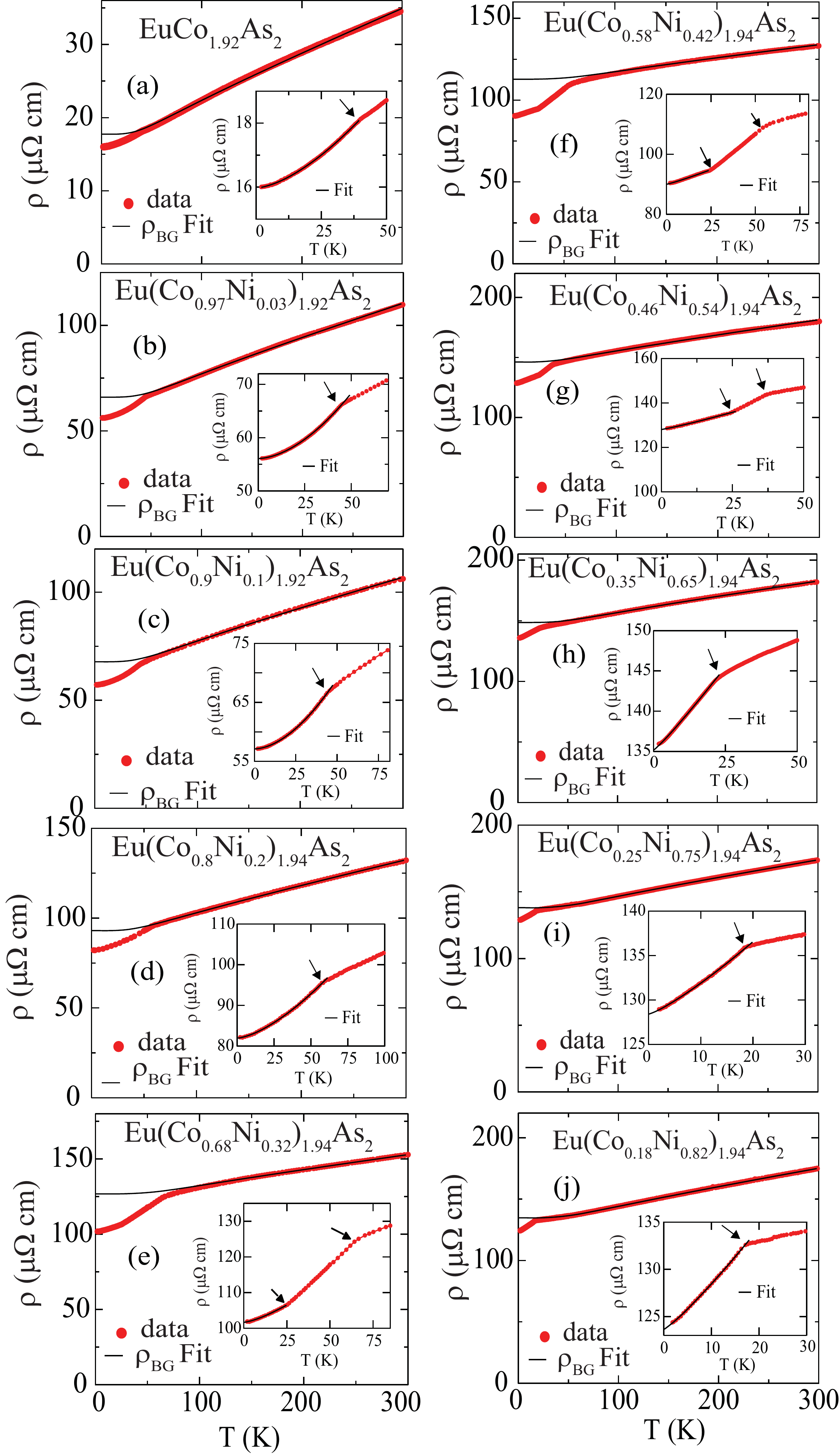}
\caption{In-plane electrical resistivity of \ecna\ single crystals ($x =$ 0, 0.03, 0.10, 0.20, 0.32, 0.42, 0.54, 0.65, 0.75, and 0.82) as a function of temperature~$T$ from 1.8 to 300 ~K measured in zero magnetic field. The black curves in the main panels are fits by the prediction of the Bloch-Gru{\"u}neisen theory in Eqs. (\ref{Eqs:rhoBG}) to the data above 70~K and are extrapolated to $T=0$. Insets: Expanded plots of $\rho(T)$ at low~$T$\@. The black line is the fit of the data by Eq.~(\ref{Eq:rhoFitTn}) over the temperature interval below min($T_{\rm C\,Co/Ni},\ T_{\rm N\,Eu}$) marked by an arrow.}
\label{Fig:Res}
\end{figure}

The in-plane ($ab$-plane) electrical resistivity $\rho$ of the \ecna\ single crystals as a function of $T$ from 1.8 to 300~K measured at zero magnetic field are shown in Fig.~\ref{Fig:Res}. The $\rho(T)$ exhibits metallic behavior for all crystals. The AFM transition at $T_{\rm N\,Eu}$  and the second transition $T_{\rm C\,Co/Ni}$ for $x=0.32$, 0.42 and 0.54 observed in the $\rho(T)$ data are marked by arrows.  The transitions are more clearly shown in the expanded plots of $\rho(T)$ at low temperatures in the insets to Fig.~\ref{Fig:Res}. The values of $T_{\rm N}$ and $T_{\rm C\,Co/Ni}$ obtained from the data are listed in Table~\ref{Tab.Tn} and are seen to be consistent with the values found from the above $\chi(T)$ data and $C_{\rm p}(T)$ measurements. Interestingly, the $\rho(T)$ data do not exhibit any significant feature at $T_{\rm N\,Eu}$ for $x=0.20$ and~0.65.

The low-$T$ $\rho_{ab}(T)$ data below $T_{\rm N\,Eu}$ are fitted by
\begin{equation}
\rho(T)=\rho_0+AT^n,
\label{Eq:rhoFitTn}
\end{equation}
as shown by the solid curves in the insets of Fig.~\ref{Fig:Res}. The fitted parameters $\rho_0$, $A$, and the power $n$ are listed in Table~\ref{Tab:Res}.  From Table~\ref{Tab:Res}, the $\rho(T)$ data follow the Fermi-liquid $T^2$ behavior for $x=0$, but with doping  $\rho(T\leq T_{\rm N})$ does not, likely because it is affected by the $T$-dependent loss of spin-disorder scattering on cooling below $T_{\rm N\,Eu}$.

\begin{table*}
\caption{\label{Tab:Res} Parameters for the electrical resistivity obtained from a low-temperature fit by Eq.~(\ref{Eq:rhoFitTn}) below $T_{\rm N\,Eu}$ and for the fit by the Bloch-Gr\"{u}neisen model  in Eqs.~(\ref{Eqs:rhoBG}) in the range 70~K~$\leq T\leq 300$~K\@.}
\begin{ruledtabular}
\begin{tabular}{c|ccc|ccc}	
				
&\multicolumn{3}{c|}{Low-$T$, $T^n$ fit} &\multicolumn{3}{c} {Bloch-Gr\"{u}neisen fit}   \\
	$x$
	& $\rho_0$ & $A$ & $n$
	&$\rho_0+\rho_{\rm sd}$ & $C$ & $\Theta_R$	
\\
	&($\mu\Omega$~cm) & ($\mu\Omega$~cm/K$^n$) & 
	&($\mu\Omega$~cm) & ($\mu\Omega$~cm) & (K)
\\
\hline
	0		&16.0			&0.0022(1)	&2		&17.7(3)		&12(1)		&213(3)		\\
	0.03		&56.07(1)		&0.0127(3)	&1.75(1)	&65(1)		&37(3)		&246(8)		\\
	0.1		&57.11(1)		&0.0115(1)	&1.77(1)	&67(1)		&32(4)		&256(9)		\\
	0.2		&81.96(3)		&0.0289(1)	&1.52(1)	&93.0(5)		&31(2)		&245(7)		\\
	0.32		&101.63(3)		&0.0581(6)	&1.38(3)	&126(2)		&27(9)		&319(6)		\\
	0.42		&89.97(4)		&0.14(1)		&1.11(2)	&112(3)		&21(4)		&311(8)		\\
	0.54		&128.03(3)		&0.21(1)		&1.12(1)	&146(1)		&26(5)		&228(9)		\\
	0.65  	&135.17(2)		&0.30(1)		&1.10(1)	&148(1)		&29(6)		&266(6)		\\
	0.75		&128.41(3)		&0.21(1)		&1.22(1)	&138.0(3)	&32(1)		&271(5)		\\
	0.82		&123.63(4)		&0.34(1)		&1.15(1)	&134.7(2)	&37.8(9)		&282(7)		\\
	1		&8.71(3)			&0.262(9)	&1.26(1)	&17.18(3)	&27.0(2)		&235(1) 	\\
\end{tabular}
\end{ruledtabular}
\end{table*}

The $\rho(T)$ in the normal state above 70~K is fitted by the Bloch-Gr\"{u}neisen (BG) model where the resistivity arises from scattering of electrons from acoustic phonons.  Our fitting function is 
\begin{subequations}
\label{Eqs:rhoBG}
\begin{equation}
\rho{\rm_{BG}}(T)=\rho_0+\rho_{\rm sd}(T) +C~f(T),
\end{equation}
where~\cite{Goetsch2012}
\begin{equation}
f(T) = \left(\frac{T}{\Theta_R}\right)^5\int_{0}^{\Theta_R/T}\frac{x^5 dx}{(1-e^{-x})(e^x-1)}dx.
\label{Eq:BGM}
\end{equation}
\end{subequations}
Here $\rho_0+\rho_{\rm sd}(T)$ is the sum of the residual resistivity $\rho_0$  due to static defects in the crystal lattice and the spin-disorder resistivity $\rho_{\rm sd}(T)$.  The constant $C$ describes the $T$-independent interaction strength of the conduction electrons with the thermally excited phonons and contains the ionic mass, Fermi velocity, and other parameters, $x=\frac{h\omega}{2\pi k_{\rm B}T}$, and $\Theta_R$ is the Debye temperature determined from electrical resistivity data. The representation of $f(T)$ used here is a fit of data calculated for the integral on the right-hand side of Eq.~(\ref{Eq:BGM}) by an accurate analytic Pad\'{e} approximant function of $T$~\cite{Goetsch2012}. The fits to the $\rho(T)$ data between 70 and 300~K by Eqs.~(\ref{Eqs:rhoBG}) are shown as the solid black curves in the main panel of Fig.~\ref{Fig:Res}. The fitted parameters $(\rho_0+\rho_{\rm sd})$, $C$, and $\Theta_R$ are listed in Table~\ref{Tab:Res}. 

\section{\label{Sec:Summary} Concluding Remarks}

\begin{figure}
\includegraphics[width=2.5in]{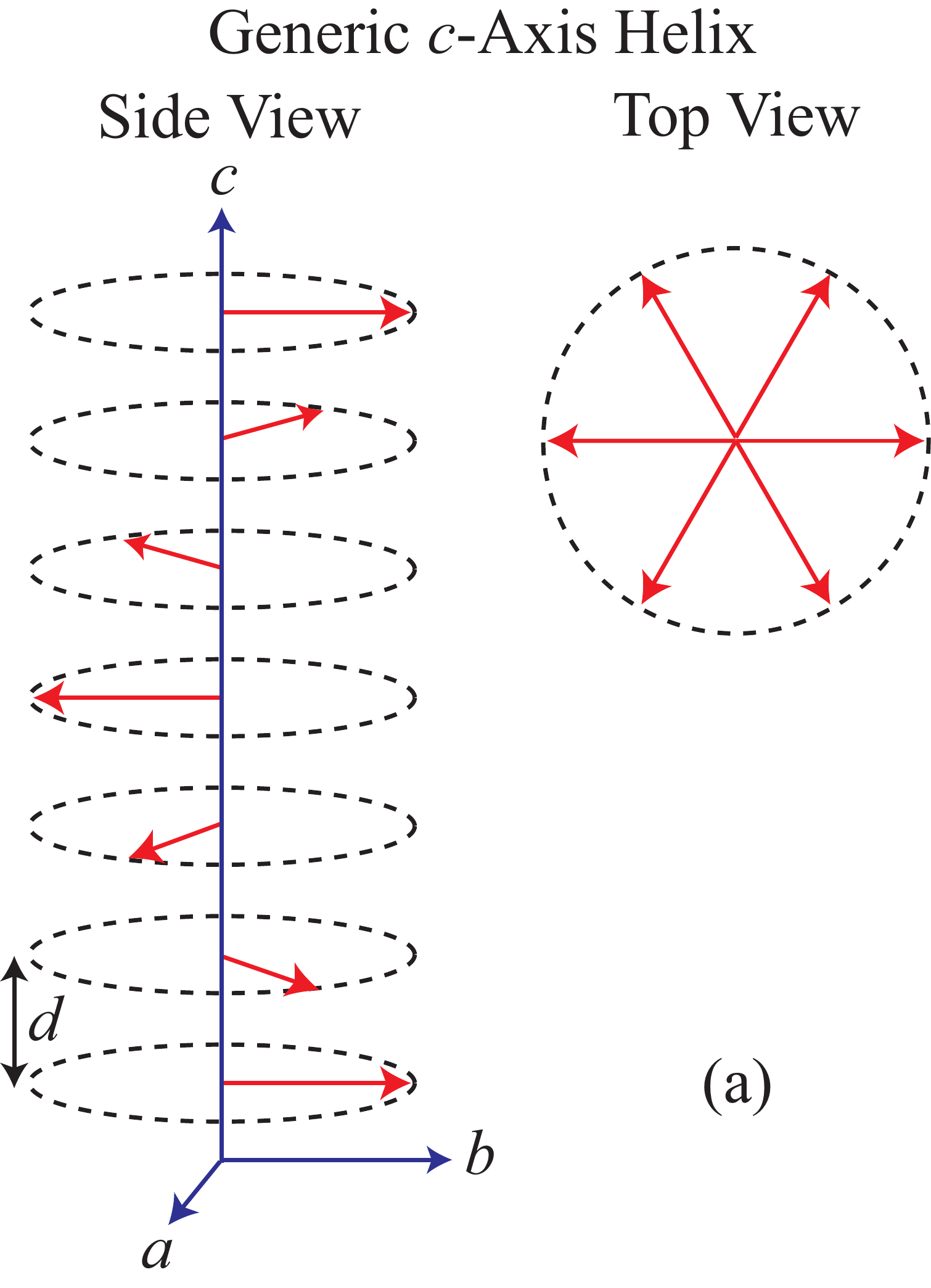}
\includegraphics[width=2.5in]{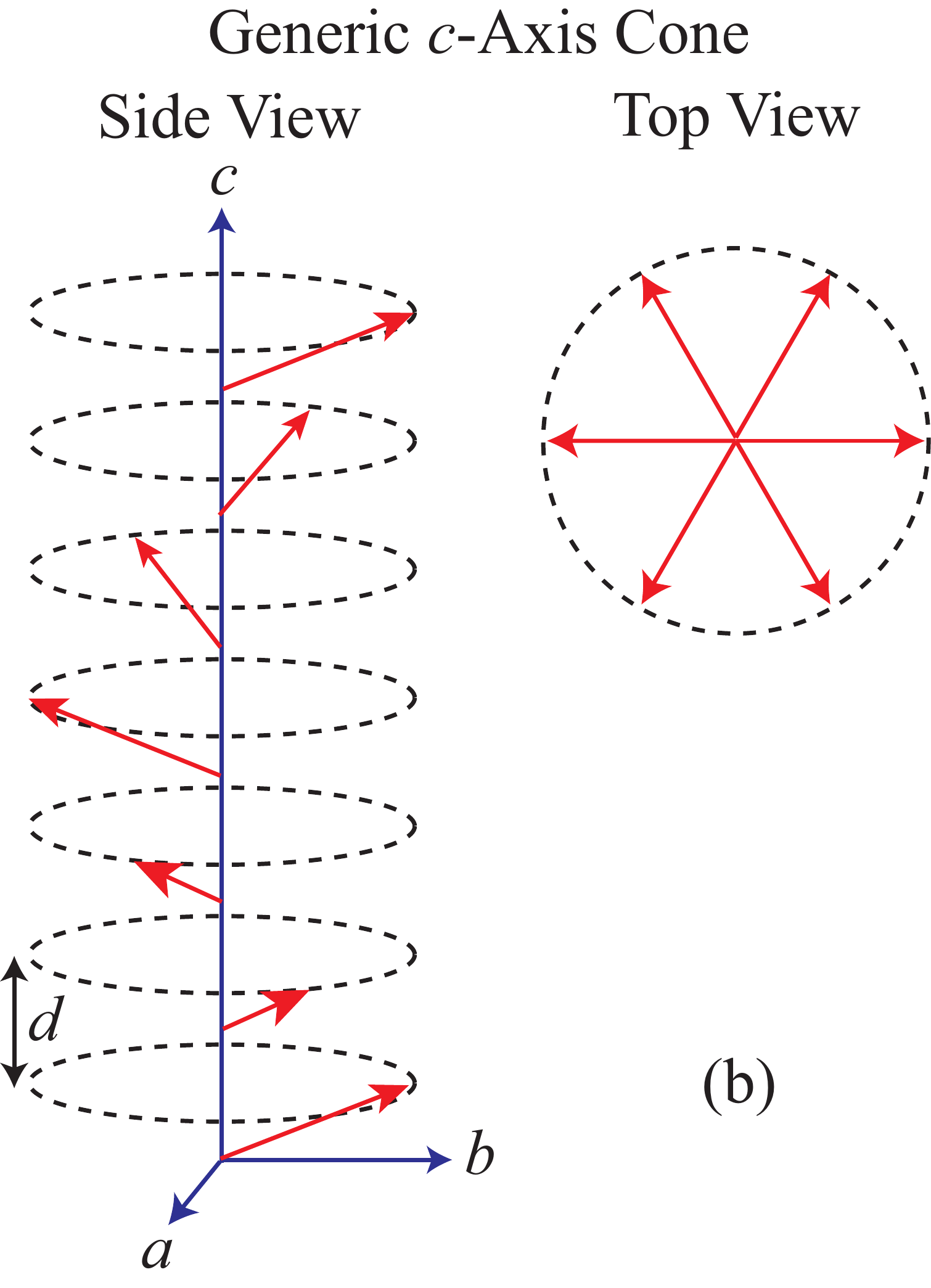}
\caption{(a) Generic magnetic structure of a $c$-axis helix, which consists of ferromagnetically (FM)-aligned moments in the $ab$~plane represented by a single arrow which rotates by an angle $q_cd$ from layer to layer along the $c$~axis, where $q_c$ is the c-axis magnetic propagation vector and $d$ is the distance between adjacent $ab$ planes of FM-aligned moments as shown (after Ref.~\cite{Johnston2012}).  (b)~Generic $c$-axis cone structure consisting of a superposition of a $c$-axis helix structure and a $c$-axis FM component of each moment. }
\label{Fig:helix_cone}
\end{figure}

\begin{figure}[t]
\includegraphics[width=3.3in]{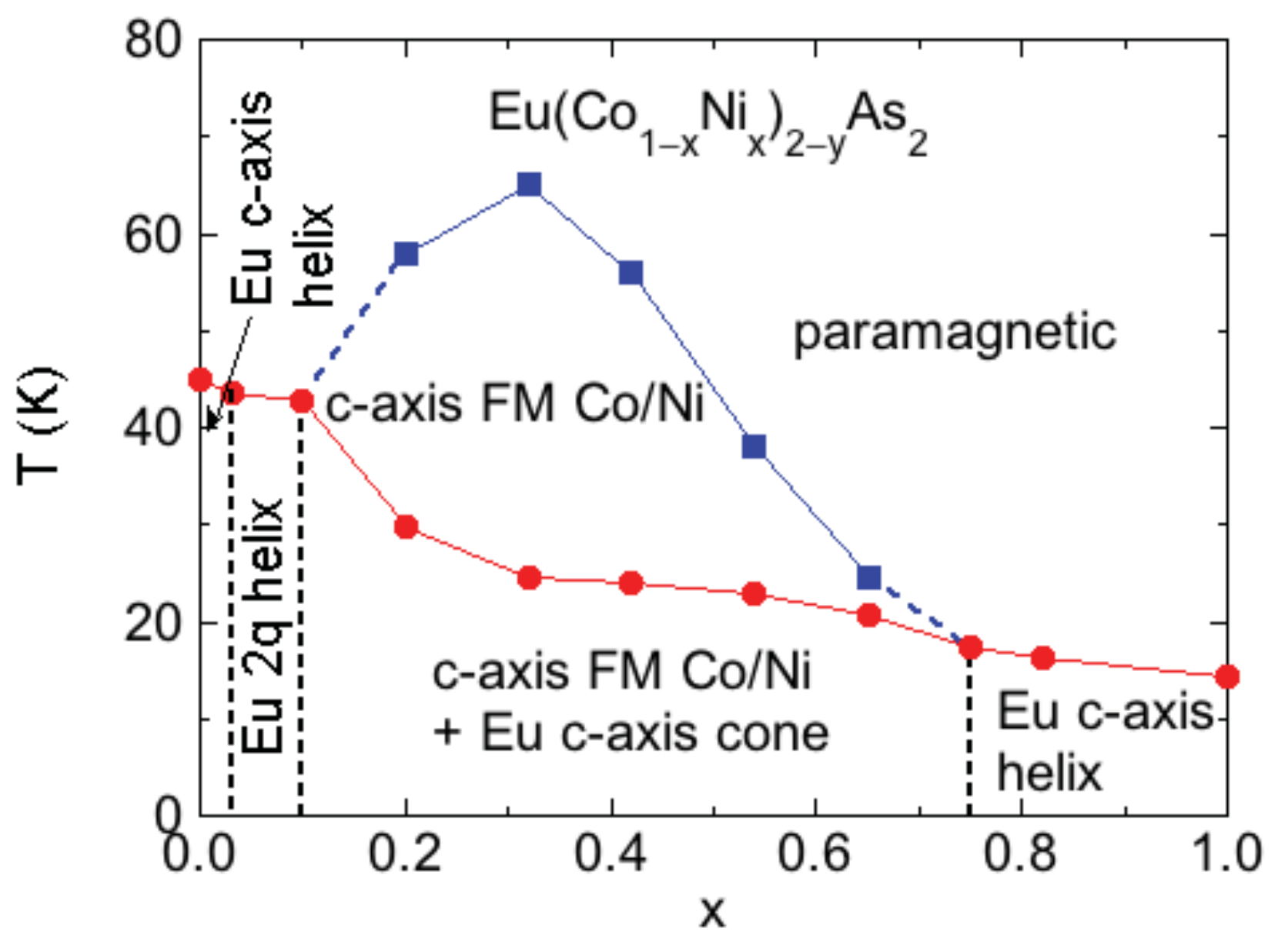}
\caption{Tentative magnetic phase diagram in the $T$-$x$ plane of the \ecna\ system.  The itinerant FM transition temperatures $T_{\rm C\,Co/Ni}$ of the Co/Ni sublattice are given by the filled blue squares, and the AFM helical local-moment transition temperatures $T_{\rm N\,Eu}$ of the Eu spins are denoted by filled red circles.}
\label{Eu(Co,Ni)2As2_T-x_PhaseDiag}
\end{figure}

Multiple magnetic phase transitions have been identified in the \ecna\ system.  The highest-temperature one is a composition-dependent  itinerant FM ordering between 25 and 60~K associated with the Co/Ni sublattice as clearly revealed from both heat capacity $C_{\rm p}(H,T)$ and magnetic susceptibility $\chi_\alpha(T)$ measurements.  This FM transition is fragile and disappears in a 3~T magnetic field according to the $C_{\rm p}(T)$ measurements.  The composition-dependent magnetic ordering of the Eu spins is clearly evident from the  $\lambda$- or step-shape anomaly in the zero-field $C_{\rm p}(T)$ measurements that is easily tracked versus composition.  Close to the end-point compositions the magnetic structure is a $c$-axis helix as illustrated generically in Fig.~\ref{Fig:helix_cone} for a 60$^\circ$ turn angle $kd$.  In the intermediate composition range $0.20\leq x\leq0.65$, low-field magnetization versus field $M_\alpha(H)$ isotherms demonstrate that the Eu magnetic structure has a rather large ferromagnetic (FM) component along the $c$~axis (0.6 to $3.5~\mu_{\rm B}$/Eu); combined with the $\chi_\alpha(T)$ data, we suggest that this structure is an Eu $S=7/2$ cone structure, which is a superposition of an $c$-axis helix and a $c$-axis FM component as illustrated in Fig.~\ref{Fig:helix_cone}(b).  In addition, in the narrow composition range $0.03 \lesssim x \lesssim 0.1$ there appears to be a $2q$ helix, which consists of a superposition of a helical structure along the $c$~axis with the moments aligned in the $ab$~plane  and a related helix structure with an axis in the $ab$~plane and the moments aligned in an $ab$-$c$ pane perpendcular to the helix axis.  These results are summarized in the tentative phase diagram in Fig.~\ref{Eu(Co,Ni)2As2_T-x_PhaseDiag}.

The most surprising result from this work is the discovery of multiple transition temperatures around the Eu transition temperature that are observed in the $^{151}$Eu M\"ossbauer measurements of \ecna\ with $x=0$, 0.2, and 0.65,  which are not included in Fig.~\ref{Eu(Co,Ni)2As2_T-x_PhaseDiag}.  These measurements indicate that ordering of the Eu moments proceeds via an incommensurate sine amplitude-modulated structure with additional transition temperatures associated with this effect.  A fit of the ordered moment versus temperature in \ecaa\ from the literature suggests that there might be a systematic deviation from the prediction of molecular-field theory on approaching $T_{\rm N\,Eu}$ from below, which is expected to be accurate for the large spin $S=7/2$.  It would be interesting to carry out higher-resolution measurements of the magnetic order parameter versus  temperature for $x=0$, and also see if there is a temperature dependence to the turn angle of the helix at this composition.

%\clearpage

\acknowledgments

This research was supported by the U.S. Department of Energy, Office of Basic Energy Sciences, Division of Materials Sciences and Engineering.  Ames Laboratory is operated for the U.S. Department of Energy by Iowa State University under Contract No.~DE-AC02-07CH11358.  Financial support for this work was provided by Fonds Qu\'eb\'ecois de la Recherche sur la Nature et les Technologies.  Much of the M\"ossbauer work was carried out while DHR was on sabbatical leave at Iowa State University and Ames Laboratory and their generous support (again under contract No.~DE-AC02-07CH11358) during this visit is gratefully acknowledged.

\end{document}